\documentclass[acmtog,screen]{./acmart}

\usepackage{graphicx,subcaption}
\usepackage{float}  %

\usepackage[normalem]{ulem}
\usepackage{color,soul}
\usepackage{xcolor}
\usepackage{pifont}
\usepackage{multirow}
\usepackage[linesnumbered,ruled,vlined,noend]{algorithm2e} %
\SetArgSty{textup}

\SetAlFnt{\small}
\SetAlCapFnt{\small}
\SetAlCapNameFnt{\small}
\SetAlCapHSkip{0pt}

\renewcommand\footnotetextcopyrightpermission[1]{}
\setcopyright{none}      %

\usepackage{siunitx}

\PassOptionsToPackage{pdfusetitle=false,pdfcreator={},pdfproducer={}}{hyperref}

\newcommand{\ignorethis}[1]{}

\newcommand{\sectnum    } [1] {\ref{#1}}

\newcommand{\fignum     } [1] {\ref{#1}}
\newcommand{\eqnnum     } [1] {\mbox{(\ref{#1})}}

\newcommand{\sect       } [1] {Section~\sectnum{#1}}

\newcommand{\fig        } [1] {Fig.~\fignum{#1}}
\newcommand{\figs       } [1] {Figures~\fignum{#1}}
\newcommand{\eqn        } [1] {equation~\eqnnum{#1}}

\newcommand{\etal       }     {{et~al.}}

\newcommand{\eg         }     {{e.g.~}}

\newcommand{\Reals      }     {{\textrm{I\kern-0.18em R}}}

\newcommand{\fourth     }     {\ensuremath{\frac{1}{4}}}

\newcommand{\change     } [1] {\mbox{{\footnotesize $\Delta$} \kern-3pt}#1}

\newcommand{\norm       } [1] {{\| #1 \|}}

\definecolor{darkred}{rgb}{0.7,0.1,0.1}
\definecolor{darkgreen}{rgb}{0.1,0.5,0.1}
\definecolor{cyan}{rgb}{0.7,0.0,0.7}
\definecolor{otherblue}{rgb}{0.1,0.4,0.8}
\definecolor{maroon}{rgb}{0.76,.13,.28}
\definecolor{burntorange}{rgb}{0.81,.33,0}

\ifdefined\ShowNotes
  \newcommand{\colornote}[3]{{\color{#1}\bf{#2 #3}\normalfont}}
\else
  \newcommand{\colornote}[3]{}
\fi

\newcommand{\namedfn}[1]{\ensuremath{\boldsymbol{\mathtt{#1}}}}
\newcommand{\methodname}{BinocMesher}
\newcommand{\treename}{binary-octree}
\newcommand{\Treename}{Binary-octree}
\newcommand{\TreeName}{Binary-Octree}
\newcommand{\treenames}{binary-octrees}

\newcommand\sbullet[1][.8]{ \mathbin{\vcenter{\hbox{\scalebox{#1}{$\bullet$}}}}}

\newcommand{\pullup}[1]{\vspace{#1}}
\newcommand{\pullupone}{\pullup{-1mm}}
\newcommand{\pulluptwo}{\pullup{-2mm}}

\newcommand{\boldparfirst}[1]{\noindent\textbf{#1}}
\newcommand{\boldpar}[1]{\vspace{1mm}\boldparfirst{#1}}
\newcommand{\bulletpar}[1]{\boldpar{$\sbullet$~\emph{#1}}}
\newcommand{\step}[2]{{\boldpar{Step~{#1}: {#2}.}}}

\newcommand{\node}{\ensuremath{\mathcal{N}}}

\newcommand{\diam}{\ensuremath{\mathcal{D}}}

\newcommand{\diamN}{\ensuremath{\diam_{\node}}}

\newcommand{\diamIN}{\ensuremath{\diam_{\node}^{(i)}}}

\newcommand{\deltat}{\ensuremath{{\delta}_t}}

\newcommand{\Si}{\ensuremath{\mathbf{S}_i}}
\newcommand{\Sim}{\ensuremath{\mathbf{S}_{i-1}}}
\newcommand{\Sip}{\ensuremath{\mathbf{S}_{i+1}}}

\newcommand{\sizecap}{\ensuremath{\mathcal{C}}}

\newcommand{\group}[1]{\ensuremath{\mathcal{G}_{#1}}}

\newcommand{\colorgroup}[2]{\colorbox[HTML]{#1}{\group{\textbf{#2}}}}

\newcommand{\colortreefn}[3]{\colorbox[HTML]{#1}{\namedfn{#2}(\group{\textbf{#3}})}}

\newcommand{\groupNodeS}{\colorgroup{f4cccc}{s}}
\newcommand{\groupNodeSm}{\colorgroup{c9daf8}{s-1}}
\newcommand{\groupNodeSp}{\colorgroup{d9ead3}{s+1}}
\newcommand{\subtreeS}{\colortreefn{fff2cc}{descendants}{s}}
\newcommand{\ancestorsS}{\colortreefn{eeeeee}{ancestors}{s}}
\newcommand{\ancestorsSm}{\colortreefn{eeeeee}{ancestors}{s-1}}
\newcommand{\ancestorsSp}{\colortreefn{eeeeee}{ancestors}{s+1}}
\newcommand{\rightbranchSm}{\colortreefn{c9daf8}{right\_branch}{s-1}}
\newcommand{\leftbranchSp}{\colortreefn{d9ead3}{left\_branch}{s+1}}

\usepackage{soul,xcolor}

\definecolor{h8}{HTML}{d3c5ed} %
\definecolor{h7}{HTML}{deffa0} %
\definecolor{h9}{HTML}{f6e1fa} %
\definecolor{h1}{HTML}{f2c4c4} %
\definecolor{h2}{HTML}{fcd8ae} %
\definecolor{h3}{HTML}{f5c1b5} %
\definecolor{h4}{HTML}{d5e2b0} %
\definecolor{h5}{HTML}{b7e7ed} %
\definecolor{h6}{HTML}{bdc8db} %

\newcommand{\highlight}[2]{#2}

\newcommand{\imagefolder}{images/}

\acmJournal{TOG}

\begin{document}
\title{Temporally Smooth Mesh Extraction for Procedural Scenes \\ with Long-Range Camera Trajectories using Spacetime Octrees}

\author{Zeyu Ma}
\affiliation{
  \institution{Princeton University}
  \country{USA}
}
\author{Adam Finkelstein}
\affiliation{
  \institution{Princeton University}
  \country{USA}
}
\author{Jia Deng}
\affiliation{
  \institution{Princeton University}
  \country{USA}
}

\begin{abstract}

The procedural occupancy function is a flexible and compact representation for creating 3D scenes. 
For rasterization and other tasks, it is often necessary to extract a mesh that represents the shape. 
Unbounded scenes with long-range camera trajectories, such as flying through a forest, pose a unique challenge for mesh extraction. 
A single static mesh representing all the geometric detail necessary for the full camera path can be prohibitively large.
Therefore, independent meshes can be extracted for different camera views, but this approach may lead to popping artifacts during transitions.
We propose a temporally coherent method for extracting meshes suitable for long-range camera trajectories in unbounded scenes represented by an occupancy function. 
The key idea is to perform 4D mesh extraction using a new spacetime tree structure called a \treename. 
Experiments show that, compared to existing baseline methods, our method offers superior visual consistency at a comparable cost.
The code and the supplementary video for this paper are available at \url{https://github.com/princeton-vl/BinocMesher}.

\end{abstract}



\begin{teaserfigure}
\captionsetup[subfigure]{justification=centering}
\includegraphics[width=\linewidth]{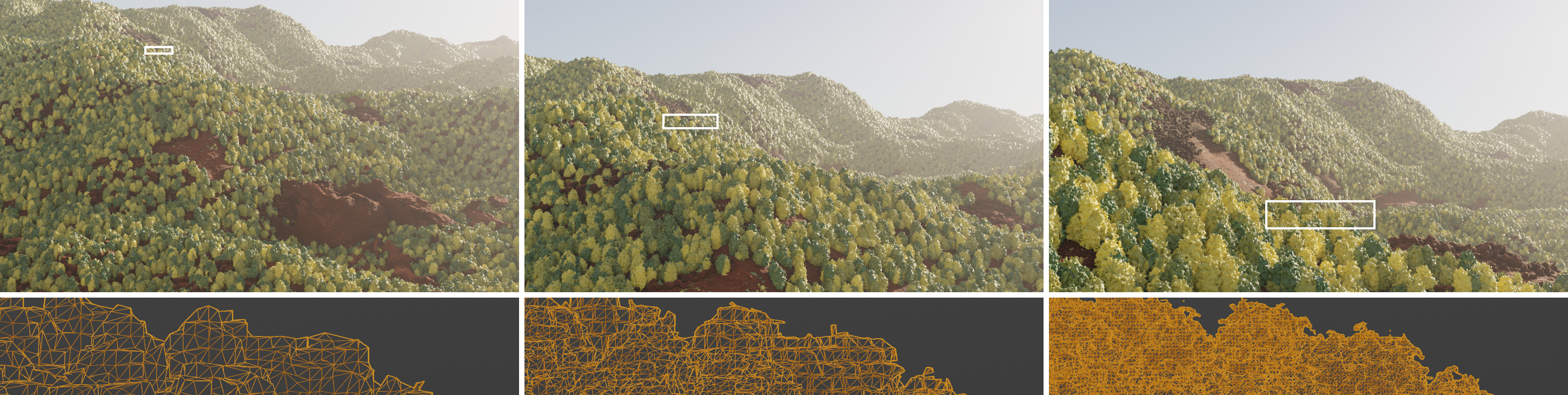}
  \caption{We propose \methodname{} to extract temporally smooth 3D meshes from occupancy functions by slicing a 4D mesh. The first row shows sample images rendered from the meshes. The second row shows how the 3D mesh in the white boxed area gradually gains geometric detail as the camera moves closer to it.}
  \label{fig:teaser}
\vspace{3mm}
\end{teaserfigure}

\maketitle

\section{Introduction}

Procedural approaches including occupancy functions are widely used in creating 3D content ranging from movies and games~\cite{ebert2003texturing, smelik2014survey,perlin2002hyper}, to synthetic datasets for 3D vision~\cite{wrenninge2018synscapes,raistrick2023infinite,greff2022kubric}. Compared with alternatives such as scenes scanned from the real-world or modeled by artists, procedural approaches have the advantage of expressing complex and unbounded geometry with compact mathematical rules. One possible way to render procedural scenes represented by occupancy functions involves ray-marching. However, ray-marching into an unbounded occupancy function requires small step sizes to avoid missing thin structures; and even so, adjacent pixels can produce inconsistent results. 
Therefore, it is often advantageous to extract and render triangle meshes. Moreover, meshes also offer compatibility with a broad range of rendering pipelines, ease of texturing, and familiarity for artists.

We address view-dependent mesh extraction for unbounded scenes with  \highlight{h8}{pre-defined} long-range camera trajectories, such as flying through a vast forest or mountainscape (\fig{fig:teaser}). Given a static occupancy function $f: \mathbb{R}^3 \mapsto \{0,1\}$ and a camera trajectory at timestamps $\{t_i, i=1, 2, \dots, L\}$, the goal is to extract a sequence of meshes $\{\mathbf{M}_i\}$. To avoid popping artifacts, these meshes should be temporally smooth. We assume the camera path is \highlight{h8}{known in advance}, as in applications like animation and synthetic data generation.

A naive solution is to extract a single global mesh using methods such as OcMesher~\cite{ma2023view}. However, unbounded scenes with long-range camera trajectories can result in meshes so large as to exceed the capacity of extraction algorithms and rendering engines. 
To address this challenge, one could extract multiple meshes, one for each subsequence along the full trajectory, using OcMesher or other methods~\cite{scholz2015real,raistrick2023infinite}. However, severe popping can occur without an excessively high mesh resolution or a way to smoothly transition between the meshes.
The \emph{progressive meshes} approach of Hoppe~\shortcite{hoppe1996progressive}
affords a smooth transition called a \emph{geomorph} between meshes of differing resolution.
While this multi-resolution approach can reduce rendering cost, it begins by decimating the full-resolution mesh.
Splitting a huge terrain into blocks and processing each block as a separate progressive mesh helps limit memory needed, but also creates challenges at block boundaries~\cite{hoppe1997view}.

\begin{figure}[t]
\captionsetup[subfigure]{justification=centering}
    \begin{subfigure}[t]{.22\linewidth}
    \includegraphics[width=.86\linewidth]{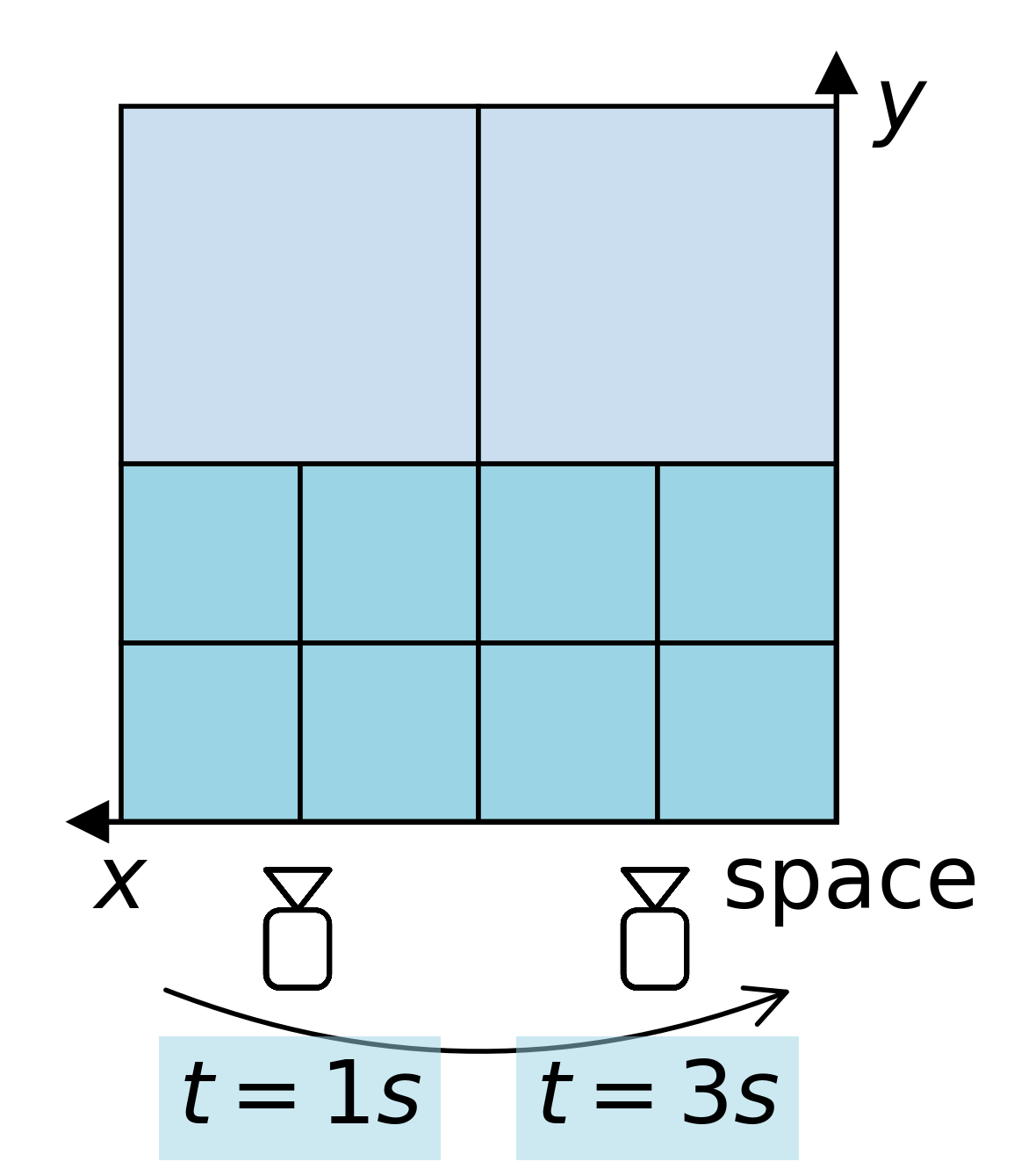}
    \caption{Single octree}
  \end{subfigure}
  \hfill
    \begin{subfigure}[t]{.44\linewidth}
    \centering
    \includegraphics[width=.43\linewidth]{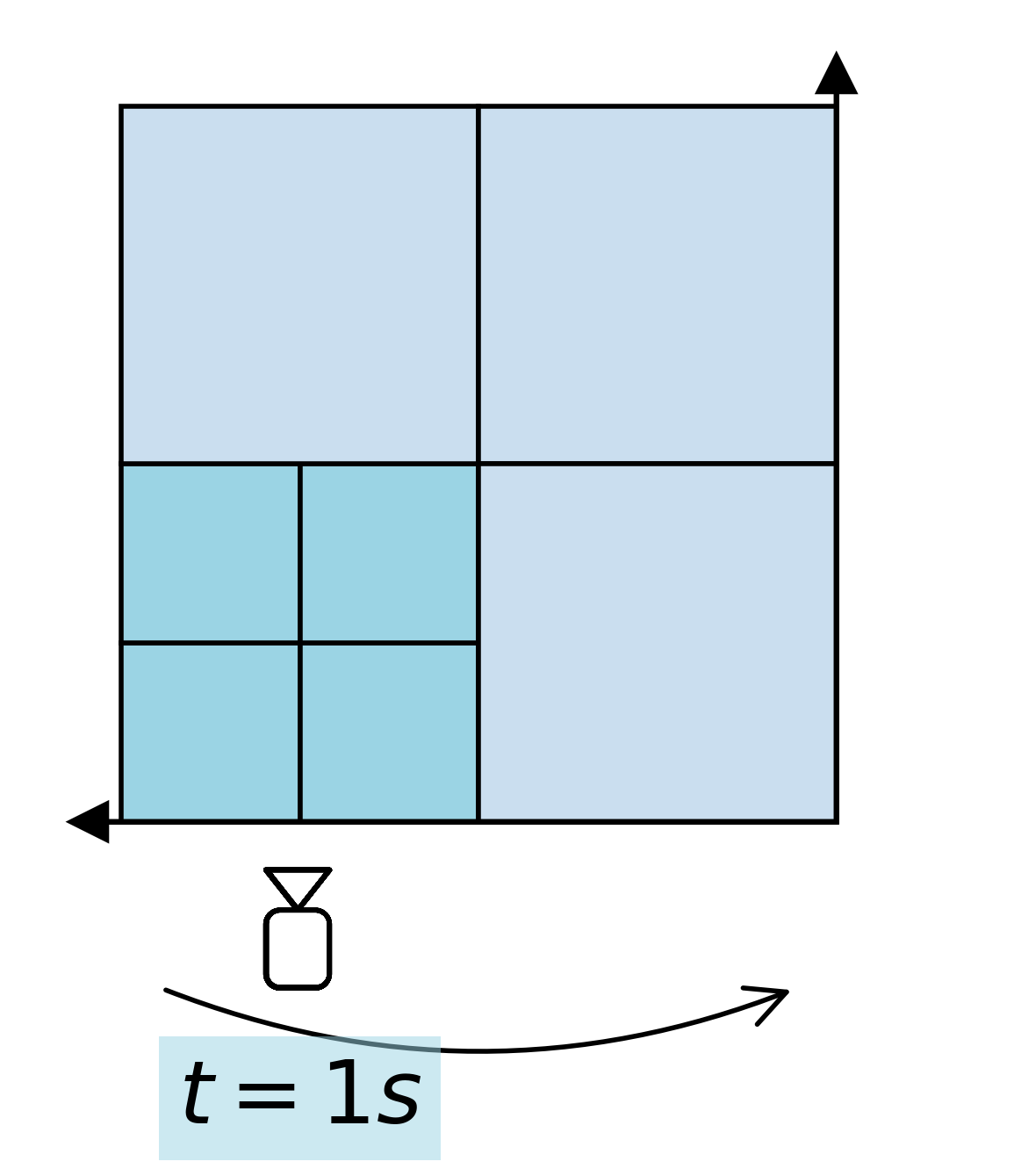}
    \includegraphics[width=.43\linewidth]{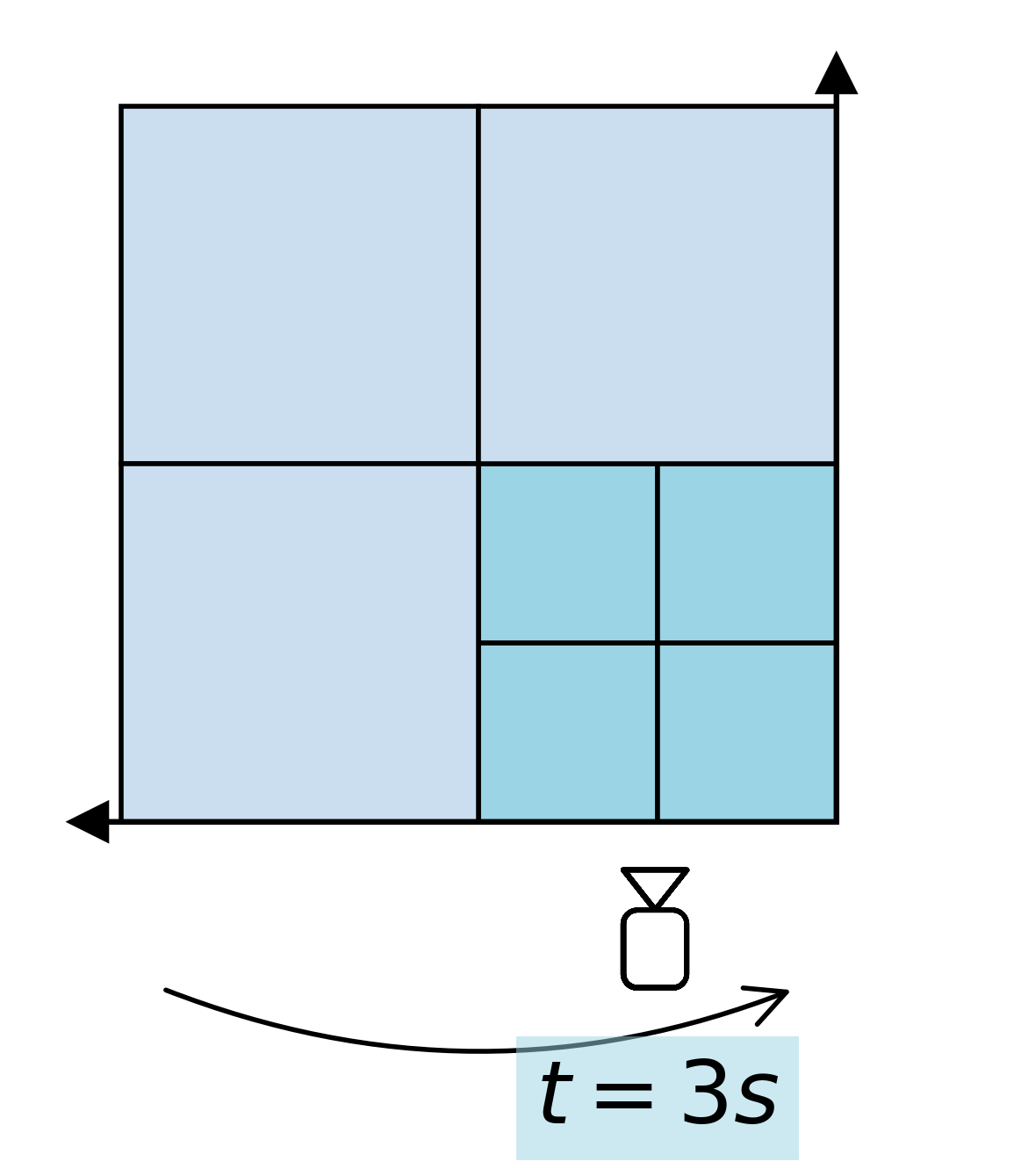}
    \caption{Differing octrees}
  \end{subfigure}
  \hfill
    \begin{subfigure}[t]{.31\linewidth}
    \includegraphics[width=0.9\linewidth]{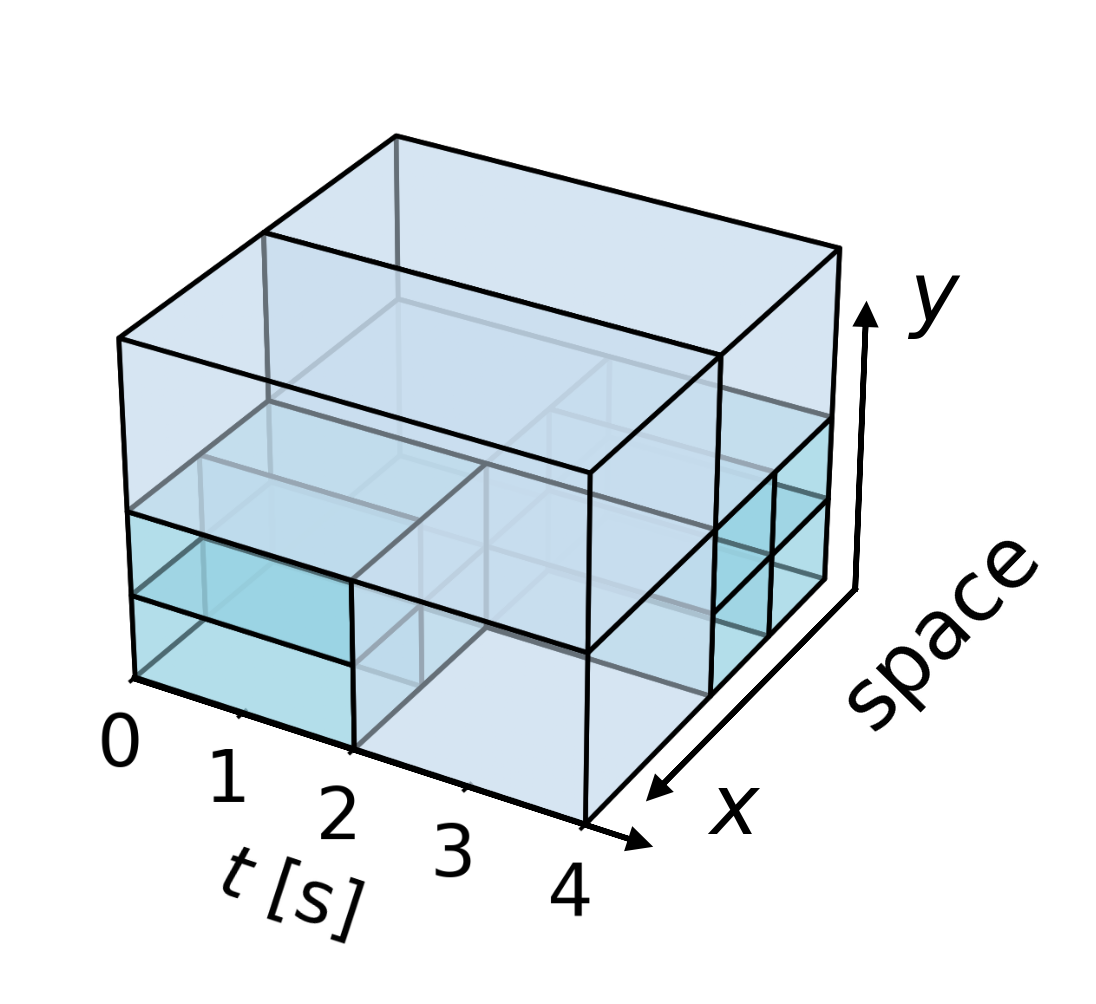}
    \caption{A \treename}
  \end{subfigure}
  \pullupone
  \caption{Several options for a camera-dependent tree structures. The $x$ and $y$ spatial axes are labeled, while $z$ is not shown. The binary-octree (c) supports the two spatial splits shown in (b) at different times.}
  \label{fig:hyper}
  \pulluptwo
\end{figure}

This paper introduces \textbf{\methodname}, a spacetime tree-based mesh extraction algorithm.
Instead of constructing a global octree~\cite{meagher1982geometric} or multiple octrees in the 3D space,  we partition the 4D spacetime into hypercubes, as shown in \fig{fig:hyper}. 
By slicing this 4D structure in the time dimension, we can extract static meshes at different times.
To avoid popping artifacts, it is crucial to perform this slicing task in a way that provides temporal coherence in the resulting meshes.
Ponchio and Hormann~\shortcite{ponchio2008interactive} observe (for uniform grids) that interpolation between two 3D meshes is equivalent to slicing a \highlight{h7}{\textit{4D mesh}, i.e., a polyhedral mesh embedded in 4D space-time}.
We extend this approach to multiresolution grids wherein we need to address the challenge of slicing neighboring regions of differing resolution.
We introduce a new tree structure called \textbf{\treename}, where each non-leaf node either splits in the time dimension (into two children) or splits in the spatial dimensions (into eight children). 
In principle one could use a 4D hyper-octree which has 16 children at internal nodes~\cite{puech1985quadtrees}, but this tree requires uniform spatial refinements at all times.
Instead the \treename\ combines aspects of a K-D tree~\cite{bentley1975multidimensional} and a $2^D$ tree, with each temporal split enabling different spatial splits in the two child nodes (\fig{fig:hyper}c). This approach is more memory-efficient for long camera trajectories.

\begin{figure}[t]
\captionsetup[subfigure]{justification=centering}
    \begin{subfigure}[t]{.24\linewidth}
    \centering\includegraphics[width=\linewidth]{\imagefolder oth3.png}
    \caption{ \Treename{} \protect\\ construction}
  \end{subfigure}
  \hfill
      \begin{subfigure}[t]{.48\linewidth}
    \centering
    \includegraphics[width=.48\linewidth]{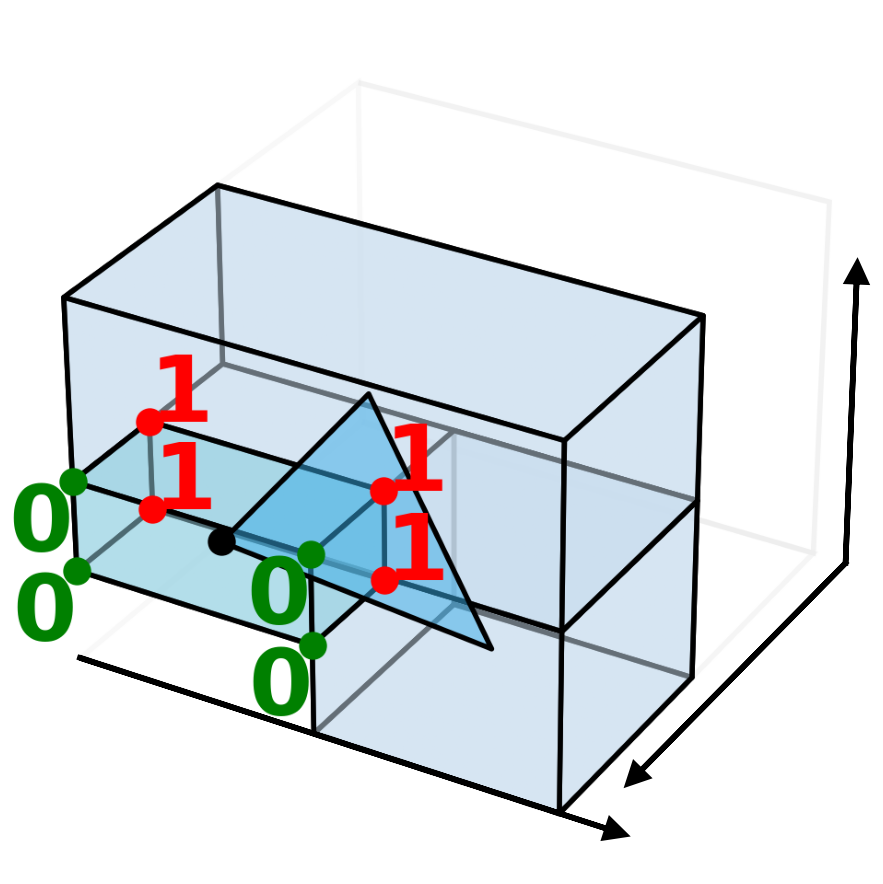}
    \includegraphics[width=.48\linewidth]{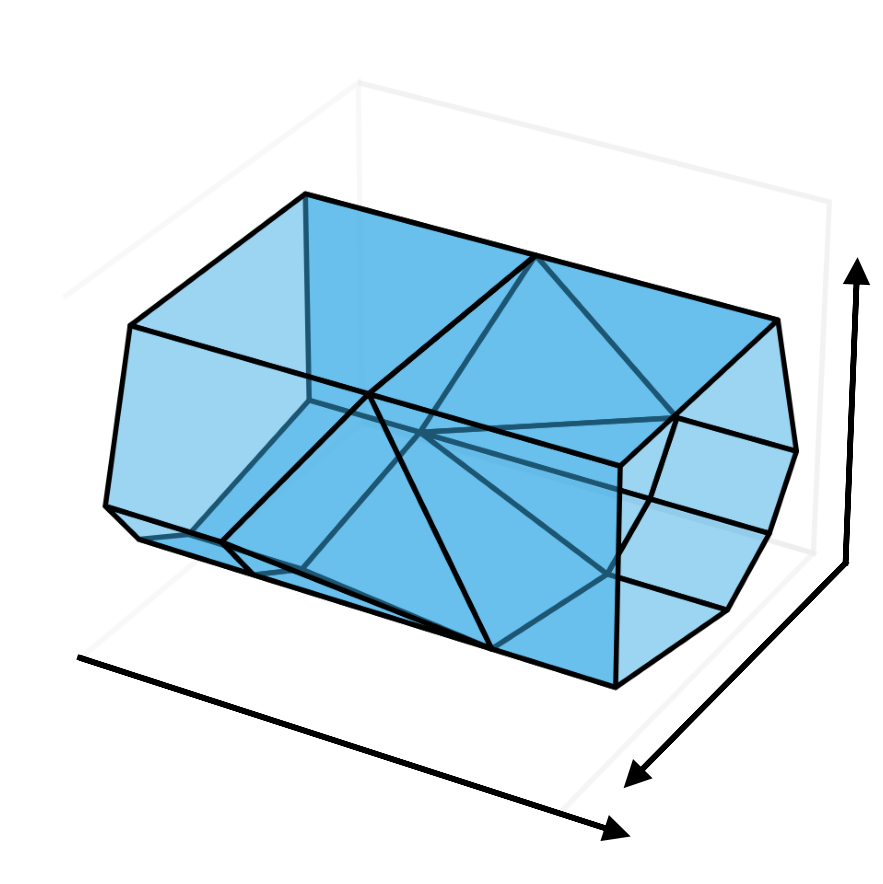}
    \caption{4D mesh extraction \protect\\ with dual contouring}
  \end{subfigure}
  \hfill
      \begin{subfigure}[t]{.24\linewidth}
    \centering\includegraphics[width=.96\linewidth]{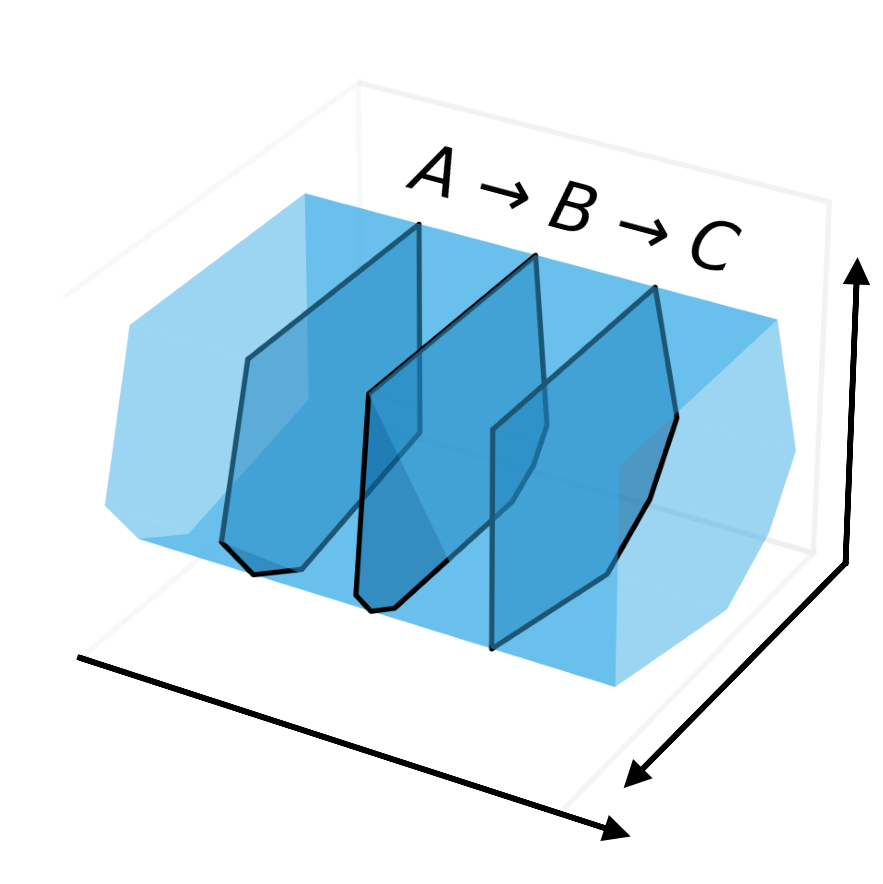}
    \caption{3D meshes \protect\\ as slices}
  \end{subfigure}
  \caption{Overview of \methodname{}}
  \label{fig:pip}
\end{figure}

Our method proceeds in three steps, as illustrated in \fig{fig:pip}. First, we construct a \treename. Next, we extract a 4D mesh from the occupancy function evaluated at the spacetime-tree ``corners'' using dual contouring~\cite{ju2002dual}. Finally, we slice the 4D mesh to generate 3D meshes at different timestamps. As the camera moves along its path, some polygons merge into a low-resolution shape, while other polygons split to provide more detail, as shown in \fig{fig:teaser}.
We control the visual consistency of the resulting 3D meshes by the size of the hypercube in the \treename. For example, if each hypercube spans at least two seconds in \fig{fig:pip}(a), then it must takes at least two seconds to transform between meshes of adjacent levels of detail (LODs), \eg meshes \textit{A-C} in \fig{fig:pip}(c). We call this duration the \emph{transition control parameter} $\deltat$, which balances between the goals of memory efficiency and temporal coherence.

Our contributions may be summarized as follows. We introduce a spacetime-octree-based mesh extraction algorithm to produce temporally smooth meshes for long-range camera trajectories. Second, we optimize the algorithm with efficient designs to minimize memory and computational costs. We also describe experiments showing that our method offers better visual consistency at a comparable cost to baseline methods.

\section{Related Work}

\boldparfirst{Mesh Extraction. }
The \emph{marching cubes} method of
Lorensen and Cline~\shortcite{lorensen1998marching} extracts a mesh as an isosurface of an implicit function,
and multiresolution methods support varying LOD for large scenes.
One class of methods considers alternate grids using spherical coordinates~\cite{raistrick2023infinite}, or constructing multiresolution grids of polyhedra~\cite{zhou1997multiresolution,gerstner2000topology} or tetrahedra~\cite{pascucci2002efficient,weber2003extraction}.
Dual contouring offers an alternate approach that operates directly on arbitrary multiresolution grids, including spherical grids ~\cite{jang2022egocentric} and octree grids in arbitrary dimensions~\cite{perry2001kizamu,ju2002dual,ju2006intersection,ma2023view,wenger2013isosurfaces}.
Existing multiresolution mesh extraction methods -- including the work of Jang et al.~\shortcite{jang2022egocentric}, which shares a similar name with ours --  focus on static meshes with fixed LODs. In contrast, \methodname\ offers temporally smooth mesh extraction from scenes with dynamic, camera-dependent LODs.

\boldpar{Geomorphs. }
Given a single mesh of the highest LOD, various approaches produce simplified meshes with smooth transitions, known as geomorphs. Early algorithms~\cite{taylor1994algorithm,lindstrom1996real} address the specific case where the mesh is from a height field. 
Hoppe's~\shortcite{hoppe1996progressive} progressive mesh representation defines a continuous sequence of meshes of varying LOD, enabling efficient geomorphs between any pair. 
Subsequent efforts~\cite{hoppe1997view,sander2006progressive,el2022unreal}, including Nanite in Unreal Engine 5, extended these LOD control methods from a single monotonic scale to view-dependent LOD. 
Applying these methods to large scenes described by procedural occupancy functions requires an intractable intermediate mesh at the highest LOD.

\boldpar{3D Anti-aliasing. }
Beyond geomorphs, various 3D anti-aliasing techniques have been explored. 
Scholz~\etal~\shortcite{scholz2015real} apply a spatially varying low-pass filter to modify the input function, which is limited to signed distance functions (and still admits aliasing). 
Giegl and Wimmer~\shortcite{giegl2007unpopping} apply LOD blending in image space without accounting for 3D geometry. 
Infinigen~\cite{raistrick2023infinite} uses extremely high-resolution meshes in order to supersample the scene, which is an expensive approach in this context.

\boldpar{Spacetime Methods. }
Graphics research has a rich history that considers animated sequences in 4D spacetime for other purposes.
For example, Glassner~\shortcite{glassner1988spacetime} uses \emph{spacetime raytracing} to render animation as a 4D scene.
Cameron~\shortcite{cameron1990collision} consider collision in 3D as a 4D intersection test.
Schmid~\etal~\shortcite{schmid2010programmable} propose a 4D structure to aggregate object motion for rendering motion blur and other effects.
Du~\etal~\shortcite{du2021video} recolor videos by slicing 4D polyhedral palettes.

\section{Preliminaries: Inputs, Octree and OcMesher}
\label{sec:ocmesher}

\begin{figure}[t!]
\centering
\captionsetup[subfigure]{justification=centering}
\begin{subfigure}[t]{.325\linewidth}\centering
    \includegraphics[width=.7\linewidth]{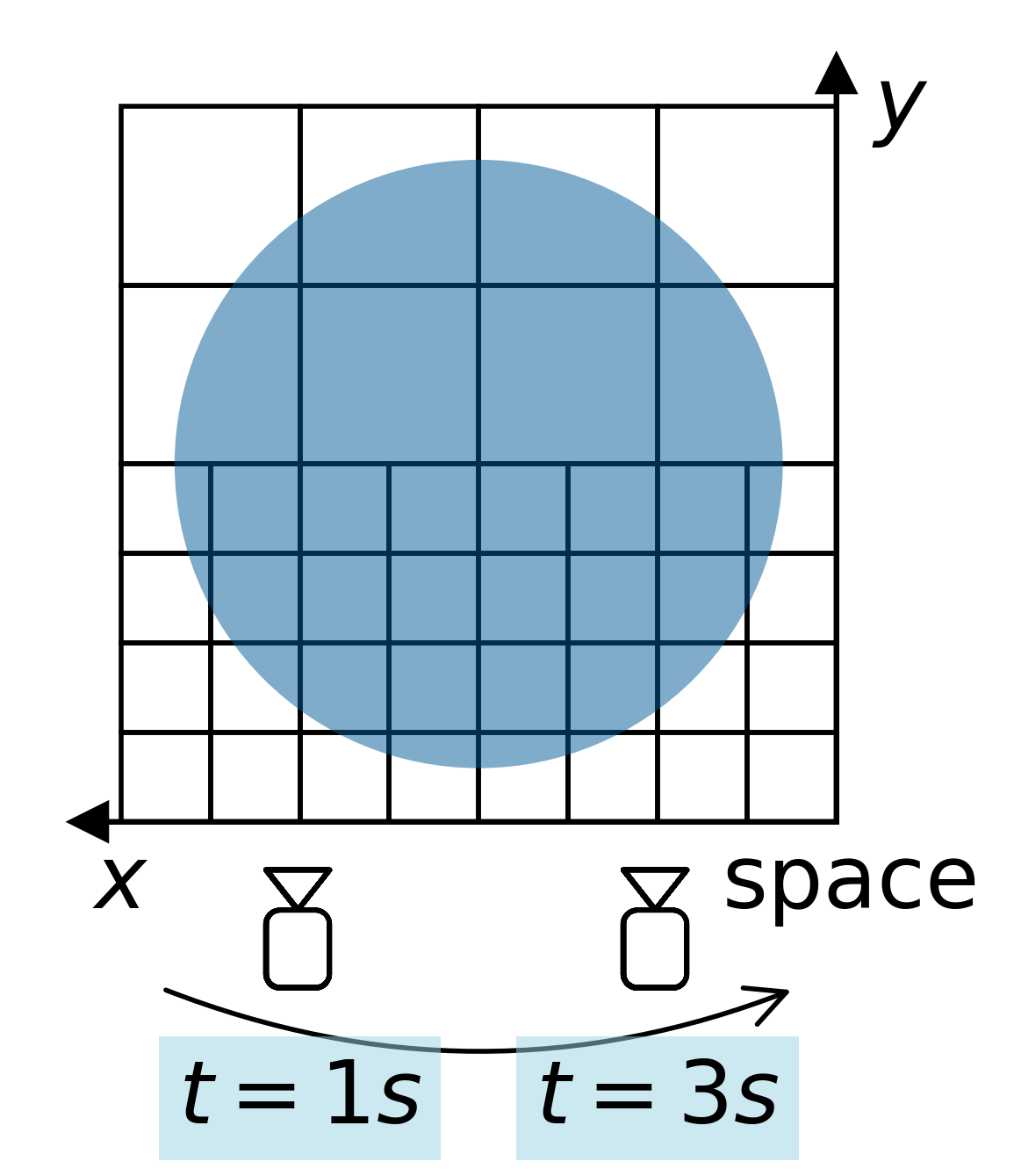}
   \label{fig:cf1}
    \caption*{(a) Coarse octree}
    \end{subfigure}
\begin{subfigure}[t]{.325\linewidth}\centering
    \includegraphics[width=.7\linewidth]{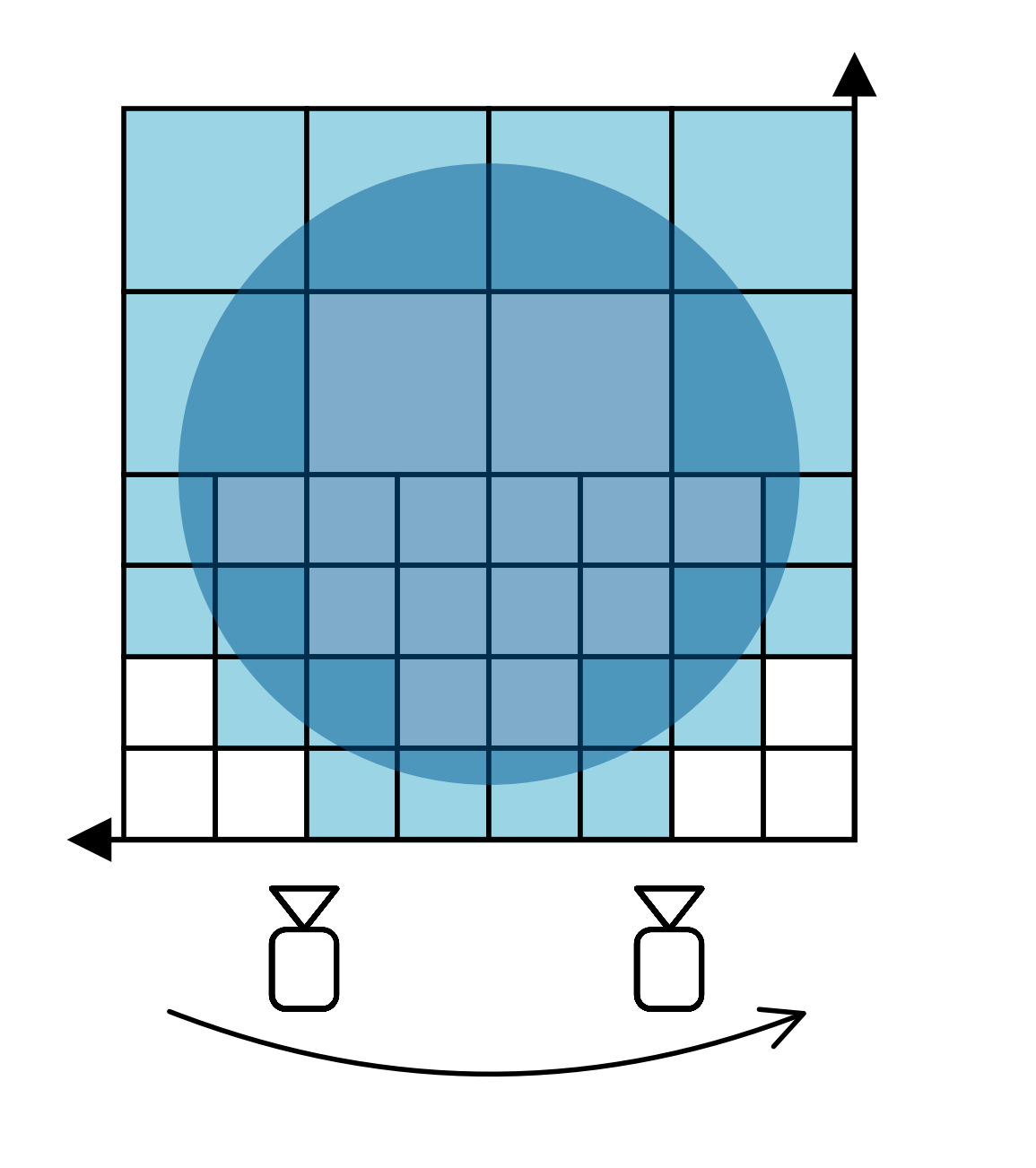}
   \label{fig:cf2}
    \caption*{(b) Intersection test}
    \end{subfigure}
\begin{subfigure}[t]{.325\linewidth}\centering
    \includegraphics[width=.7\linewidth]{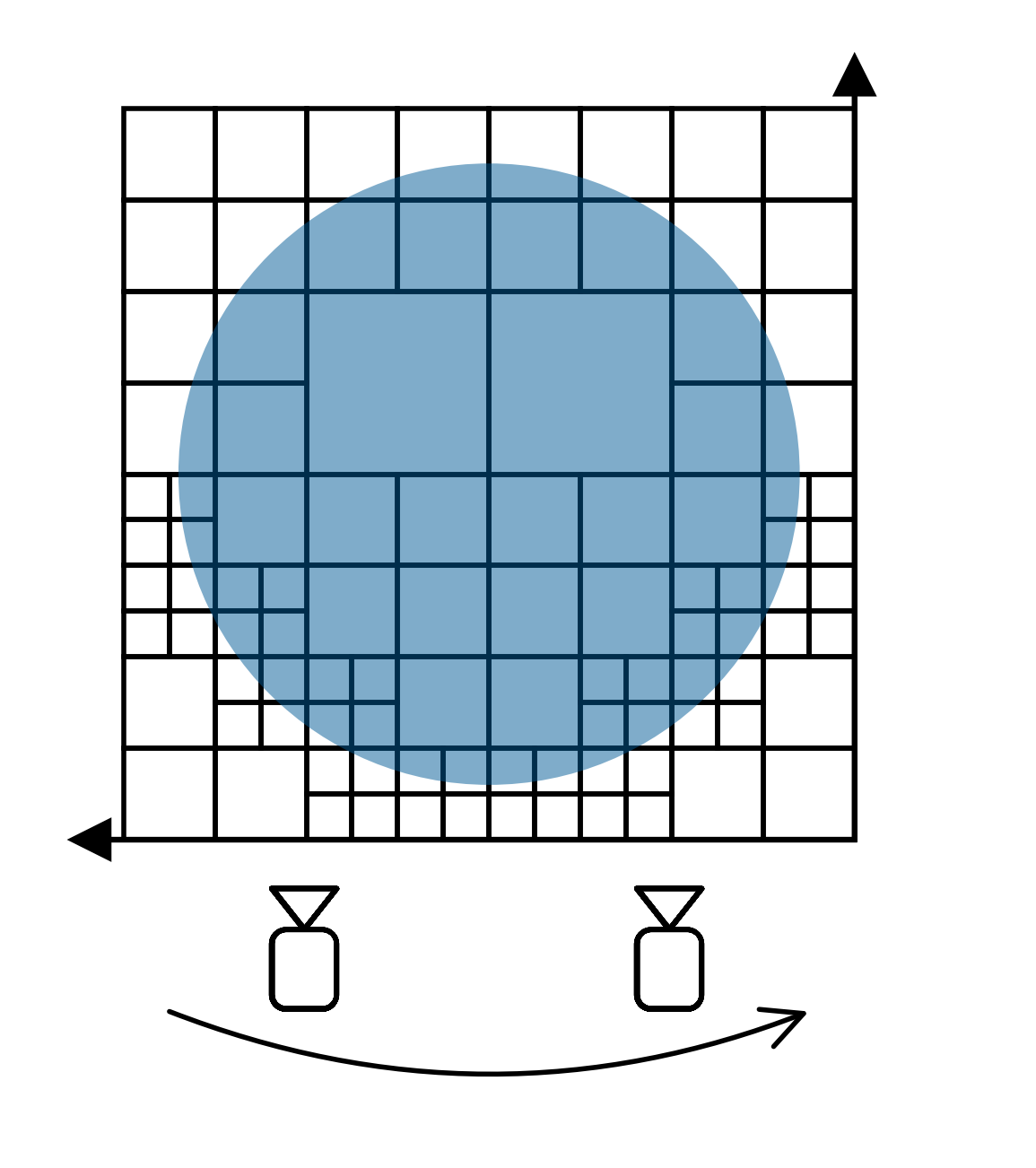}
   \label{fig:cf3}
    \caption*{(c) Refined octree}
    \end{subfigure}
   \caption{OcMesher builds the octree from coarse to fine. Two of the spatial axes $x$ and $y$ are shown. (a) Cubes closer to the camera have larger projected diameter, and are therefore refined. (b) Nodes intersecting the surface are found. (c) Tree is refined around these intersections.
   }
   \label{fig:ocmesher}
\end{figure}

\begin{figure}[b!]
\centering
\captionsetup[subfigure]{justification=centering}
\begin{tabular}{c c c}
    \includegraphics[width=.3\linewidth]{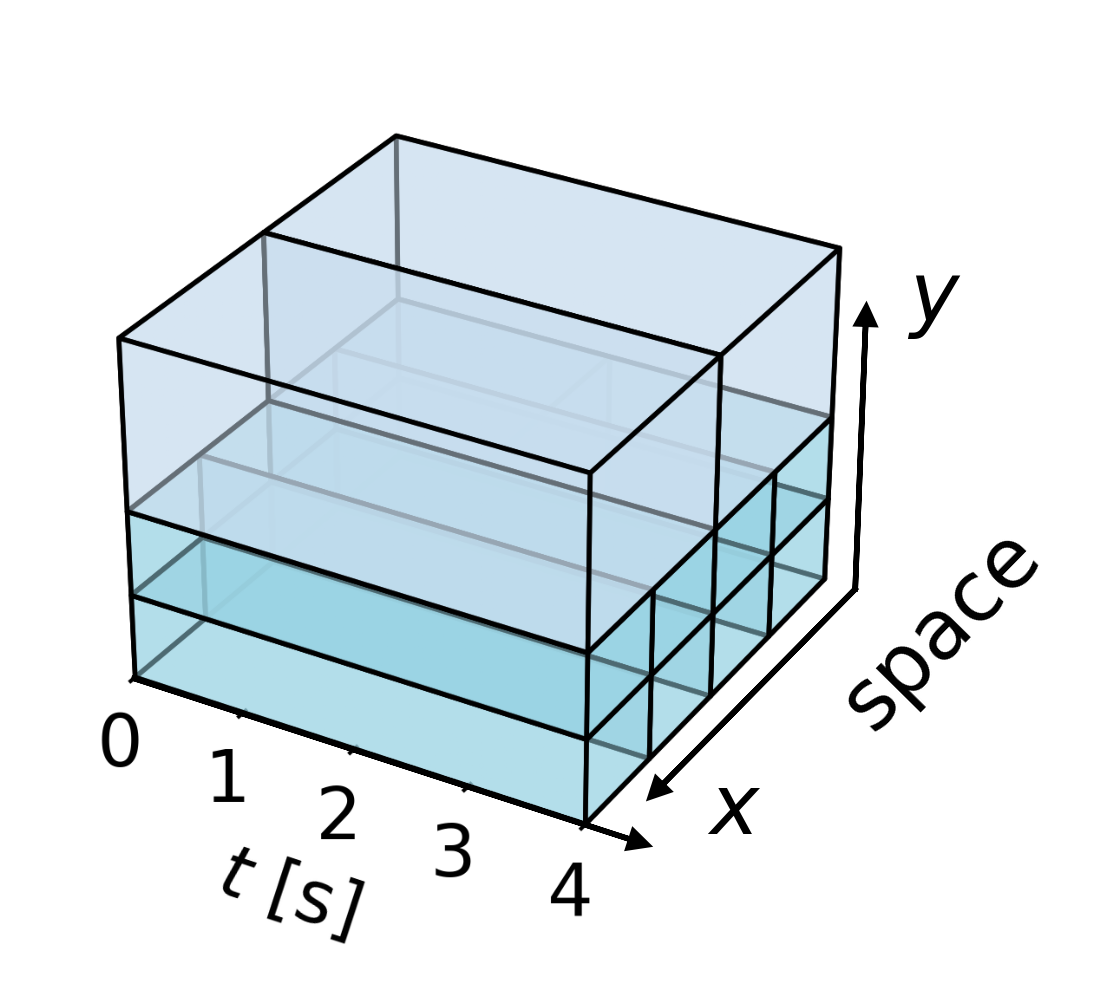} &
    \includegraphics[width=.3\linewidth]{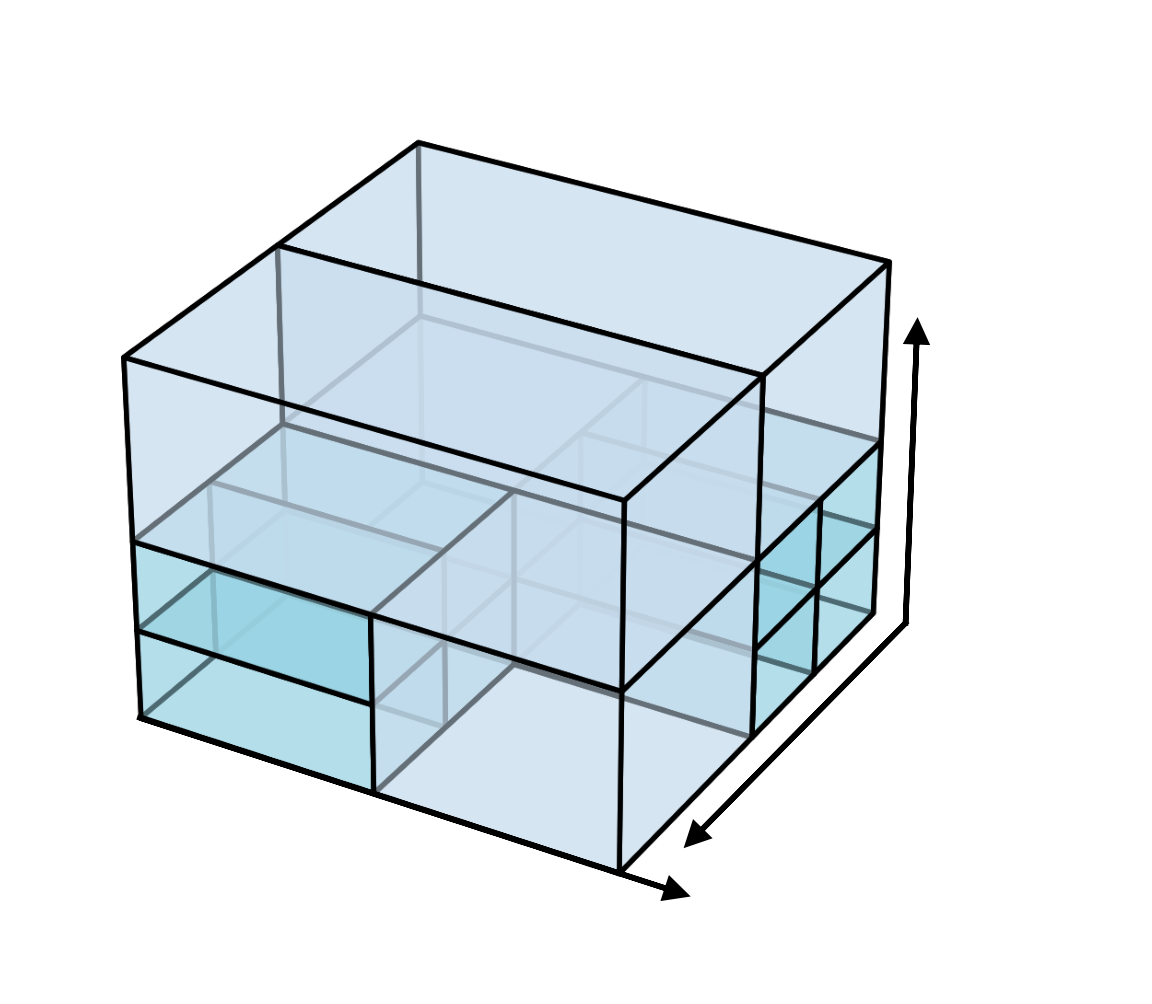} &
    \includegraphics[width=.3\linewidth]{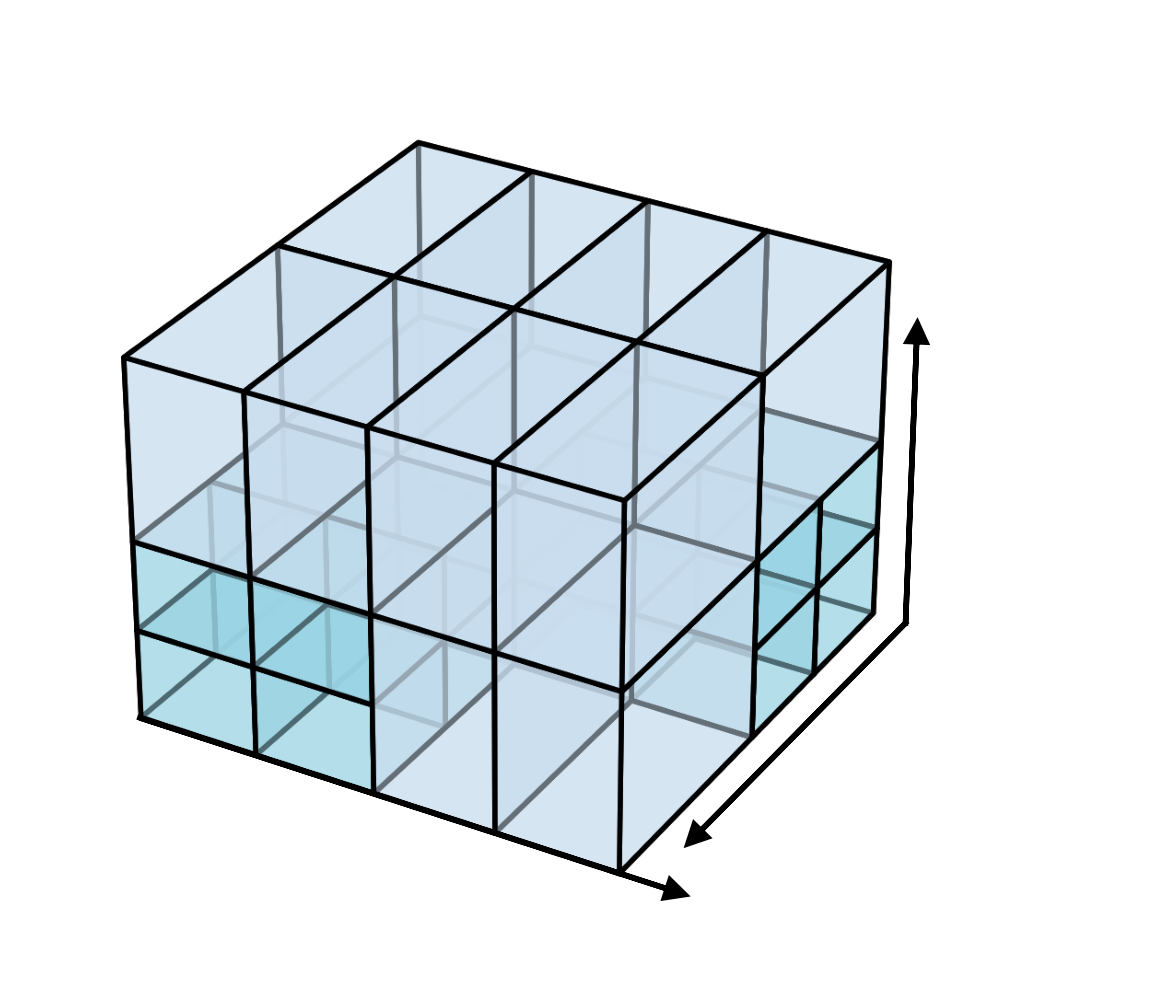}
    \\
    {\small (a) Dense in space} &
    {\small (b) Balanced } &
    {\small (c) Dense in time}
\end{tabular}
  \caption{Possible \treenames\ using different split decisions. A block of 3D spacetime is shown with two spatial axes $x$ and $y$.}
  \label{fig:binoc}
\end{figure}

Our \treename\ construction process builds on the octree refinement approach of OcMesher~\cite{ma2023view}, which generates a minimal number of polygons using a coarse-to-fine approach (\fig{fig:ocmesher}). Therefore, we review the OcMesher framework before diving into the details of the \treename\ and its construction.

\boldpar{Inputs.}
OcMesher takes a pre-defined camera path and an occupancy function $f: \mathbb{R}^3 \mapsto \{0,1\}$ as inputs.
For example, the occupancy function for a sphere is \( \mathbf{1}[x^2+y^2+z^2<r^2] \), where \( \mathbf{1}[\cdot]\) is the indicator function. With more complex techniques, a dense forest is modeled by collections of ellipsoids where the radii are perturbed by adding multi-octave Perlin noise \(g(x, y,z)\) over the \(r^2\) above.

\step{a}{Coarse Octree}
OcMesher first constructs a coarse octree based on the projected angular diameter of each octree node \node\ as seen by the camera at each timestamp $t_i$. The angular diameter is approximated as \diamIN, which is defined to be the ratio of the node size to the distance between the node and the camera:
\begin{equation}
\diamIN = S_{\node} / \norm{x_i-X_{\node}}, i=1,2,\dots,L
\label{eq:dia}
\end{equation}

\noindent where $x_i$ is the location of the camera at time $t_i$,
$X_{\node}$ is the location of the cube center, and
$S_{\node}$ is its side length.

A max-priority queue is initialized with a root node that is large enough to cover the whole scene. Throughout the splitting process, each queue entry corresponds to a single leaf node in the tree, prioritized by its maximum projected diameter over the entire trajectory:
\begin{equation}
{\diamN} = \max_{1\le i \le L} \diamIN
\label{eq:maxdia}
\end{equation}
\noindent
The algorithm iteratively extracts the top node, splits it into eight children, and re-inserts them \highlight{h9}{into the queue until either the maximum diameter falls below a coarse threshold ($\diamN < \hat D_1$) or the total number of nodes reaches a cap. The priority queue allows subdividing the largest nodes.} Details appear in Sec.~\ref{sec:eff}.

\step{b}{Intersection Test}
OcMesher next identifies nodes intersecting the isosurface by evaluating the occupancy function at the corners of each node, seeking nodes that straddle ``inside/outside''.

\step{c}{Refined Octree}
OcMesher finally splits all nodes intersecting the surface, similar to Step (a), but with a smaller threshold.

\section{Method: \TreeName\ and \methodname}
\label{sec:binoc}

This section provides an overview of how to extend the 3D approach of OcMesher to 4D, by splitting in the time dimension and thereby allowing for different spatial subdivisions at different times in a \treename. 
More details regarding efficient implementation are described in \sect{sec:details}.

\subsection{\TreeName}

Internal \treename\ nodes split in one of two cases:

\bulletpar{Spatial Splits:} The node contains eight children that uniformly partition the spatial range and span the same temporal range.

\bulletpar{Temporal Splits:} The node contains two children that span the same spatial range but split the temporal range into halves.

\fig{fig:binoc} shows a block of 3D spacetime, with only two spatial axes. There are many possible \treenames, based on different splitting decisions. In Option~(a) there are no temporal splits, which reduces to the octree solution and produces a dense spatial splitting. Option~(c) shows
dense temporal splits with too many tree nodes. Option~(b) strikes balance between the 3D mesh size and the tree size.
To achieve such a balance, we propose the following goals for deciding when to apply temporal splits:

\bulletpar{Minimal Temporal Splits:} We apply temporal splits only when needed to allow for differing spatial splits in the two children.

\bulletpar{Temporal Coherence:} A node should span at least a minimum duration set by the transition control parameter $\deltat$. We should avoid a temporal split that produces nodes of duration $<\deltat$.

\begin{figure}[b]

\captionsetup[subfigure]{justification=centering}
      \begin{subfigure}[t]{.48\linewidth}
    \centering\includegraphics[width=\linewidth]{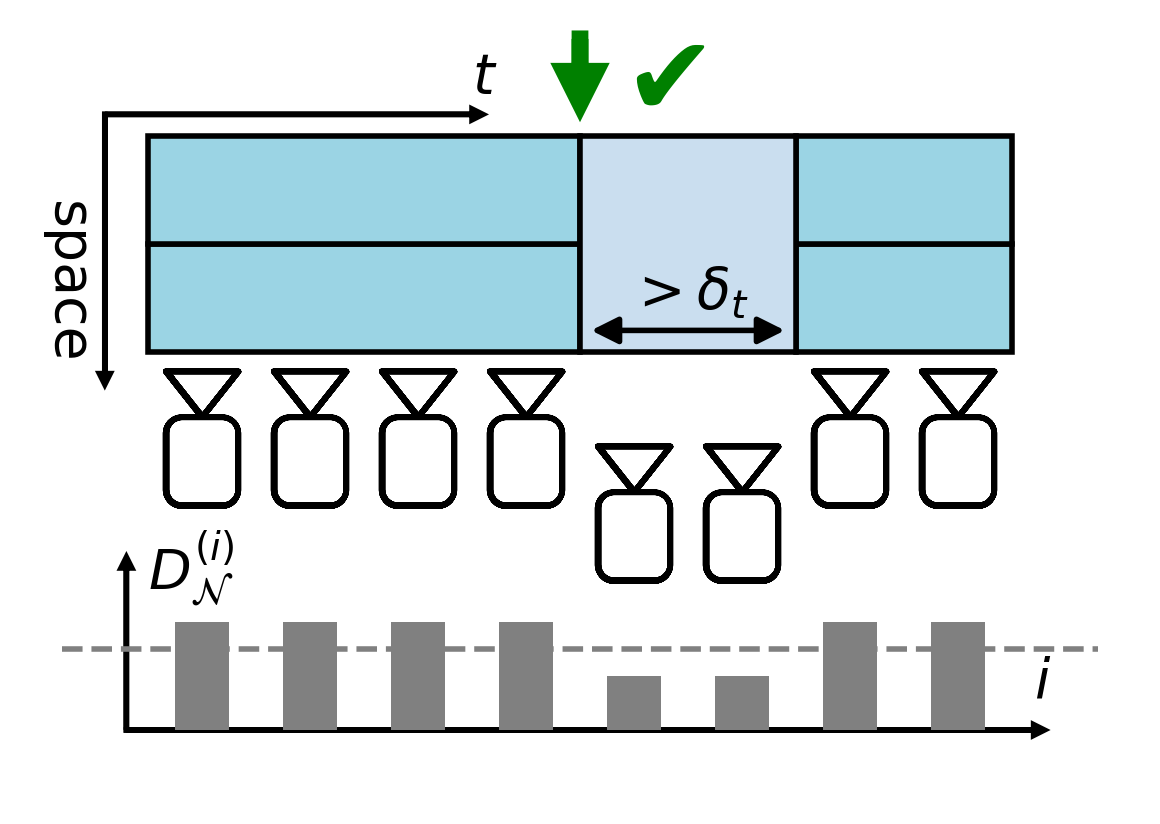}
    \caption{Small diameter subsequence \\
    of duration $> \deltat$ permits splitting. }
    \label{fig:trule1}
  \end{subfigure}
    \begin{subfigure}[t]{.48\linewidth}
    \centering\includegraphics[width=\linewidth]{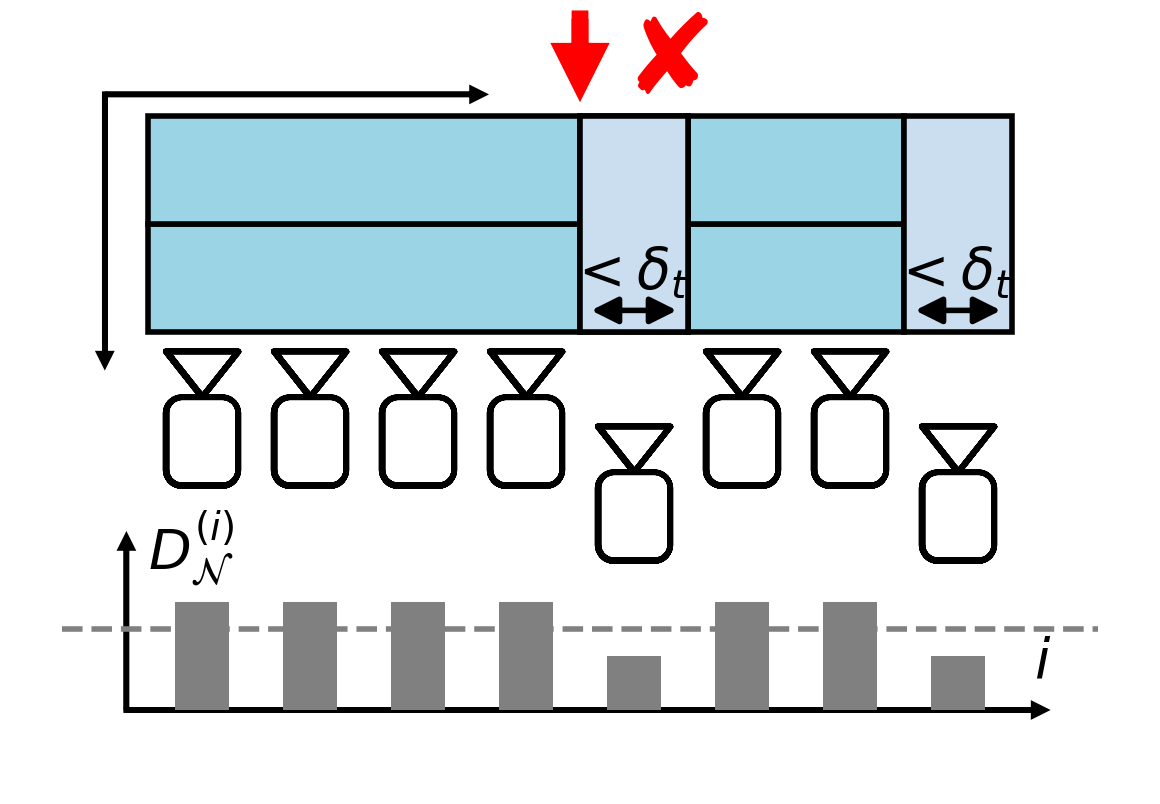}
    \caption{Small diameter subsequence \\ 
    too short to permit a split. }
    \label{fig:trule2}
  \end{subfigure}

  \caption{Proposed temporal coherence criterion.}
  \label{fig:trule}

\end{figure}

For a node $\node$ with its time window $[T_{\node}^0, T_{\node}^1]$, instead of considering all the cameras, we compute its diameter only for cameras within its window.
As illustrated in \fig{fig:trule}, we examine the list of $\diamIN,T_{\node}^0 \le t_i  \le T_{\node}^1$. If there is a subsequence that consists of diameters smaller than a threshold (half of the maximum diameter), and it spans at least duration $\deltat$, we may separate this sequence via temporal splits (\fig{fig:trule1}).
This would avoid equivalent spatial splitting over the whole sequence, saving overall memory.
Otherwise, if either diameters are similar across the sequence or the subsequences are too short (\fig{fig:trule2}), temporal splits are not applied.

A max-priority queue is used to construct the \treename, prioritized by \diamN, generalizing OcMesher and \eqn{eq:maxdia} to consider only a subsequence of camera frames.
Temporal and spatial splits are performed alternately, with temporal split tests preceding spatial splits for newly split nodes, as described in Algorithm~\ref{alg:alt}.

\begin{algorithm}[t]
\SetAlgoNoLine
\KwIn{Root spacetime node $\mathcal{N}_0$; Diameter threshold $\hat D$.}
\KwOut{\Treename\ with nodes satisfying threshold $\hat{D}$.}

$Q \gets \{\} $ \tcp{create an empty max-priority queue}

$\mathcal{N} \gets \mathcal{N}_0\ $

\SetKwBlock{Repeat}{repeat}{}

\While {$\diamN > \hat D$} {
    
    \If {$\namedfn{temporal\_split\_test}(\mathcal{N})$ \tcp{see \fig{fig:trule}} } {  
        $\{\mathcal{N}_1, \mathcal{N}_2\} \gets \namedfn{temporal\_split}(\mathcal{N})$
        
        \For{$i=1,2$} {
            $\namedfn{enqueue}(Q, \mathcal{N}_i)$
        }
    }
    \Else {
        $\{\mathcal{N}_1, \mathcal{N}_2, \ldots, \mathcal{N}_8\} \gets \namedfn{spatial\_split}(\mathcal{N})$
        
        \For{$i=1,2,\ldots,8$} {
            $\namedfn{enqueue}(Q, \mathcal{N}_i)$
        }
    }

    $\mathcal{N} \gets \namedfn{dequeue}(Q)$
}
\Return $ {\node}_0 $

\caption{Alternating Temporal and Spatial Splits}
\label{alg:alt}
\end{algorithm}

\subsection{Coarse-to-fine Algorithm}

Similar to OcMesher, a coarse-to-fine algorithm is used to construct a refined \treename.

\step{a}{Coarse \TreeName}
First, we apply Algorithm~\ref{alg:alt} to a root spacetime node with a coarse diameter threshold $\hat D_1$. The root node encompasses a large spatial extent and all camera timestamps.

\step{b}{Intersection Test}
Next, we identify nodes intersecting the isosurface by evaluating the occupancy function at the 16 corners of the hypercube represented by each node. 
Checking every node is expensive, so following OcMesher, we accelerate this step using flood-fill approach.
We select a subset of the nodes as seeds and iteratively propagate from any surface-intersecting node to both spatial and temporal neighbors. Spatial flooding ensures no spatial discontinuities occur, while temporal flooding prevents temporal inconsistencies that would lead to popping effects.

\step{c}{Refined \TreeName}
Finally, we refine the \treename\ by reapplying Algorithm~\ref{alg:alt} independently to each surface-intersecting node, now using a smaller threshold 
$\hat D_2 < \hat D_1$.

\subsection{4D Mesh Extraction}

\begin{figure}[t]
\centering
\captionsetup[subfigure]{justification=centering}
\begin{tabular}{c c c}
    \includegraphics[width=.26\linewidth]{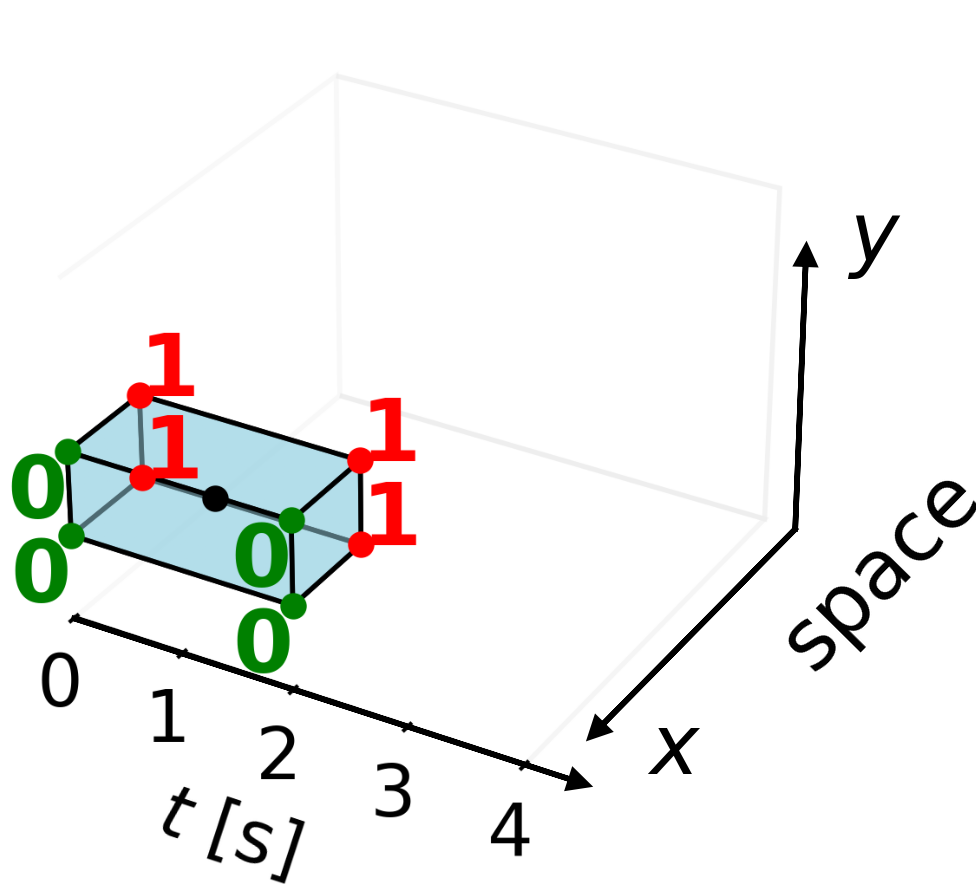} &
    \includegraphics[width=.26\linewidth]{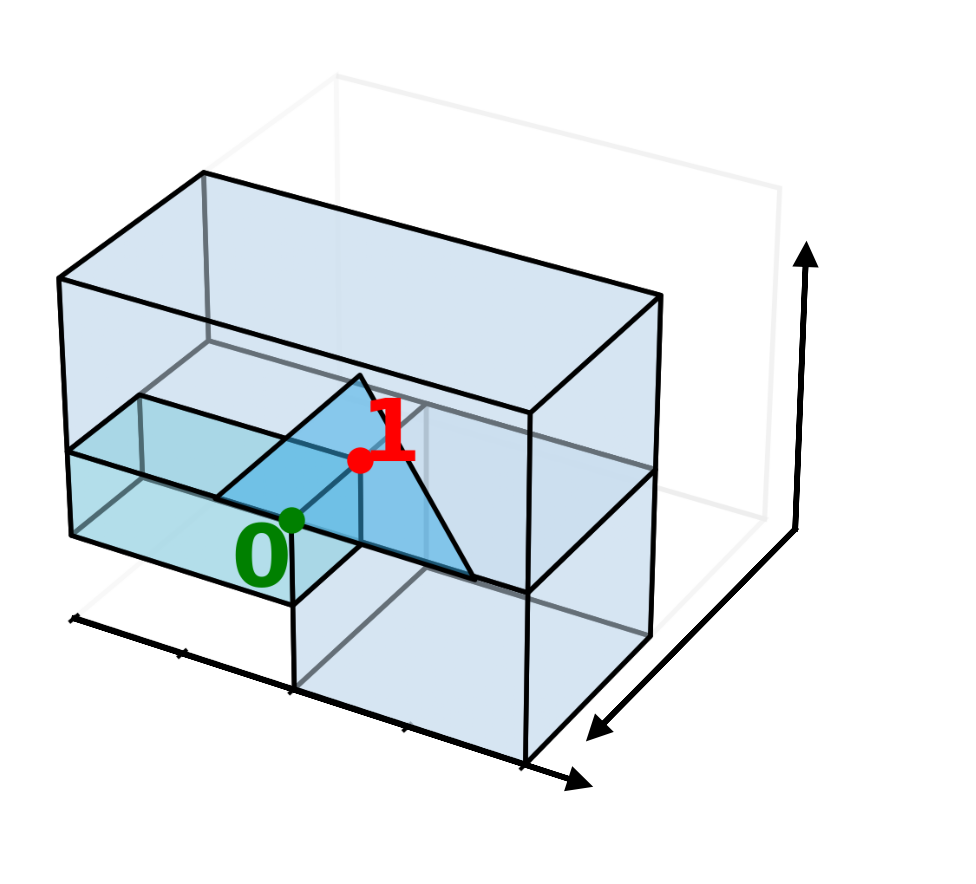} &
    \includegraphics[width=.26\linewidth]{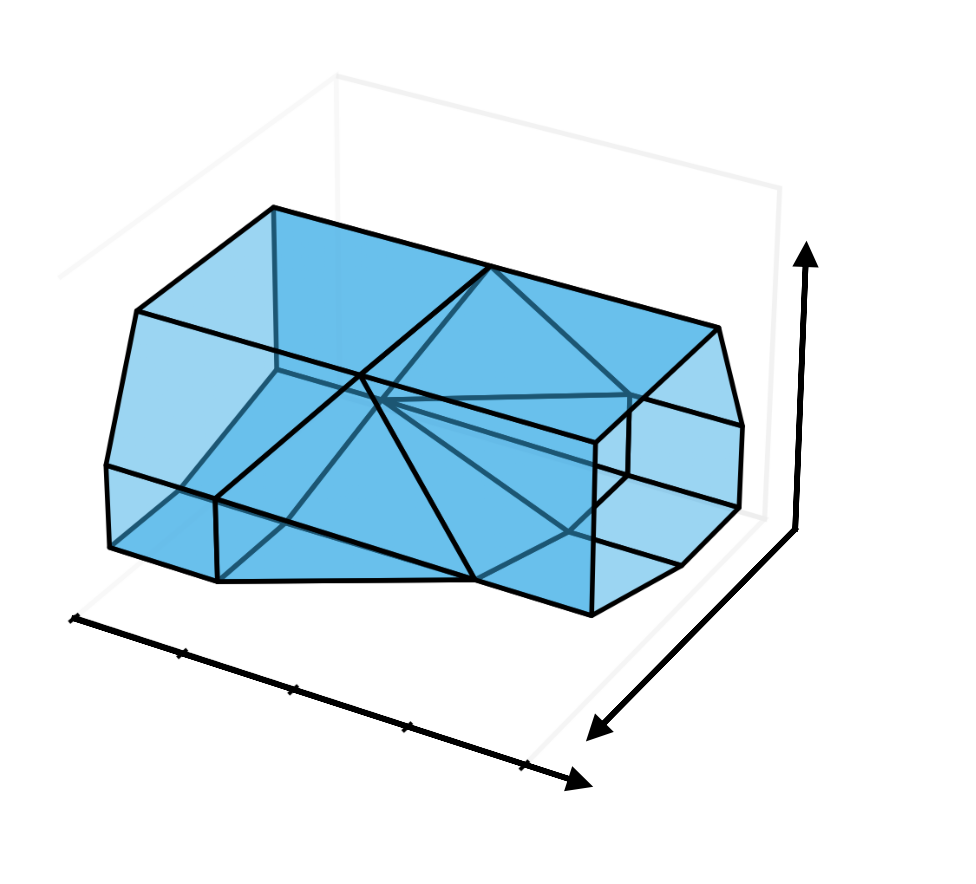} \\
    {\small (a) Spacetime cube} & 
    {\small (b) Bipolar edge \& dual } & 
    {\small (c) Extracted mesh}
\end{tabular}
  \caption{Block of 3D spacetime with the two spatial axes $x$ and $y$. (a)~Dual contouring associates each surface-intersecting node with a vertex, and (b)~each bipolar edge with a dual polygon. Extracted mesh shown in (c).}
  \label{fig:dual}
\end{figure}

\boldparfirst{Dual Contouring. } In $K$-dimensional dual contouring, each \emph{surface-intersecting} hypercube (having different corner values) corresponds to a mesh vertex. Each \emph{bipolar edge} (connecting different values) is associated with a $(K-1)$-dimensional \highlight{h7}{polyhedron (polytope)} embedded in $K$-dimensional space. 
In the 3D case illustrated in \fig{fig:dual},  each bipolar edge is associated with a polygon and a 3D mesh consists of many polygons. 
In the 4D case for the \treename, each bipolar edge is associated with a \highlight{h7}{polyhedron} embedded in 4D space. Such a \highlight{h7}{polyhedron} can be a hexahedron with 8 vertices in general, or other forms with fewer vertices. A 4D mesh consists of many \highlight{h7}{polyhedra}.

\begin{figure}[t]
\captionsetup[subfigure]{justification=centering}
\begin{tabular}{c c c}
    \raisebox{1mm}{\includegraphics[width=.20\linewidth]{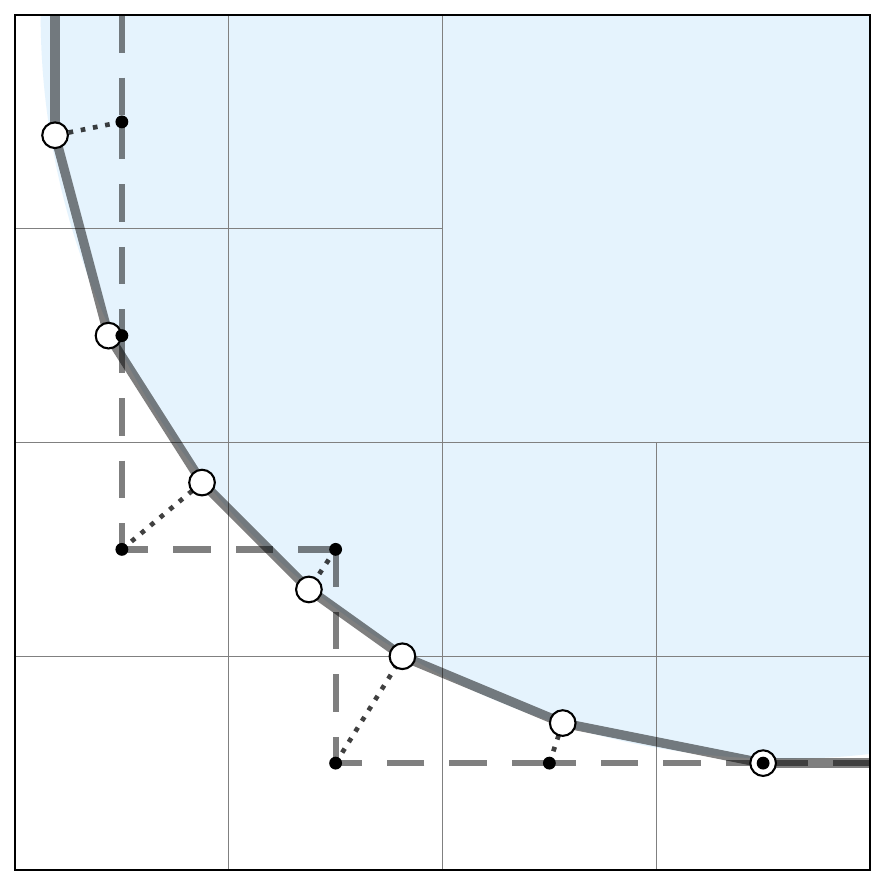}} &
    \includegraphics[width=.26\linewidth]{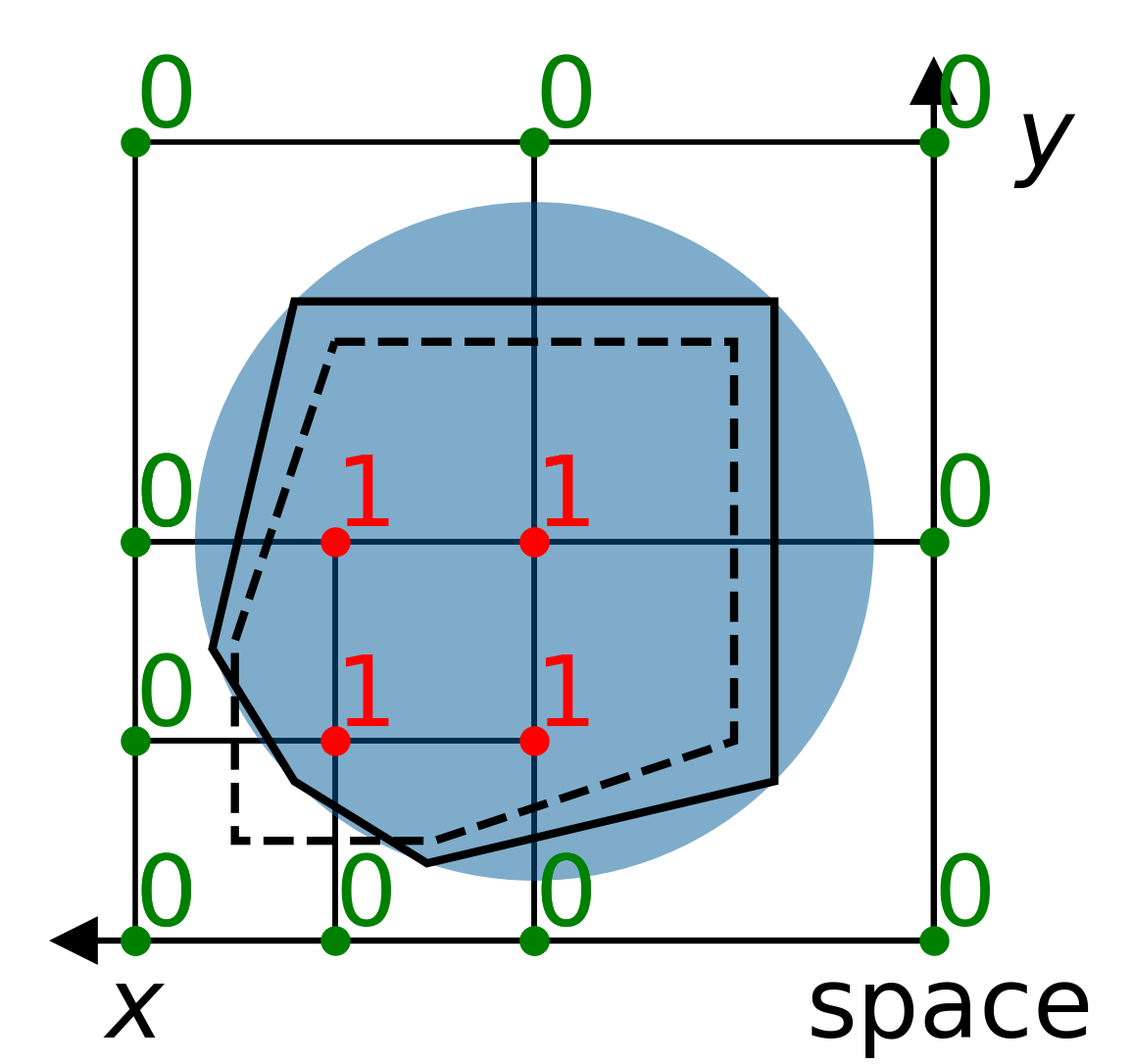} &
    \includegraphics[width=.26\linewidth]{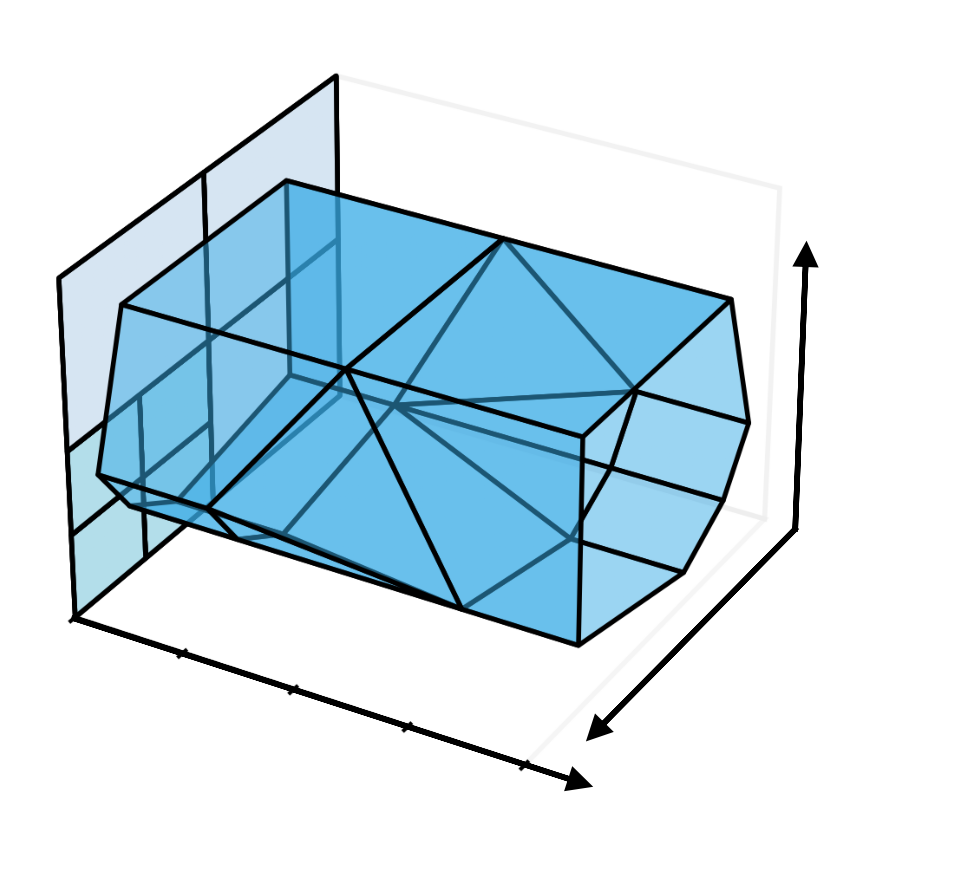} \\
    {\small (a) Bisection closeup} & 
    {\small (b) Refining 2D shape} & 
    {\small (c) Refined 3D shape}
\end{tabular}
  \caption{Placing vertices at hypercube centers leads to ``staircase'' artifacts (dashed lines). We use bisection to project these vertices closer (solid lines) to the true geometry. (a)~A close-up 2D example, where the cube centers (filled dots, dashed lines) after bisection move close to the true shape (hollow dots, solid lines). (b)~Extracting the full 2D shape -- a blue circle, on a coarser grid than in~(a). (c)~The 3D shape from Fig. 7c refined after bisection search.}
  \label{fig:bis}
\end{figure}

\boldpar{Vertex Computation by Bisection: } We extract one vertex per hypercube. 
Placing it at the center would lead to ``staircase'' artifacts illustrated by the dashed lines in \fig{fig:bis}(a-b).
For discrete scalar data, interpolation-based methods~\cite{lorensen1998marching} are suitable for placing mesh vertices; however, they are not effective for a point-wise binary occupancy function over $\mathbb{R}^3$.
Instead, we use bisection.
For a binary function in 1D, bisection reduces to binary search to pinpoint a 0-1 transition.
Similarly, we use bisection in 3D to project the vertex close to the actual surface (shown as solid lines).

\subsection{Mesh Slicing}

\begin{figure}[t]
\centering\includegraphics[width=.24\linewidth]{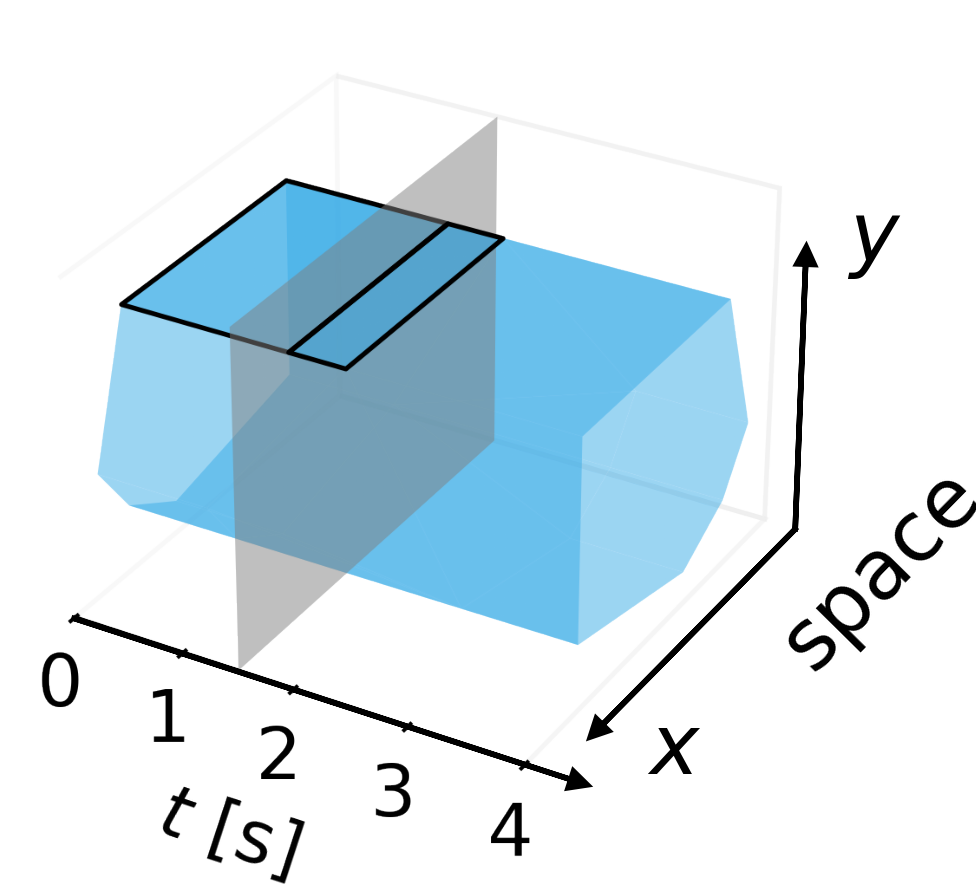}\hspace{-2mm}
\centering\includegraphics[width=.24\linewidth]{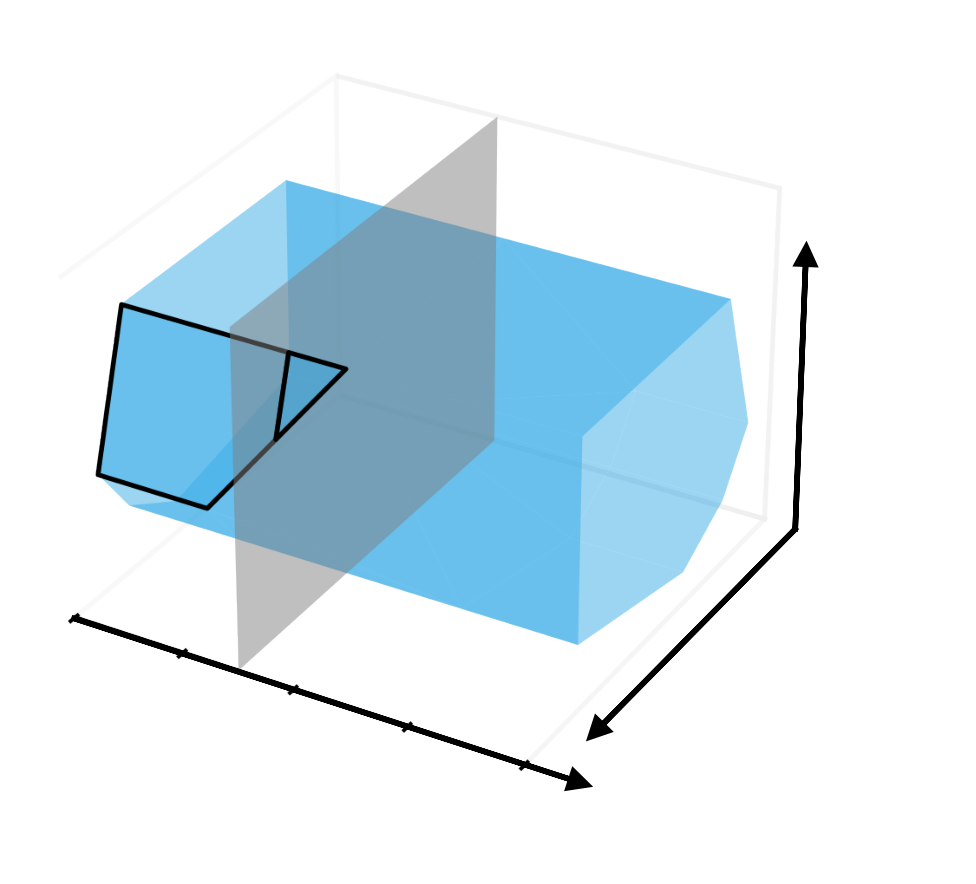}\hspace{-2mm}
\centering\includegraphics[width=.24\linewidth]{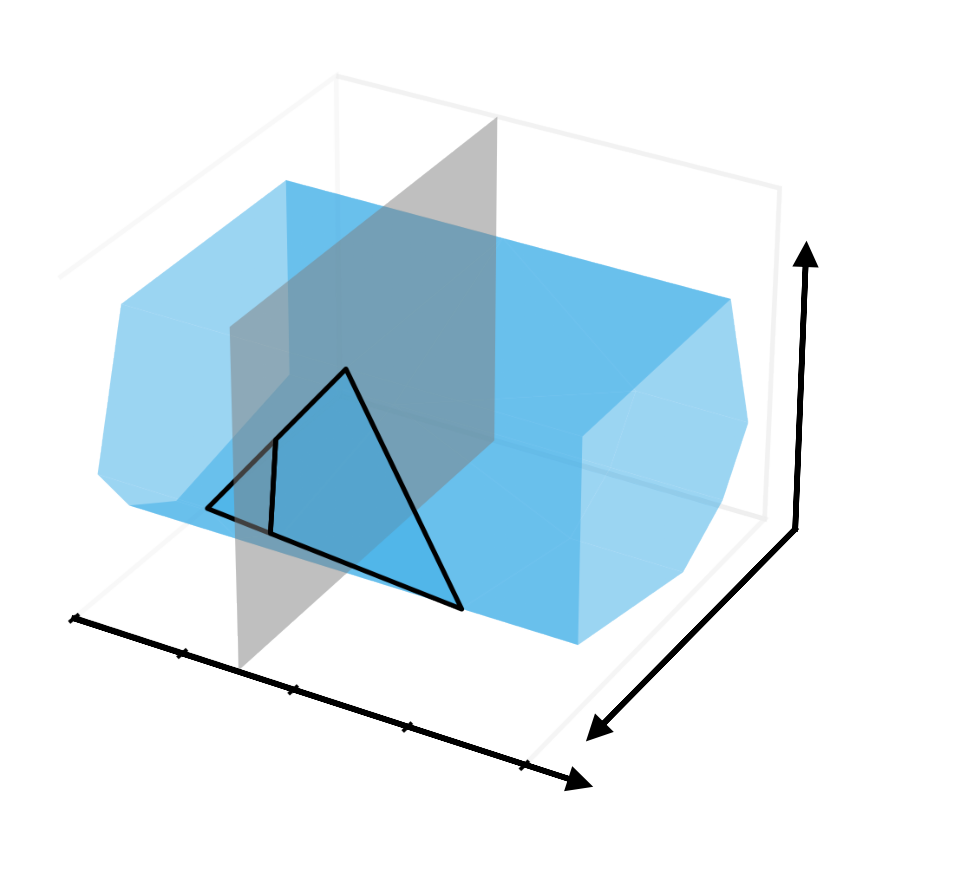}
\centering\includegraphics[width=.24\linewidth]{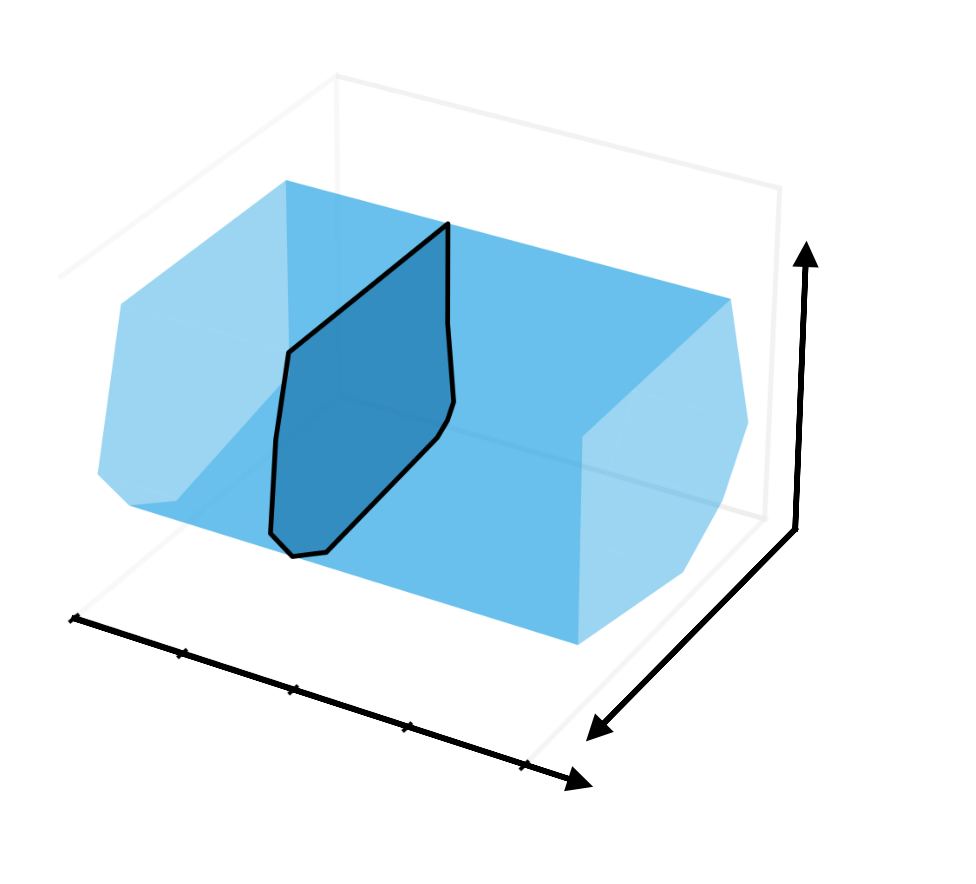}
\caption{In 3D spacetime, we extract line segments by slicing polygons, and the output segments form a set of polygons.}
  \label{fig:pslice}
\centering\includegraphics[width=.25\linewidth]{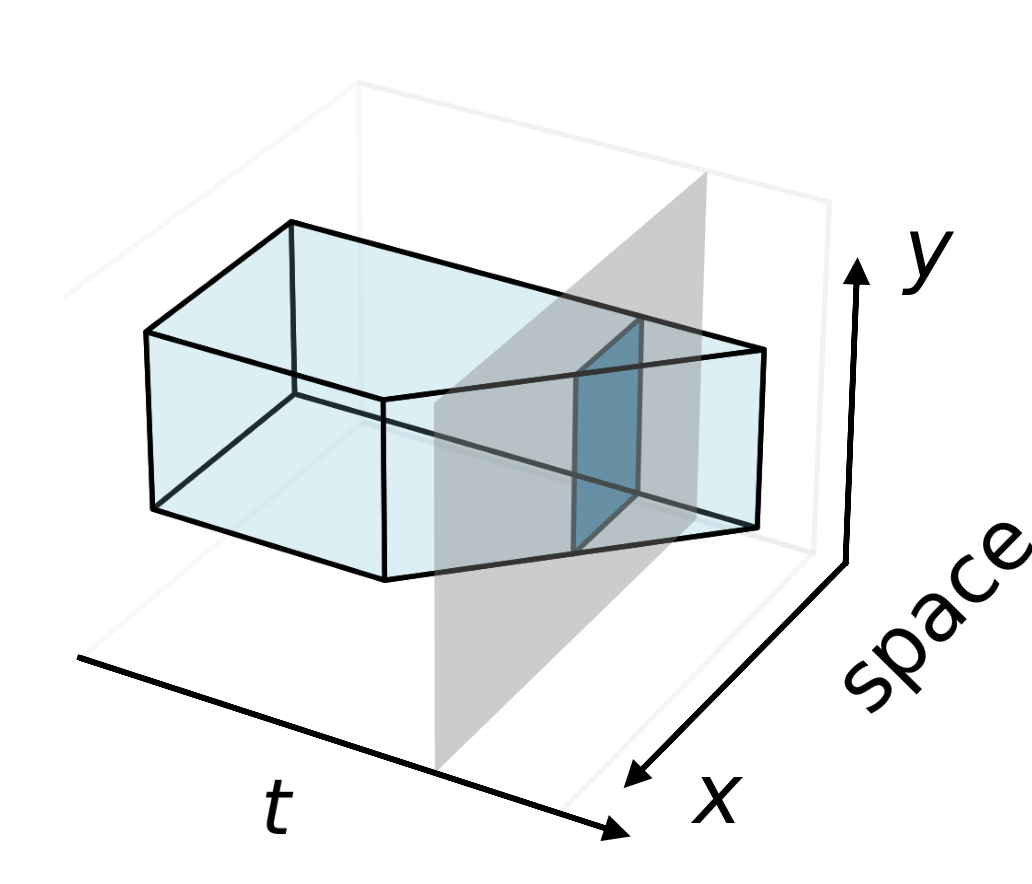}
\centering\includegraphics[width=.25\linewidth]{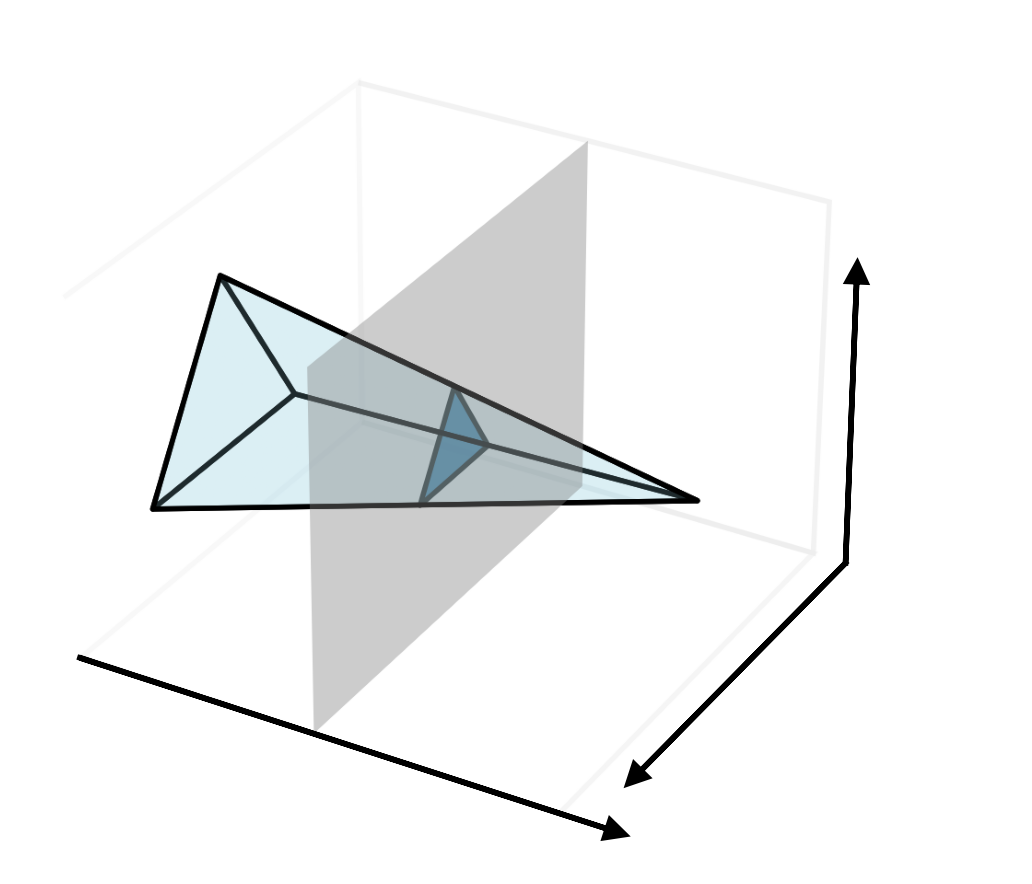}
\centering\includegraphics[width=.25\linewidth]{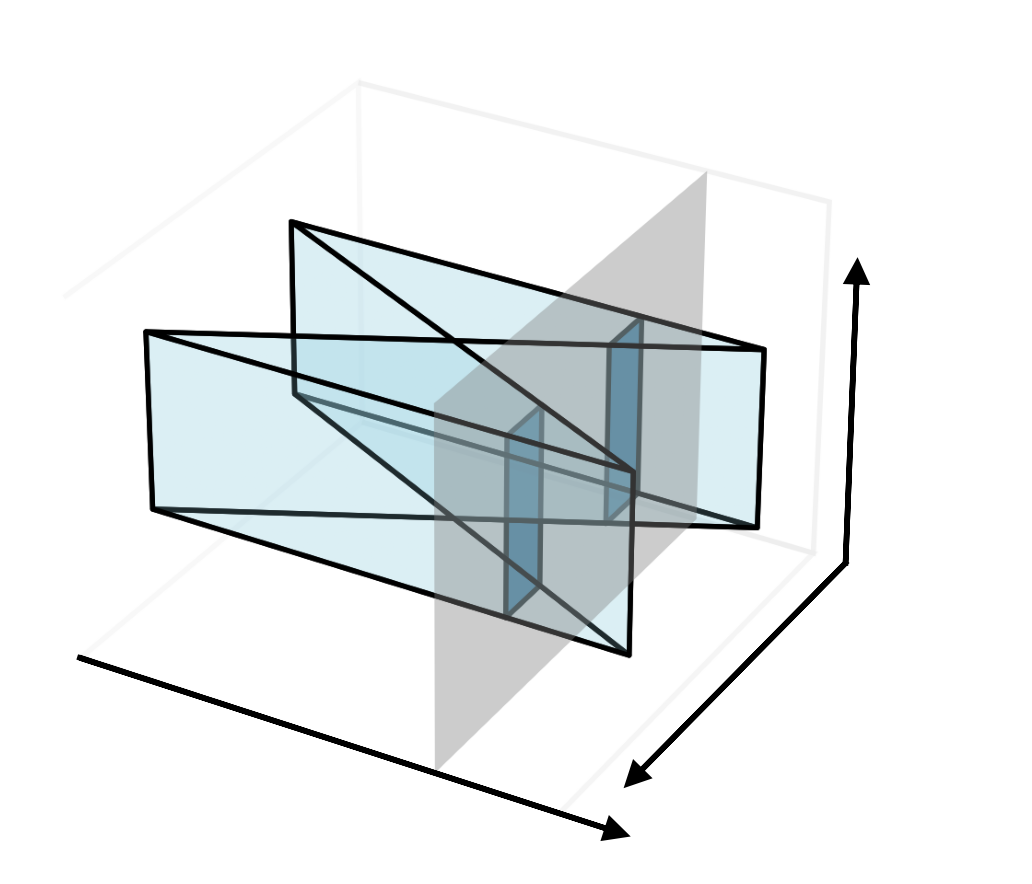}
  \caption{In 4D spacetime (drawn here in 3D), we extract polygons by slicing \highlight{h7}{polyhedra}, and the output polygons form a mesh. Shown are several different \highlight{h7}{polyhedron} types, resulting in different shapes and numbers of polygons.}
  \label{fig:slice}
\end{figure}

\figs{fig:pslice} and \fignum{fig:slice} illustrate by analogy mesh slicing in 3D and 4D.
\fig{fig:pslice} shows at a given timestamp finding polygons intersecting the slicing plane, producing line segments at the intersections. 
This set of line segments forms a 2D ``mesh'' (set of polygons).
\fig{fig:slice} shows how in 4D spacetime we find \highlight{h7}{polyhedra} intersecting the slicing plane. A \highlight{h7}{polyhedron} can be a hexahedron embedded in 4D with eight vertices (or other forms with fewer vertices) and its vertices can follow a complex temporal ordering (\fig{fig:slice}). 
To slice such a \highlight{h7}{polyhedron} with a plane at $t=t_1$, we find edges $(u,v)$ where $t_u \leq t_1 < t_v$ and place an intersection vertex for each edge. Each face of the \highlight{h7}{polyhedron} has 0, 2, or 4 intersection vertices and we connect them to form one or more polygons. These polygons form a mesh in 3D. This mesh gradually transforms as the slicing plane moves through time.

\section{Experiments}
The supplementary video shows several scenes as animations.

{\boldpar{Input Scenes.}}
We use the procedural terrain from Infinigen~\cite{raistrick2023infinite}, along with additional modeling techniques~\cite{sdftrees}~\cite{sdfcity}, to create camera animations of complex scenes, including \textit{Forest}, \textit{Mountain}, \textit{Arctic}, \textit{Cave}, \textit{Beach}, and \textit{City}.

{\boldpar{Parameters.}}
Each video spans 20 seconds at 24 frames per second and a resolution of 960$\times$540. We set the transition control parameter $\deltat=1$ s and the diameter threshold $\hat D_2 = 3$ px. The user can adjust these parameters according to their needs. Additional efficiency-related parameters in Sec.~\ref{sec:details} are available but set to defaults.

{\boldpar{Baselines.}}
We compare \methodname\ against two baselines:

\bulletpar{Spherical:} Spherical Mesher is used by default in Infinigen and constructs uniform grids in spherical coordinates and applies the marching cubes algorithm. A mesh is extracted every 8 frames.

\bulletpar{OcMesher:} OcMesher divides the camera trajectory into subsequences, and extracts a single mesh for each subsequence. We experiment with subsequence lengths of 24 and 96 frames, referred to as \textbf{OcMesher-24} and \textbf{OcMesher-96}, respectively.

\begin{figure}[t!]

\centering
\includegraphics[width=0.98\linewidth]{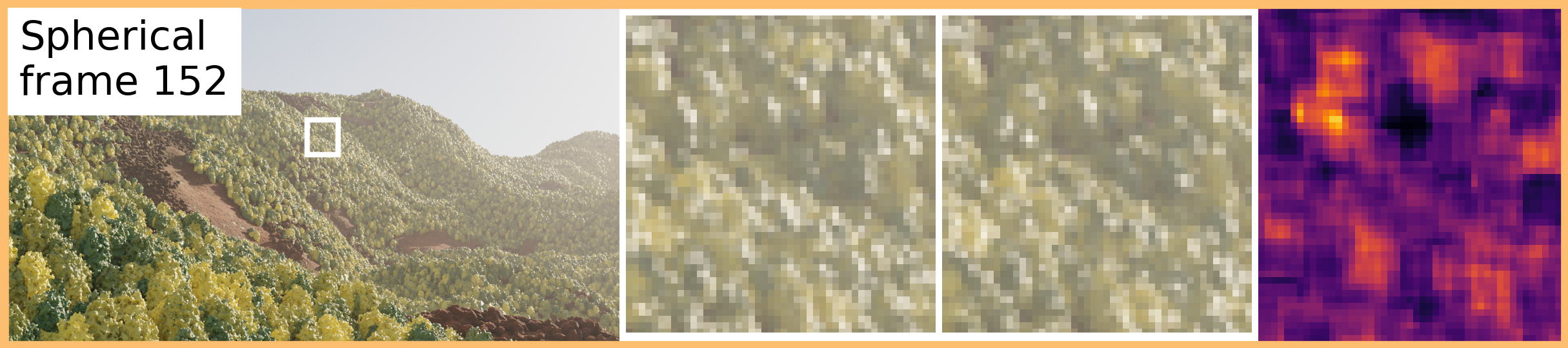}
\includegraphics[width=0.98\linewidth]{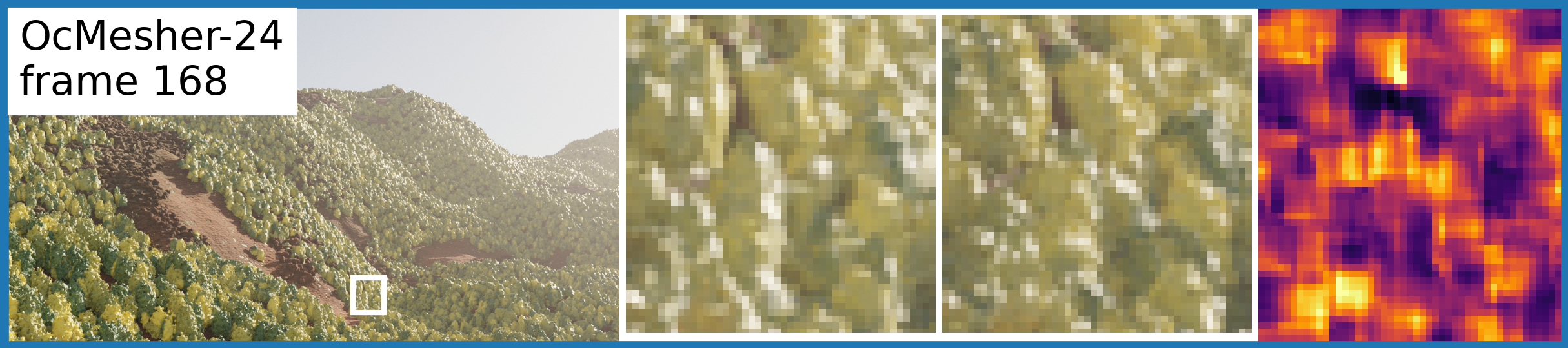}
\includegraphics[width=0.98\linewidth]{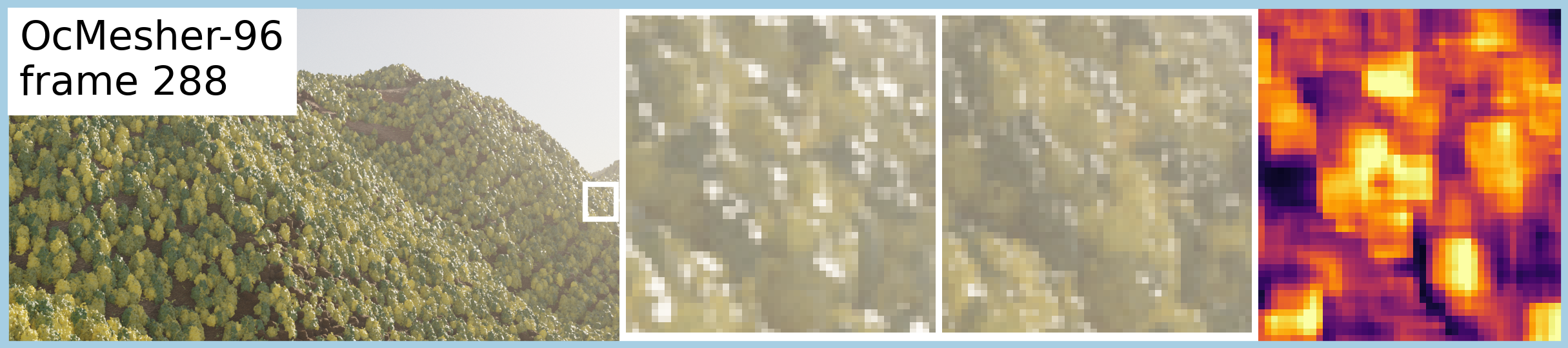}
\includegraphics[width=0.98\linewidth]{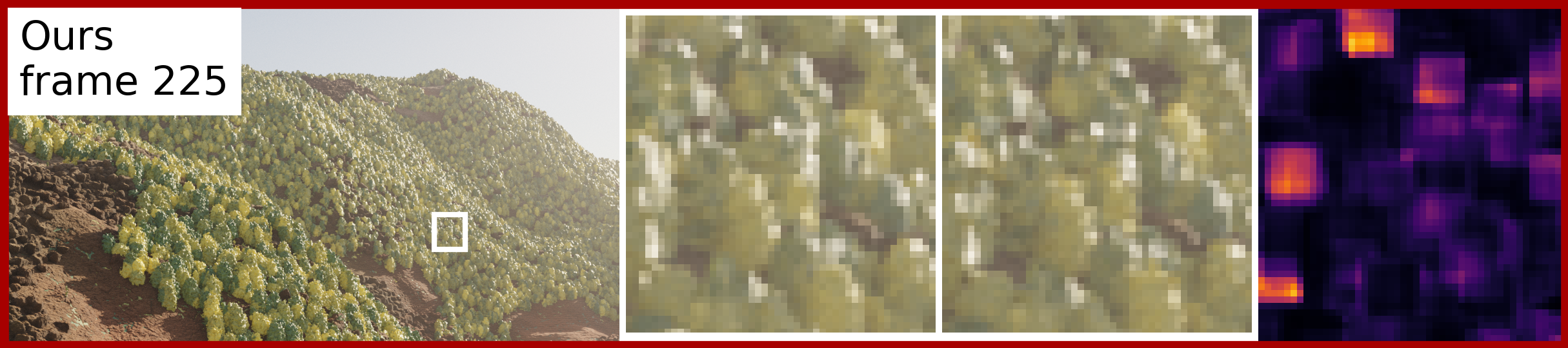}
\includegraphics[width=0.98\linewidth]{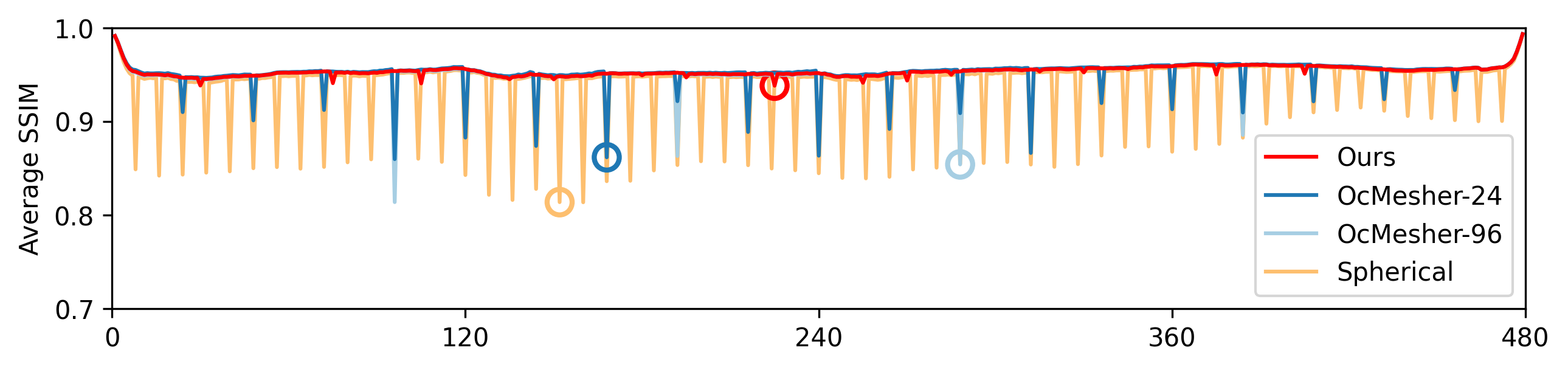}

\caption{\emph{Above:} For each method (color coded), the second worst frame according to SSIM view consistency scores. The zoomed-in squares compare this frame to the subsequent one, followed by their SSIM heatmap (brighter is worse).
These squares highlight the worst (SSIM) region in each frame.
\emph{Below:} Average SSIM for each frame, graphed across the entire sequence, for the four methods. Lower is worse. Valleys indicate popping. Circles note the SSIM at the four frames selected above.
}

\label{fig:ssim}
\end{figure}

\subsection{Visual Consistency}
\label{ssec:ssim}

\fig{fig:ssim} shows frames of the  \textit{Forest} scene comparing the four methods. Zoomed-in squares show the region boxed in white, comparing it with the same region in the next frame of the animation. Each frame is selected as the \emph{second worst} from the whole animation for each method, according to a visual consistency score. This score is visualized in the heatmap on the right (bright is worse) and computed as follows.
The score of frame $i$, denoted $\mathbf{S}_{i}$, measures the consistency between frame $i$ and the next frame $i+1$:
\begin{equation}
\Si = \namedfn{SSIM}(\mathbf{I}_{i}, \namedfn{warp}(\mathbf{I}_{i+1}, \mathbf{F}_{i\rightarrow i+1})) \label{eq:ssim}
\end{equation}
\noindent
where $\mathbf{F}_{i\rightarrow i+1}$ is the ground-truth optical flow, $\mathbf{I}_{i}$ and $\mathbf{I}_{i+1}$ are consecutive rendered images, function \namedfn{warp} warps a given image via the inverse map of the forward flow (to avoid resulting gaps), and \namedfn{SSIM} is the Structural Similarity Index Measure~\cite{wang2004image}.

The plot in \fig{fig:ssim} shows the average score \Si\ for each frame, with circles identifying \Si\ for the four frames above, color coded by method. 
Spherical, Ocmesher-24 and OcMesher-96 update the mesh for each subsequence of frames, resulting in the periodic ``valleys'' in both plots that signify ``popping'' events. 
While a larger subsequence length in OcMesher-96 reduces the frequency of popping relative to Ocmesher-24, it increases the severity of each occurrence. In contrast, our method maintains consistently high \Si and small normal difference, with only minor valleys (and in other scenes, our method exhibits even fewer valleys).
Each valley's severity is evaluated by 
$\Sim + \Sip - 2\Si$, and this is the measure by which these circled valleys are the second worst for each method in the sequence.
The worst frames for each sequence are shown for all four methods in all six scenes in 
\figs{fig:extra1} and \fignum{fig:extra2}.
Alternatively, \figs{fig:sameframe} shows a direct comparison at the same frame.

\begin{figure}[t!]
\centering
\includegraphics[width=0.98\linewidth]{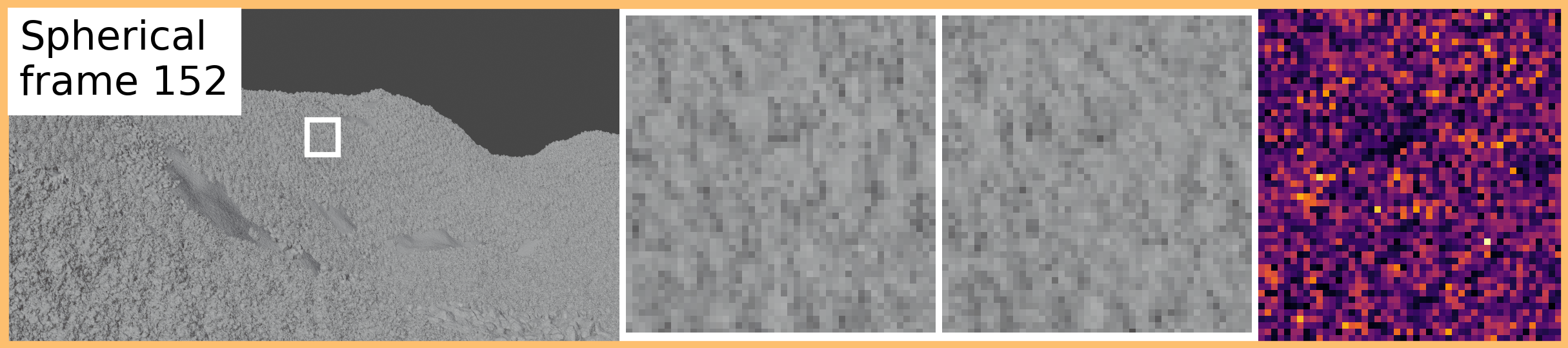}
\includegraphics[width=0.98\linewidth]{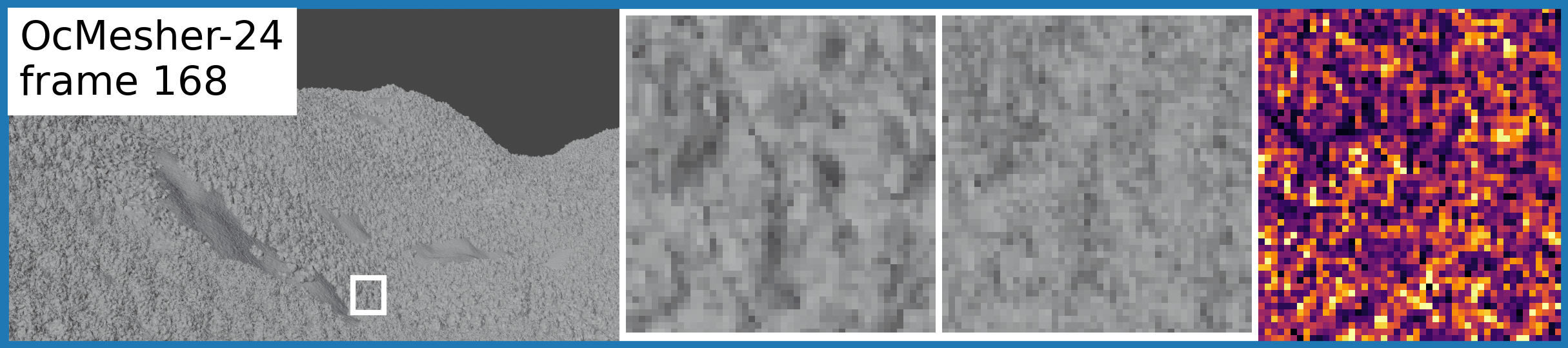}
\includegraphics[width=0.98\linewidth]{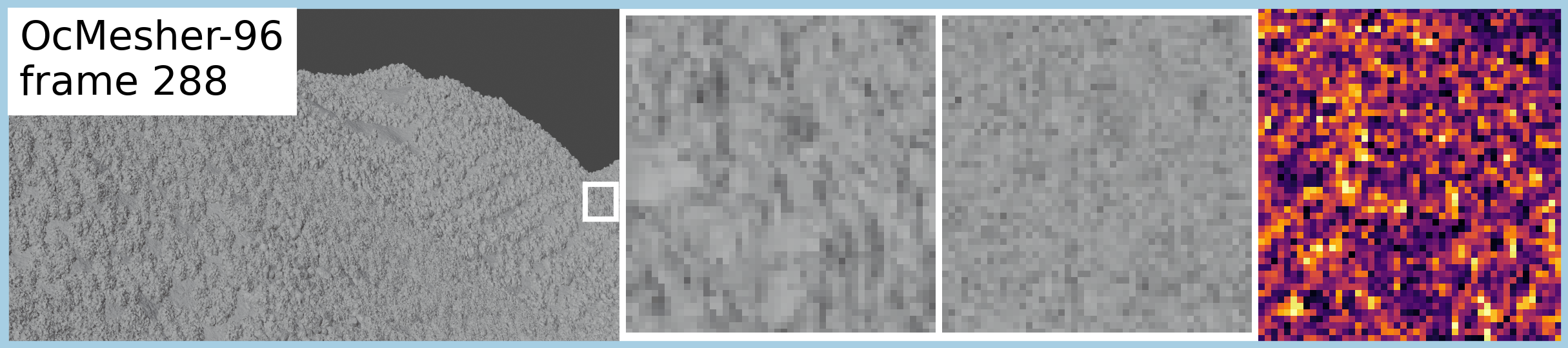}
\includegraphics[width=0.98\linewidth]{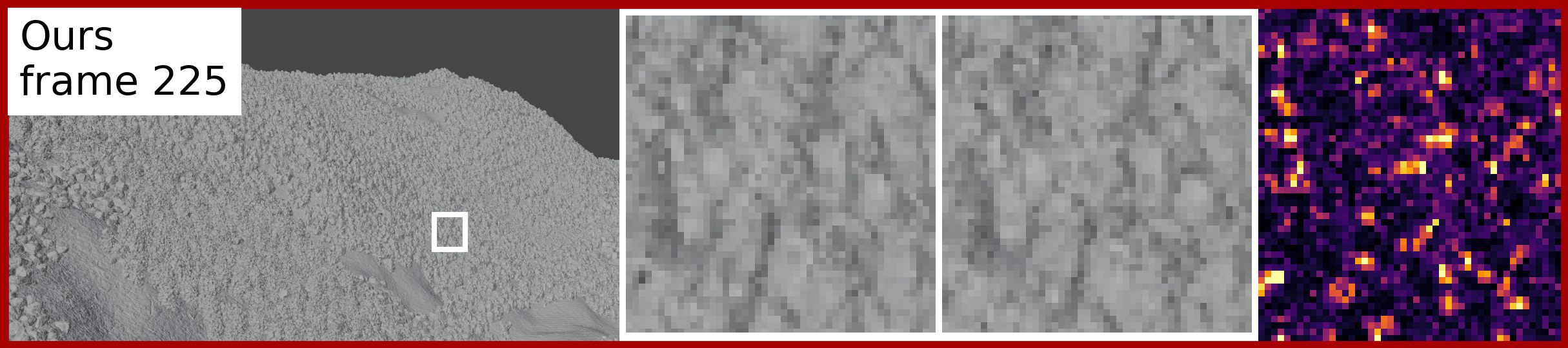}
\includegraphics[width=0.98\linewidth]{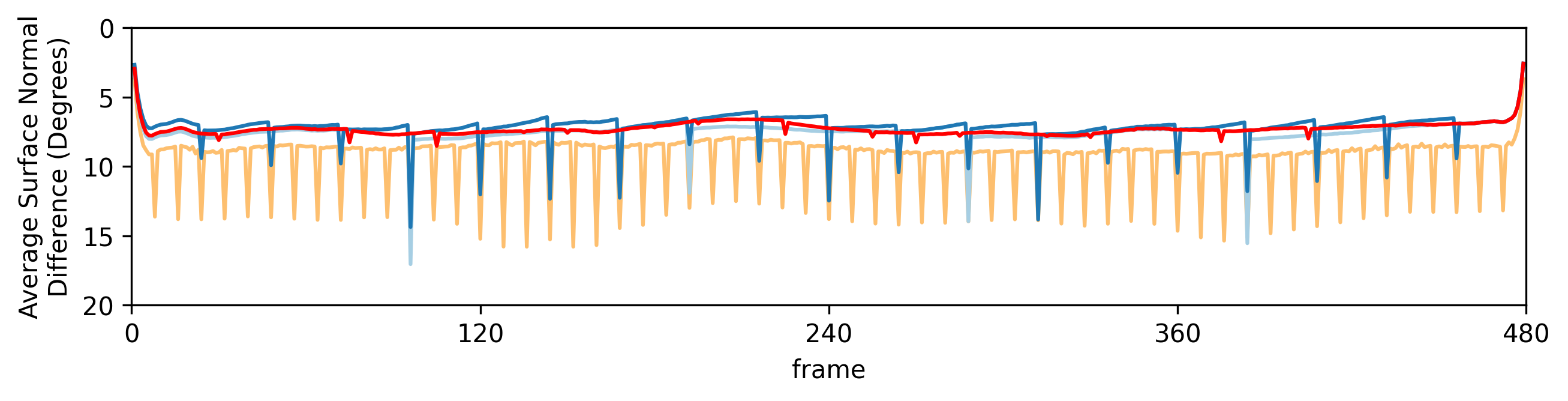}

\caption{\highlight{h1}{\emph{Above:} For each method, geometry-only (\emph{clay-style}) rendering of the frames in Fig.}~\ref{fig:ssim}\highlight{h1}{, followed by their normal difference heatmap (brighter is worse).
\emph{Below:} Plot of average normal difference, which is highly correlated with the SSIM plot shown in in Fig. 11.
}}

\label{fig:normdiff}
\end{figure}

Fig.~\ref{fig:normdiff} \highlight{h1}{illustrates geometric consistency of the four methods using geometry-only (\emph{clay-style}) renderings, with corresponding normal difference heatmaps (where brighter is worse).
The plot shows average surface normal differences (in degrees).
The normal difference is highly correlated with $\namedfn{SSIM}$ (Fig. 11), and slightly noisier, so we only report the latter throughout the rest of the paper.}

\subsection{Computational Cost}

This section presents experiments comparing the performance of the four methods; Sec.~\ref{sec:time_comp} describes the theoretical complexity of our algorithm.
All methods run on a multi-core CPU, with occupancy queries performed on the GPU. Table~\ref{tab:cost} compares the runtime and the vertex count of the four methods for the \textit{Forest} scene. The runtime is divided into two parts: mesh generation (meshing) and rendering. Meshing cost is reported as amortized per frame. While rendering is not part of these mesh extraction methods, it remains the bottleneck.

Spherical Mesher has the longest overall runtime due to a high mesh resolution that mitigates popping, but it still has much more popping than the other methods. OcMesher generate meshes faster than our method. However, as the subsequence length increases, though OcMesher-96 does less work for the entire sequence than OcMesher-24, it creates a larger mesh for each frame. This increases the risk of exceeding the memory limits and results in longer rendering time than ours. Overall, our method has a comparable cost, while offering better view consistency.

The components of meshing time for our method are roughly: 3\% for constructing the \treename{}, 32\% for extracting the 4D mesh, and the remaining 65\% for slicing the mesh. Greater GPU utilization is possible but not a current bottleneck.

\begin{table}[t!]
\caption{Computational cost of the four methods. Measured on 64-core 12th Gen Intel Core i7 CPU with RTX 2080. Our method offers better view consistency (Fig.~\ref{fig:ssim}) with comparable overall cost.}

\label{tab:cost}
\centering
\resizebox{0.92\linewidth}{!}{
\small
\begin{tabular}{lllll}
\hline
&  \multicolumn{1}{c}{\begin{tabular}[c]{@{}c@{}}Spherical\end{tabular}} 
& \multicolumn{1}{c}{\begin{tabular}[c]{@{}c@{}}OcMesher-24 \end{tabular}} 
& \multicolumn{1}{c}{\begin{tabular}[c]{@{}c@{}}OcMesher-96\end{tabular}} 
& \multicolumn{1}{c}{Ours} \\ \hline
\textbf{Runtime (sec/frame)} & & & & \\ \hline
\begin{tabular}[c]{@{}l@{}}Meshing (Amortized)  \end{tabular} & \multicolumn{1}{c}{21} & \multicolumn{1}{c}{7} & \multicolumn{1}{c}{3} & \multicolumn{1}{c}{23} \\ \hline

Rendering  &\multicolumn{1}{c}{254} & \multicolumn{1}{c}{193} & \multicolumn{1}{c}{220} & \multicolumn{1}{c}{199} \\ \hline
Total  &\multicolumn{1}{c}{275} & \multicolumn{1}{c}{200} & \multicolumn{1}{c}{223} & \multicolumn{1}{c}{222} \\ \hline \hline

\textbf{Vertex Count} & & & & \\ \hline 

Total from all Blocks & \multicolumn{1}{c}{565 M}  &  \multicolumn{1}{c}{105 M}  & \multicolumn{1}{c}{55 M} & \multicolumn{1}{c}{81 M} \\ \hline
Average Per Frame & \multicolumn{1}{c}{9.1 M} & \multicolumn{1}{c}{5.2 M} & \multicolumn{1}{c}{10.9 M} & \multicolumn{1}{c}{9.3 M} \\ \hline
\end{tabular}
}

\end{table}

\begin{table}[t!]

\caption{Larger values of $t_0$ have similar total runtime, but produce larger meshes, which increases the risk of exceeding the memory limits.}

\resizebox{0.75\linewidth}{!}{
\centering
\small
\begin{tabular}{lllll}
\hline
Parameter \deltat\ & \multicolumn{1}{c}{\begin{tabular}[c]{@{}c@{}}0.5\end{tabular}} 
&  \multicolumn{1}{c}{\begin{tabular}[c]{@{}c@{}}1.0\end{tabular}} 
& \multicolumn{1}{c}{\begin{tabular}[c]{@{}c@{}}2.0 \end{tabular}} 
& \multicolumn{1}{c}{\begin{tabular}[c]{@{}c@{}}4.0\end{tabular}}  \\ \hline
\textbf{Runtime (sec/frame)} & & & \\ \hline
\begin{tabular}[c]{@{}l@{}}Meshing (Amortized)  \end{tabular} & \multicolumn{1}{c}{27} & \multicolumn{1}{c}{23} & \multicolumn{1}{c}{22} & \multicolumn{1}{c}{16} \\ \hline
\begin{tabular}[c]{@{}l@{}}Rendering \end{tabular} & \multicolumn{1}{c}{195} &\multicolumn{1}{c}{199} & \multicolumn{1}{c}{200} & \multicolumn{1}{c}{205} \\ \hline
Total  &\multicolumn{1}{c}{222} & \multicolumn{1}{c}{222} & \multicolumn{1}{c}{222} & \multicolumn{1}{c}{221} \\ \hline \hline

\textbf{Vertex Count} & & & \\ \hline

Total & \multicolumn{1}{c}{103 M} & \multicolumn{1}{c}{81 M}  &  \multicolumn{1}{c}{66 M}  & \multicolumn{1}{c}{54 M}  \\ \hline
Averaged Per Frame & \multicolumn{1}{c}{7.0 M} & \multicolumn{1}{c}{9.3 M} & \multicolumn{1}{c}{13.8M} & \multicolumn{1}{c}{20.5M} \\ \hline
\end{tabular}
}

\label{tab:time}
\end{table}

\begin{figure}[t!]
    \centering
    \includegraphics[width=\linewidth]{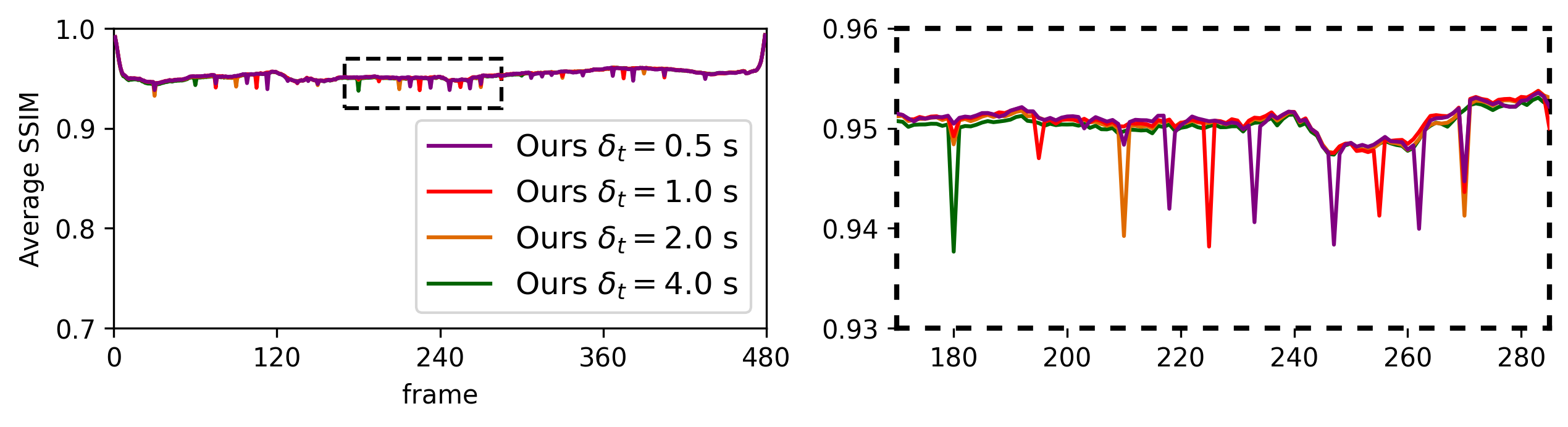}

    \caption{Effect of \deltat\ on \Si. Larger \deltat\ has less frequent valleys (but all \deltat\ have relatively small valleys). Dashed box enlarged on right.}
    \label{fig:time}

\end{figure}

\subsection{Transition Control Parameter Settings}

The transition control parameter $\deltat$ balances between the goals of memory efficiency and temporal coherence. Here we evaluate this tradeoff for different values of $\deltat$.
In \fig{fig:time}, the SSIM score plots for larger $\deltat$ have less frequent 
valleys. 
In Table~\ref{tab:time}, larger $\deltat$ values produce larger 3D meshes per frame, leading to increased risk of exceeding the memory limits.
Considering these factors, $\deltat=1$s remains a balanced choice with relatively infrequent and small consistency score fluctuations as well as lower memory usage. Therefore we use this value in our other experiments.
However, users with a larger memory budget can opt for a larger \deltat.

\section{Implementation Details}
\label{sec:details}

This section describes some strategies for efficient implementation of the \TreeName\ algorithms introduced in \sect{sec:binoc}.

\subsection{Efficient Implementation of Tree Construction}
\label{sec:eff}

One goal of the coarse-to-fine \treename\ construction algorithm is to prioritize allocating resources to regions of the scene relevant to rendering. To this end we employ several implementation strategies.

\boldpar{Virtual Grid. }
Given a fixed diameter threshold $\hat D_2$, the intermediate (coarse) threshold $\hat D_1$ plays a crucial role in the coarse-to-fine tree construction algorithm. A large $\hat D_1$ reduces the coarse tree's memory usage; while a smaller value produces finer surface-intersecting nodes, reducing subsequent refining costs. To balance these trade-offs, we adopt the virtual grid technique from OcMesher.

\begin{figure}[t]
\centering
\captionsetup[subfigure]{justification=centering}
    \includegraphics[width=0.93\linewidth]{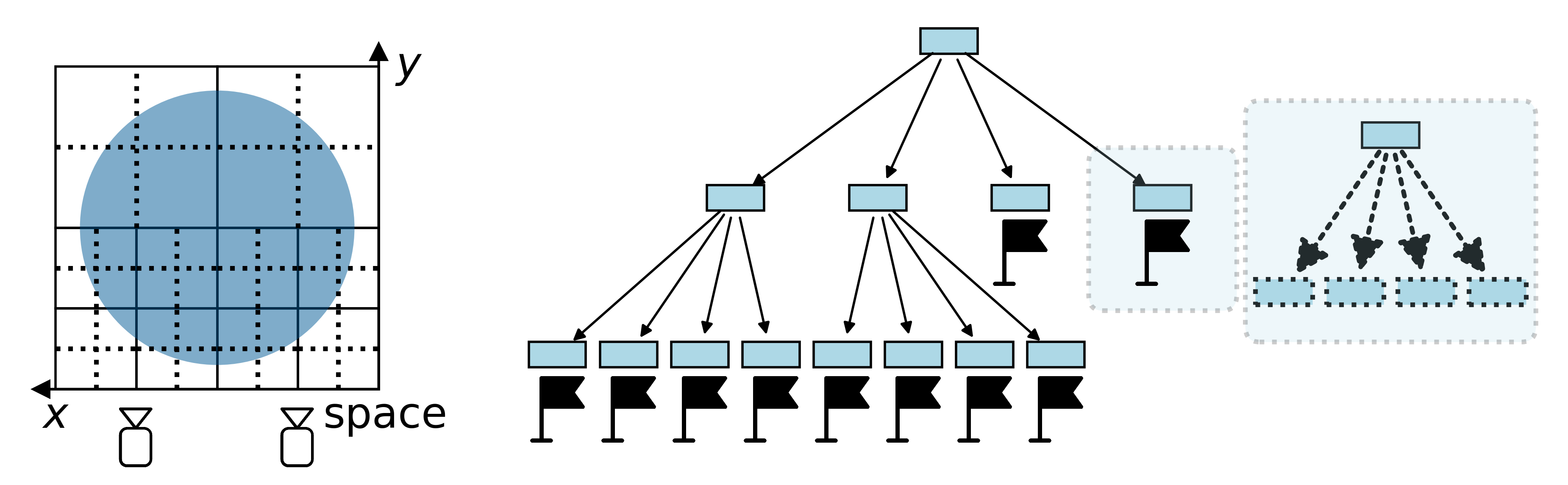}\pulluptwo
   \caption{When the coarse tree reaches a certain size cap, nodes are flagged as \emph{virtual grids} to save computation. Only two spatial axes are shown.}\pulluptwo
   \label{fig:eff}
\end{figure}

In 4D spacetime, when a coarse binary-ocree reaches a certain size cap \sizecap, we halt refinement and simply flag any node $\node$ that requires further splits (i.e., $\diamN > \hat D_1$), as illustrated in \fig{fig:eff}. The flag marks the node as a ``virtual'' grid of size $V^3 \times 1$, where $V$ is the smallest power of 2 such that $\diamN < V \hat D_1$. By combining the virtual grid with the flood-fill algorithm, surface intersection tests are performed at fine granularity without the memory overhead of actually refining the tree everywhere. During refinement, only surface-intersecting virtual nodes are instantiated.
Our experiments use $\hat D_1=30\text{ px}$ and size cap $\sizecap=10M$.

\boldpar{Contraction Out of Frustum. }
To avoid unneeded detail outside the camera frustums, we contract the computed camera-specific node diameter \diamIN of any node outside the frustum by a small factor throughout tree construction. Our experiments use factor \fourth.

\boldpar{Visibility Test with Depth Buffers. }
To further avoid unnecessary detail, we optionally use a depth buffer for each camera during the surface-intersection test. We omit portions of the surface that are known to be occluded in all camera views. For each surface-intersecting node, we project a proxy onto each depth buffer (in the node's time window) to identify visible nodes for subsequent refinement. While this optimization fails for scenes with transparency or intricate structure near silhouettes, we find it effective in many scenarios and include it in our experiments.

\subsection{Node Groups and Dual \highlight{h7}{Polyhedron} Search}

For long-range camera sequences, the refined \treename\ contains too many nodes to fit into RAM.
Therefore, we organize the coarse nodes into groups based on their {\bf temporal range}.
In the subsequent steps, bipolar edge search and \highlight{h7}{polyhedron} extraction, coarse nodes are loaded group by group with their refined child nodes.

\begin{figure}[t!]
\captionsetup[subfigure]{justification=centering}
  \centering
    \includegraphics[width=\linewidth]{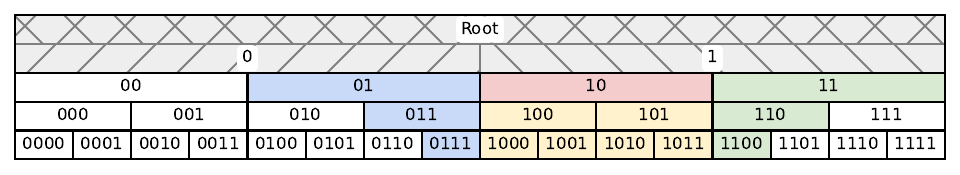}\pulluptwo
   \caption{We group coarse nodes / hypercubes with binary encodings of its temporal range. For a bipolar edge in group ``10'' (in pink), it may have neighboring hypercubes in  colored groups as formulated in Eq.~\ref{eq:group}.}\pullupone
          \label{fig:group}
\end{figure}

We name the groups with \textit{binary encoding}, as shown in \fig{fig:group}.
The group \group\ represented by the root node spans the entire time range.
The left and right children of \group, each spanning half that range, are \group{0} and \group{1}.
In general, the left child of \group{\textbf{s}} is \group{\textbf{s}0}, and the right child is \group{\textbf{s}1} -- by appending either 0 or 1 to string \textbf{s}.

The dual \highlight{h7}{polyhedron} of a bipolar edge has its vertices in the neighboring hypercubes of the edge (\fig{fig:dual}b). 
First, we note which groups neighbor the edge, then we propose efficient algorithms to find bipolar edges and extract dual \highlight{h7}{polyhedra}.

\boldpar{Neighboring Relation of Groups. }
Given a bipolar edge located in a coarse node in \group{\textbf{s}} (the edge may lie on some refined hypercube edges, either on time boundary or in the middle). 
Consider, \eg, $\textbf{s}=``10"$ -- the pink cube in \fig{fig:group}. Its neighboring cubes can be in its children (yellow), or ancestors (gray), or on the left (blue), or on the right (green), or ancestors of the left and right (gray). Denoting the set of groups neighboring  \group{\textbf{s}} as $ \mathfrak{N}_{\textbf{s}} $, we have:
\begin{equation}
\begin{aligned}
\mathfrak{N}_{\textbf{s}} =  
&\{ \groupNodeS \} \cup \subtreeS \cup \ancestorsS \\
\cup \, & \{ \groupNodeSm \} \cup \rightbranchSm \cup \ancestorsSm \\
\cup \, & \{ \groupNodeSp \} \cup \leftbranchSp \cup \ancestorsSp
\end{aligned}
\label{eq:group}
\end{equation}

\noindent where 
\namedfn{right\_branch}(\group{\textbf{s}})
are children of 
\group{\textbf{s}} labeled by 
appending a sequence of 1's to
\textbf{s}; and \namedfn{left\_branch}(\group{\textbf{s}}) by appending 0's.
We use the notation $\textbf{s}-1$ and $\textbf{s}+1$ to mean: convert the encoding to integers, subtract or add one, then convert back to the binary string of the same length (possibly with leading 0's).

\boldpar{Dual \highlight{h7}{Polyhedron} Search. }
We visit all groups in lexicographic order, and find bipolar edges by checking each hypercube edge in the group if it connects different values ($+/-$).
When we find a new edge while visiting group $\group{\textbf{s}}$, we visit the set $\mathfrak{N}_{\textbf{s}} $ to find its neighboring hypercube.
Because of the lexicographic traversal, we only need to visit the subset
\begin{equation}
\tilde{\mathfrak{N}}_{\textbf{s}} = \{\group{\textbf{r}} \in {\mathfrak{N}}_{\textbf{s}} | \textbf{r} \le \textbf{s} \}
\label{eq:subset}
\end{equation}

\noindent because larger ones are visited later.
This omits 
the descendants and the third row in \eqn{eq:group} --
the yellow, the green, and potentially some gray groups. 
This process is summarized in Algorithm~\ref{alg:ext1}.

\begin{figure}[b!]
    \centering
    \includegraphics[width=0.30\linewidth]{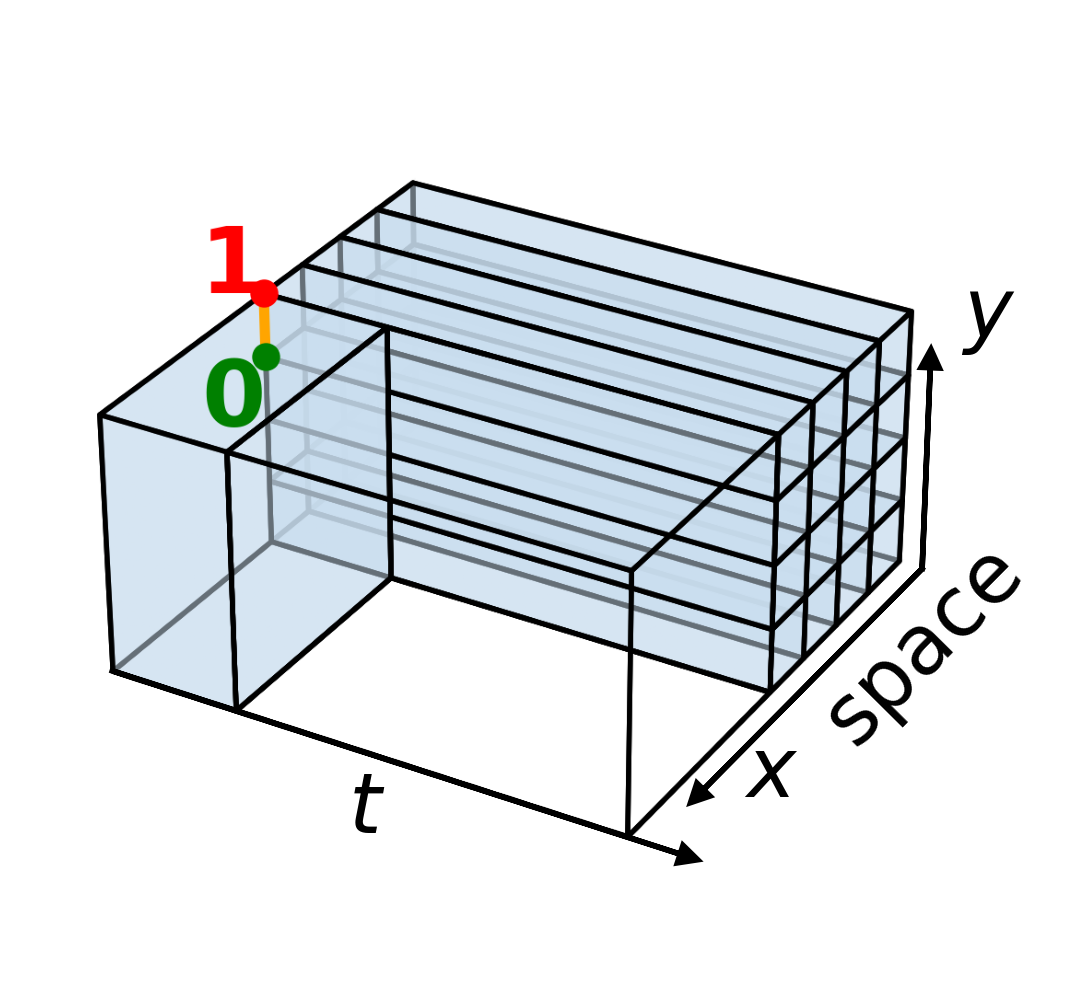}
    \includegraphics[width=0.30\linewidth]{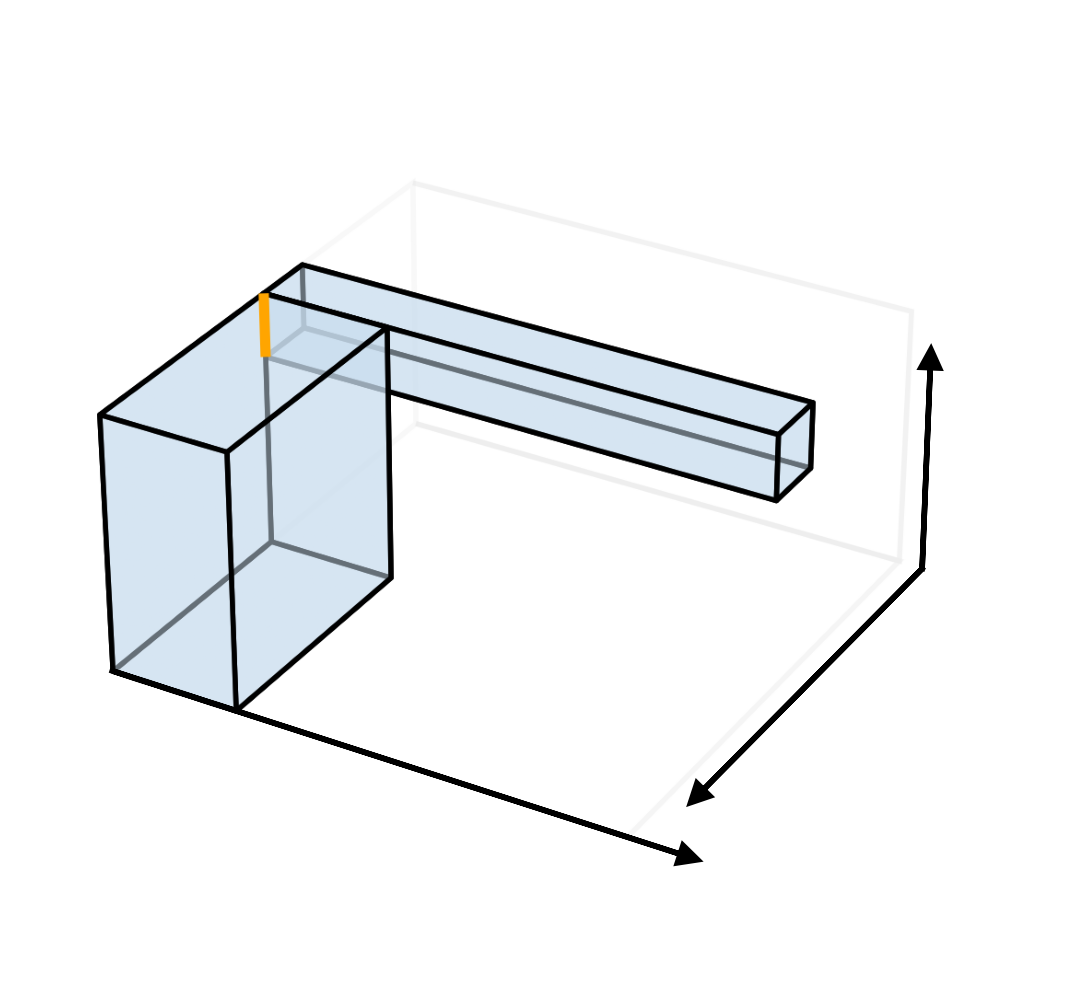}
    \includegraphics[width=0.30\linewidth]{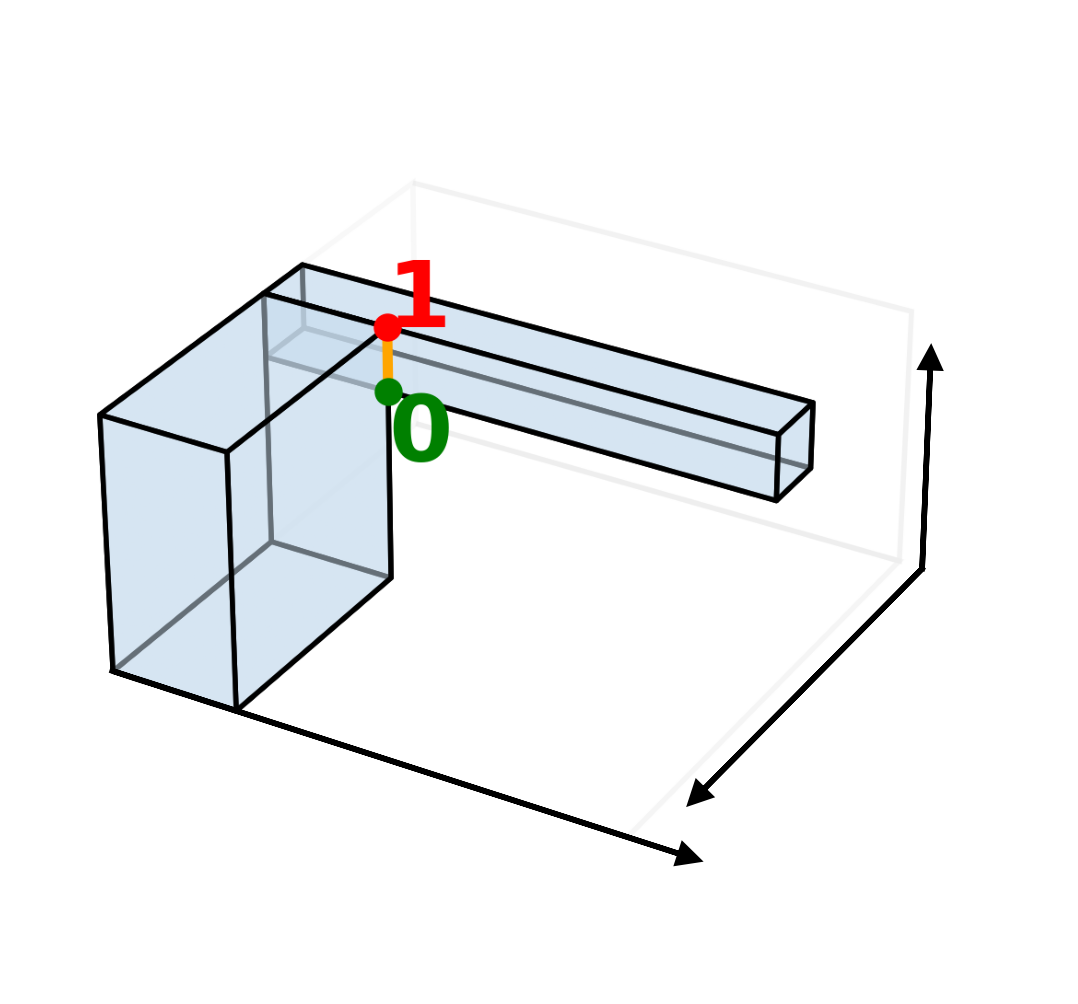}
   \pulluptwo
       \caption{\emph{Left: } A type 1 bipolar edge lie on one of the cube edges; \emph{Middle: }The neighboring nodes of this type 1 edge; \emph{Right: } we offset the type 1 edge based on its neighbors to get a type 2 bipolar edge.}
   \label{fig:bip}
   \pullupone
\end{figure}

\boldpar{Bipolar Edge propagation. }
\fig{fig:bip}-\emph{left} shows a 3D spacetime scenario where a bipolar edge lying on an edge of the narrow cubes (referred to as \emph{type~1}) will be found in line~5.
However, this may overlook bipolar edges of other types: \fig{fig:bip}-\emph{right} shows another bipolar edge not on any cube edge (referred to as \emph{type~2}).
To find bipolar edges of type~2, we need to propagate from an existing type~1 edge $e$ (line~12 in Algorithm~\ref{alg:ext1}).
After we found the neighboring hypercubes of $e$, we offset the time coordinate by the smallest window size of its neighbors and add a new bipolar edge.

\subsection{Memory and Time Complexity}
\label{sec:time_comp}

Similar to OcMesher, the memory footprint of the \methodname\ depends on the complexity of the scene and how fast the camera rotates. Faster camera movement increases temporal division in the tree. \highlight{h2}{However, memory requirements are insensitive to the length of the sequence, because Algorithm}~\ref{alg:ext1} \highlight{h2}{only loads into memory the needed binary-octree node groups (not the entire binary-octree).}

In terms of time complexity, the outer loop in Algorithm~\ref{alg:ext1} considers all binary strings, which is $O(2^d)$ for \emph{temporal tree depth} $d$ (considering only temporal splits in the coarse \treename).
The inner loop loads 
neighbors $\tilde{\mathfrak{N}}_{\textbf{s}}$,
of which there are $O(d)$.
Thus, the algorithm performs loading operations a total of $O(d \cdot 2^d)$ times. 
Moreover, in practice we find it maintains a bounded set of unresolved bipolar edges, making lines 6 and 10 take constant time.
Depth $d = O(\log T)$, where $T$ is the camera sequence duration.
Therefore, the algorithm has overall time complexity $O(T \log T)$ with respect to  duration, and thus scales gracefully to long sequences.

\begin{algorithm}[t]
\SetAlgoNoLine
\KwIn{A \treename{} with temporal tree depth $d$.}
\KwOut{The set of bipolar edges with dual \highlight{h7}{polyhedra}.}

$\namedfn{bipolar\_edges} \gets \{\}$

$\namedfn{all\_strings} \gets \text{lex-ordered list of all binary strings of length } \le d$

\For{\textbf{s} \text{in} \namedfn{all\_strings}  } {
    $\namedfn{load\_group} (\group{\textbf{s}}) $
    
    $\namedfn{new\_edges} \gets \text{bipolar edges at cube edges of \group{\textbf{s}}}$

    $\namedfn{bipolar\_edges} \gets \namedfn{bipolar\_edges} \cup \namedfn{new\_edges} $

    $\text{Compute } \tilde{\mathfrak{N}}_{\textbf{s}}  \text{ as in Eq.~(\ref{eq:group})~(\ref{eq:subset})} $

    \For{$\group{\textbf{r}} \in \tilde{\mathfrak{N}}_{\textbf{s}}$} {
        $\namedfn{load\_group} (\group{\textbf{r}}) $
        
        \For{$e \in\namedfn{bipolar\_edges}$} {
            $\text{Find neighboring nodes of }e\text{ within } \group{\textbf{r}} $

            $\text{Bipolar edge propagation}$ \tcp{see text}
        }
    }
}
\Return {\namedfn{bipolar\_edges} }
\caption{Dual \highlight{h7}{Polyhedron} Search}
\label{alg:ext1}
\pullupone
\end{algorithm}

\section{Discussion, Limitations and Future Work}

\methodname\ addresses the temporal coherence challenge for mesh extraction in procedural scenes with long camera paths by slicing a 4D mesh. It offers better view consistency than baselines at similar cost.
The proposed approach has a number of limitations, some of which suggest areas for future work.

\boldpar{Inputs. }
Our method is limited to predefined camera trajectories \highlight{h3}{and fuzzy regions around them (discussed later)}, thereby excluding interactive applications like games. \highlight{h4}{Nonetheless, offline rendering applications are prevalent in practice, including animation, video production, and synthetic data generation.}

\boldpar{Dynamic Scenes. }
We can add dynamic effects into the static scene through animated displacement maps, like the ocean waves in the \textit{Beach} scene. 
Moreover, since we ultimately render from a polygon mesh, it is also easy to add any animated characters or scene elements independent of the \treename\, like the crabs in the \textit{Beach} scene. (See the supplementary video.)
However, if the occupancy function itself is dynamic, applying LOD transition algorithms via \methodname\ reduces to one octree per frame.

\boldpar{Ray-Marching. }
We extract a polygon mesh for rendering.
In principle, it might be possible ray-march directly into the procedural occupancy function, and thereby obviate the need for mesh extraction.
But ray marching algorithms are typically constrained to a finite volume~\cite{aaltonen2018gpu, crassin2009gigavoxels, gobbetti2008single, wald2016ospray} or dependent on distance values~\cite{galin2020segment, soderlund2022ray, hart1996sphere, musgrave2002qaeb, seyb2019non}, and adjacent pixels can produce inconsistent results.
Nevertheless, one could imagine a hybrid approach that might use ray marching, but leveraging a spatio-temporal tree akin to the \TreeName, to allow for varying step sizes.

\boldpar{\highlight{h3}{Extension to Fuzzy Camera Paths. }}
\highlight{h3}{Our algorithm naturally extends to fuzzy camera paths because it takes a list of keyframe cameras with timestamps as input. Thus, instead of one camera for each keyframe, we could sample cameras covering the fuzzy region for each keyframe. Then the algorithm itself would remain unchanged, with runtime depending on the size of the fuzzy region rather than the number of sampled cameras due to GPU parallelization. In this way, a suitably high-resolution mesh could be provided so long as the actual camera path stays near the fuzzy path. Rendering from a camera path outside this region would result in graceful degradation in quality as the camera moves further away. }

\begin{figure}[t!]
    \centering
    \includegraphics[width=\linewidth]{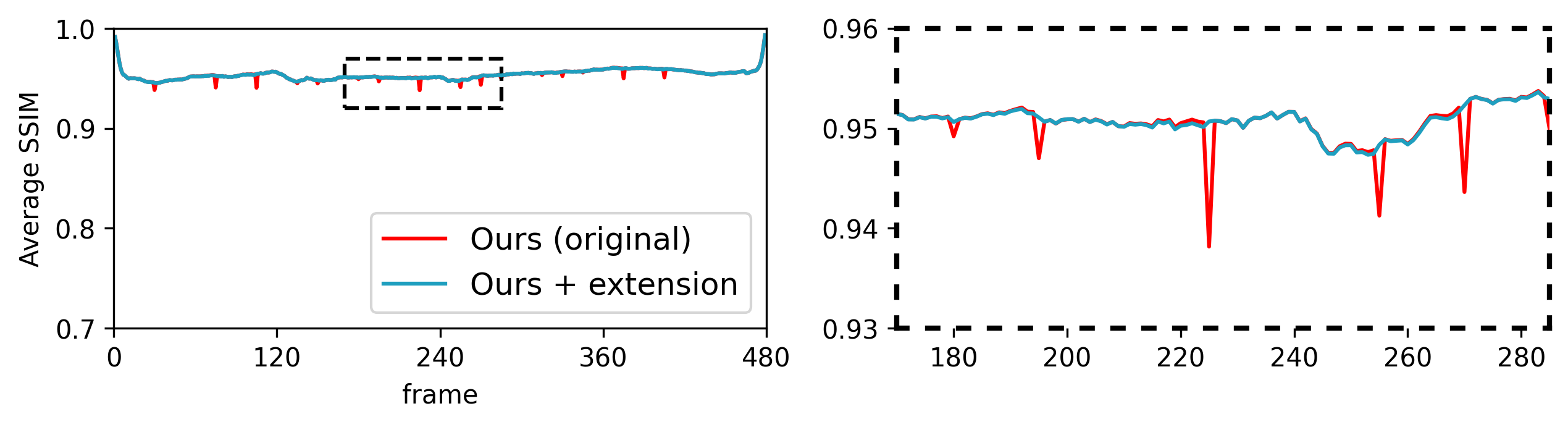}\pulluptwo
    \caption{\highlight{h5}{An extension to our algorithm ameliorates minor popping artifacts in our method (small valleys in plot). Dashed box enlarged on right.}}\pullupone
    \label{fig:fur}
\end{figure}

\boldpar{\highlight{h5}{Extension to Ameliorate Popping Artifacts. }} \highlight{h5}{The minor popping artifacts of our method revealed as small valleys
in the SSIM plots
are caused by new polygonal structures emerging discontinuously as zero-volume, double-sided slices. These structures come from faces in the polyhedra that are perpendicular to the time axis. To ameliorate such popping, we extend the algorithm to extrude these time-orthogonal faces into pyramidal volumes.} Fig.~\ref{fig:fur} \highlight{h5}{shows the effectiveness of this extension in the \textit{Forest} scene.}

\begin{acks}

This work was partially supported by the National Science Foundation under Award 2450506.
Icons from Noun Project (CC BY 3.0): ``Video Camera'' by JS; ``Flag'' by Mike Zuidgeest.

\end{acks}

\begin{figure*}[!htbp]

\begin{minipage}{0.49\textwidth}
{
\centering
\includegraphics[width=\linewidth]{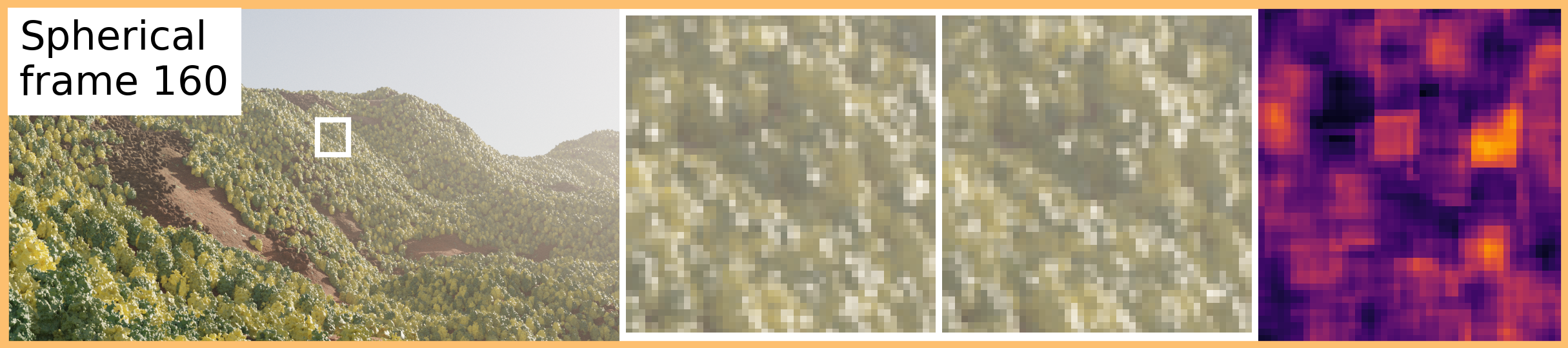}
\includegraphics[width=\linewidth]{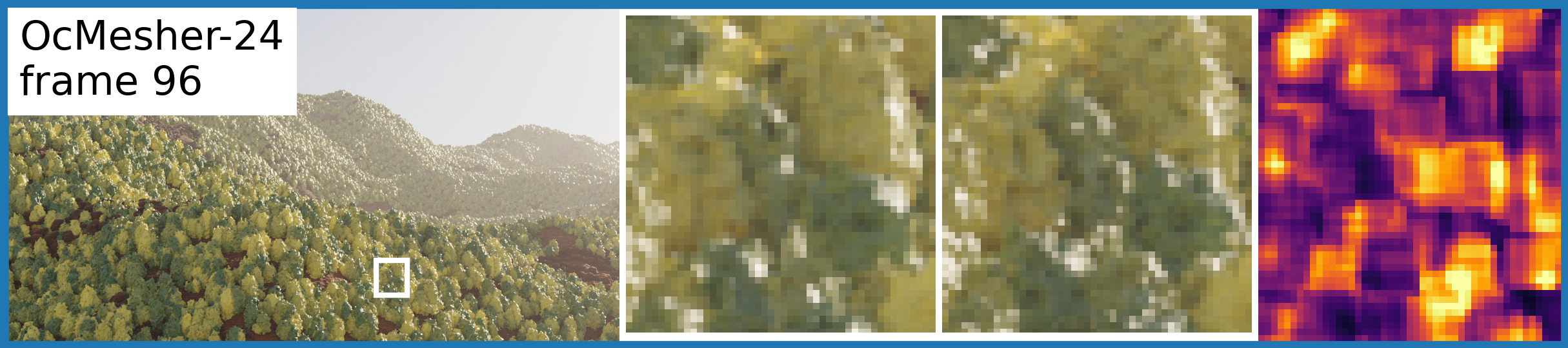}
\includegraphics[width=\linewidth]{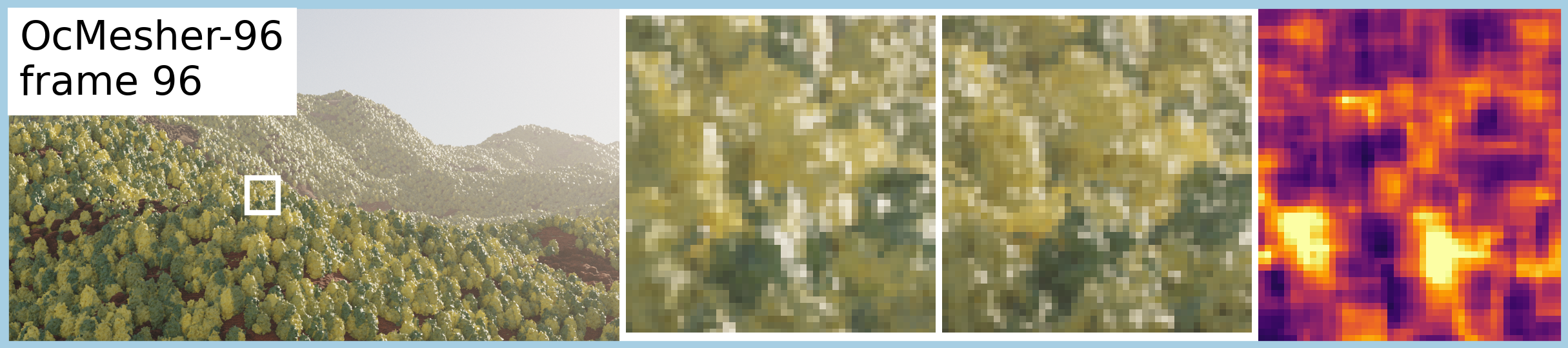}
\includegraphics[width=\linewidth]{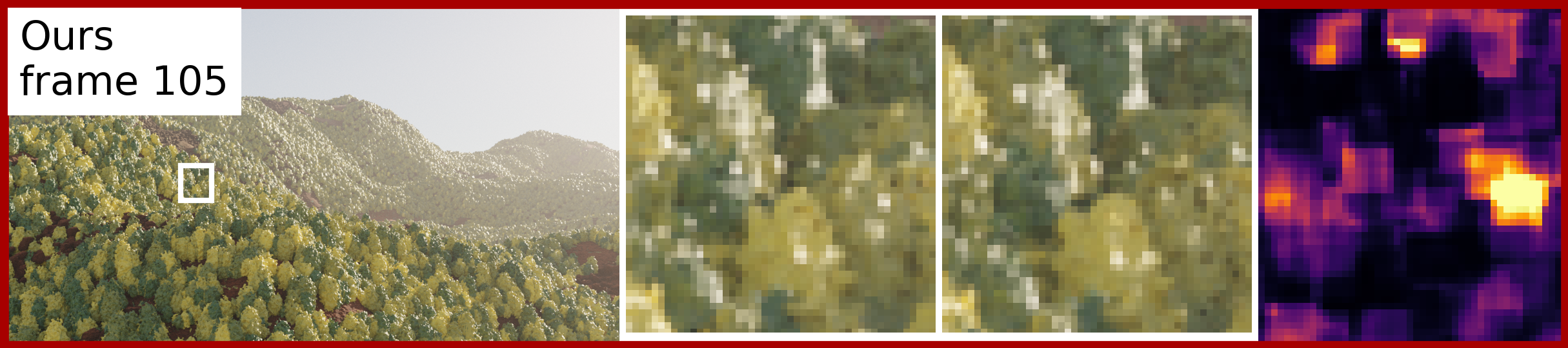}
}
\includegraphics[width=\textwidth]{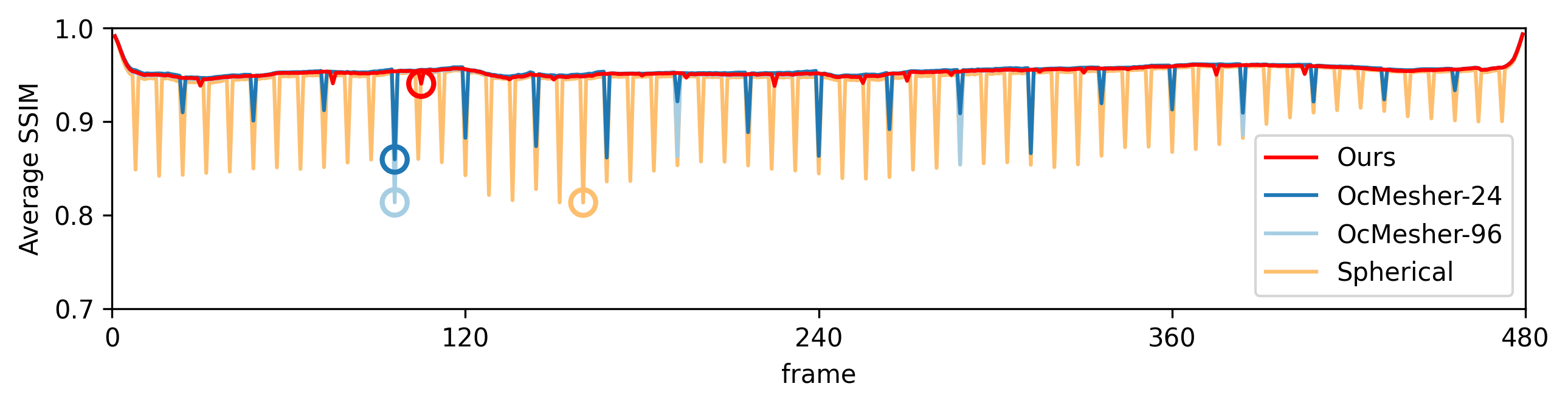}
{
\centering
\includegraphics[width=\linewidth]{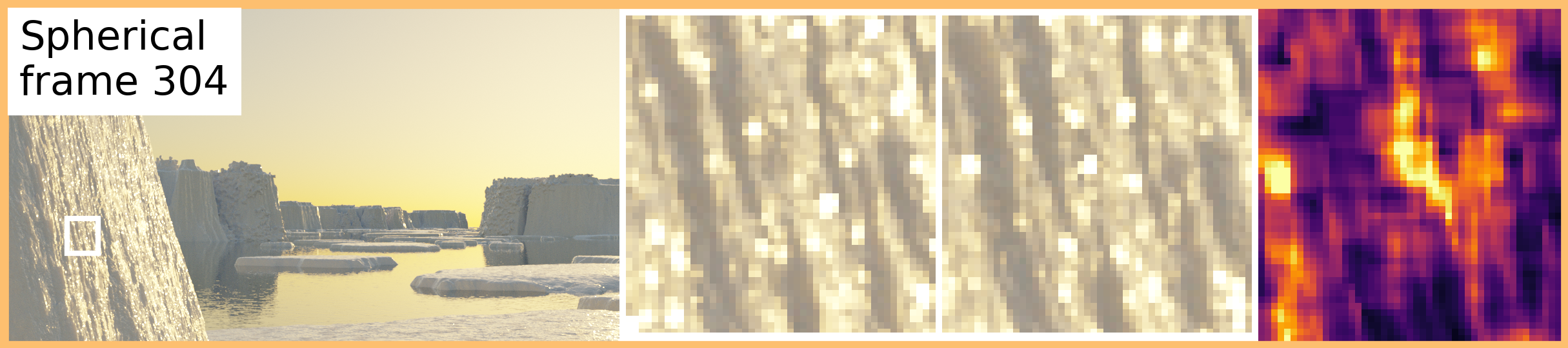}
\includegraphics[width=\linewidth]{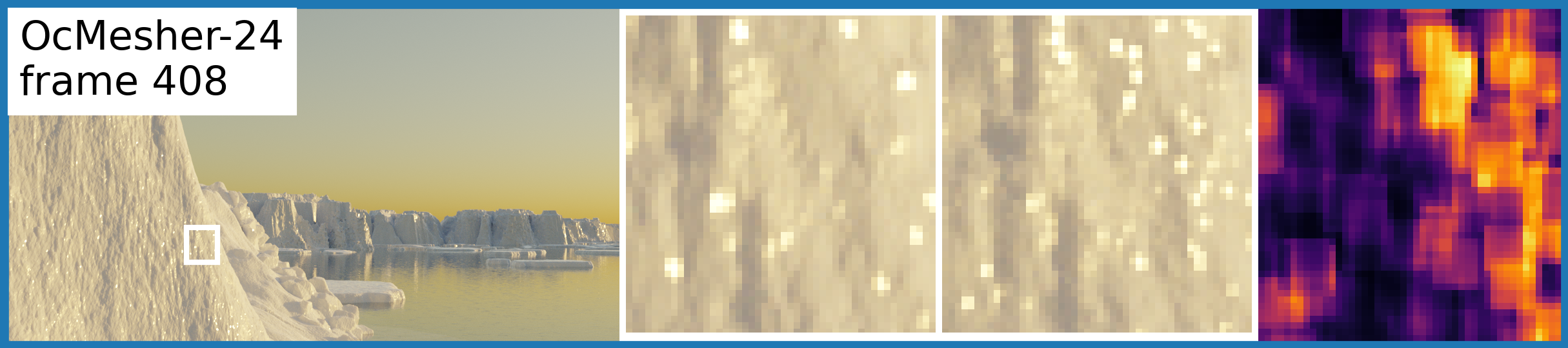}
\includegraphics[width=\linewidth]{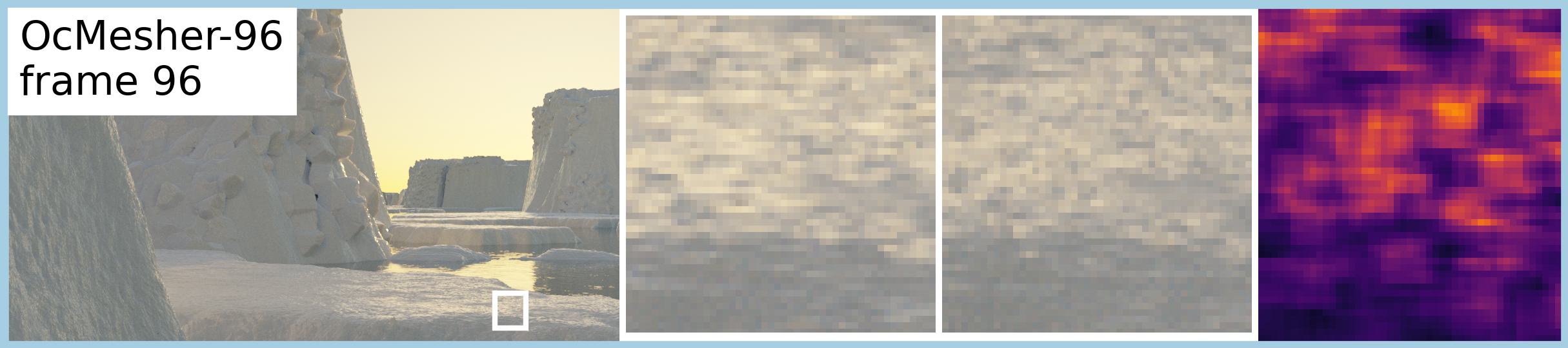}
\includegraphics[width=\linewidth]{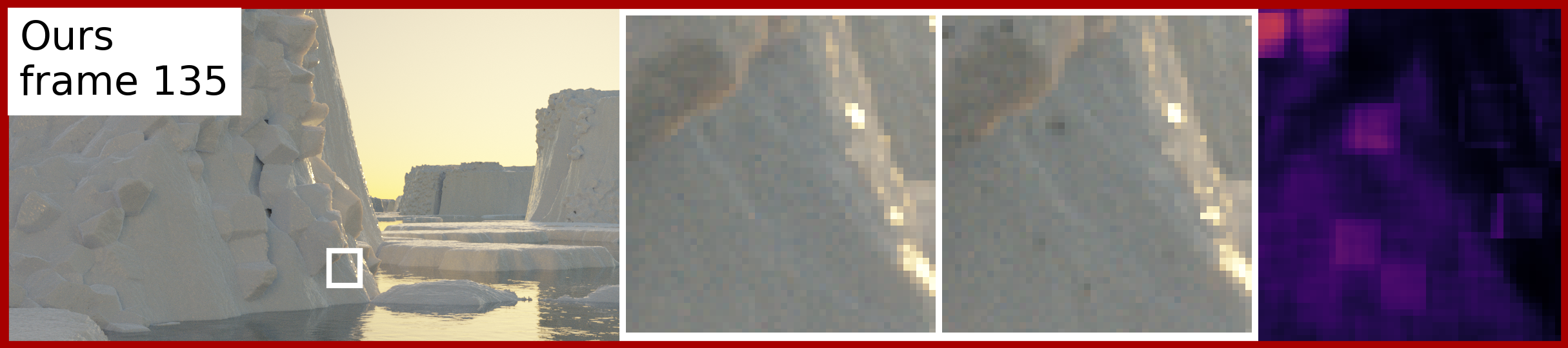}
}
\includegraphics[width=\textwidth]{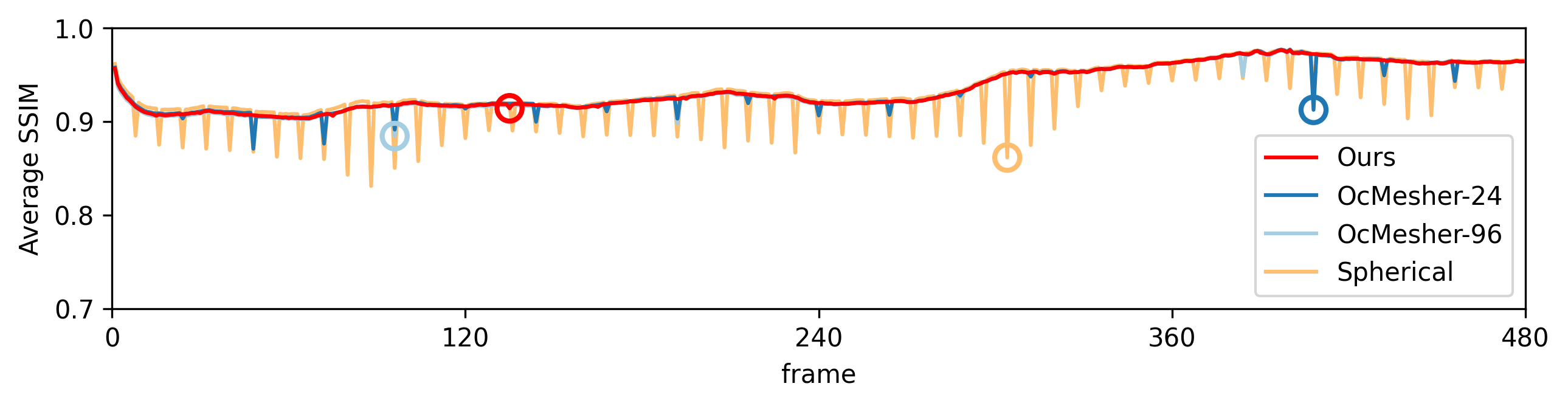}
\end{minipage} \hfill \begin{minipage}{0.49\textwidth}
{
\centering
\includegraphics[width=\linewidth]{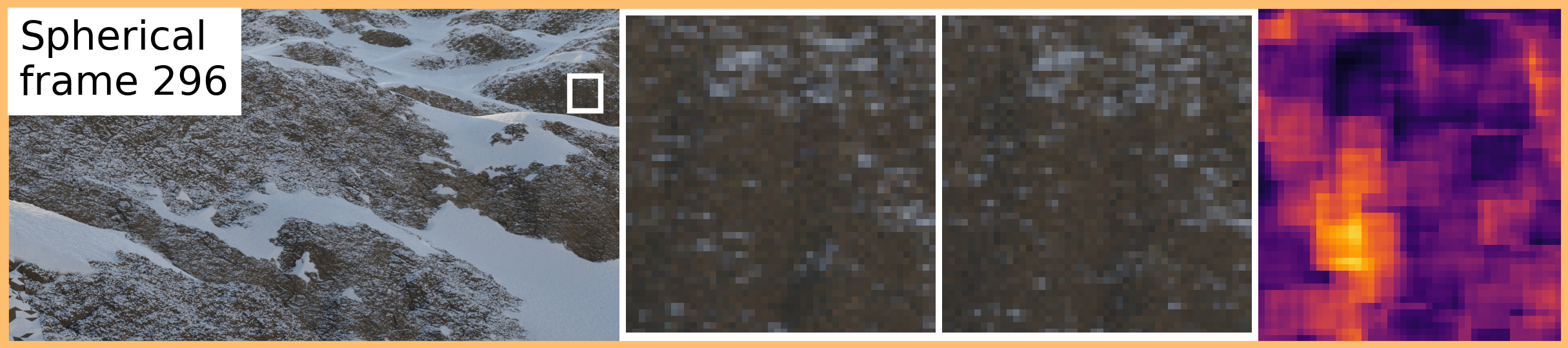}
\includegraphics[width=\linewidth]{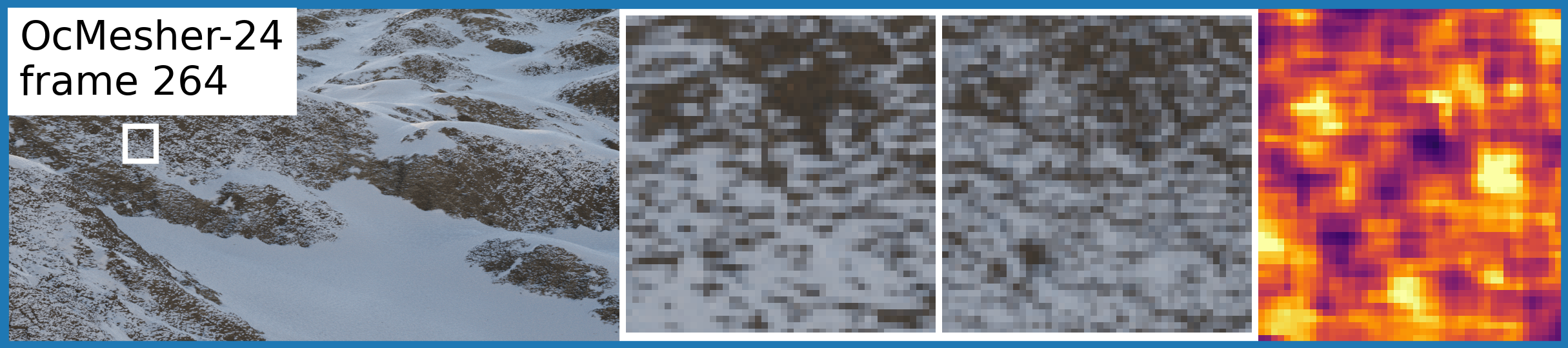}
\includegraphics[width=\linewidth]{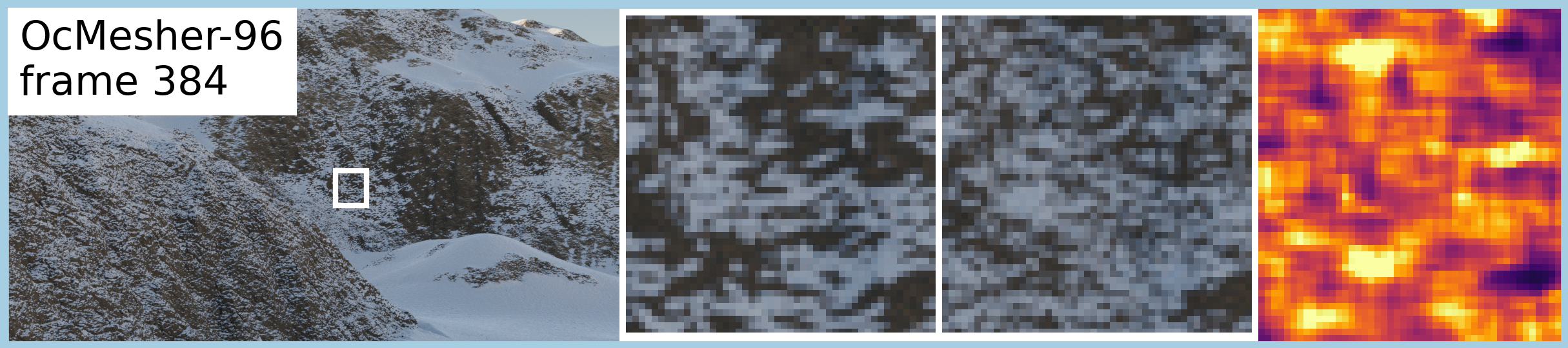}
\includegraphics[width=\linewidth]{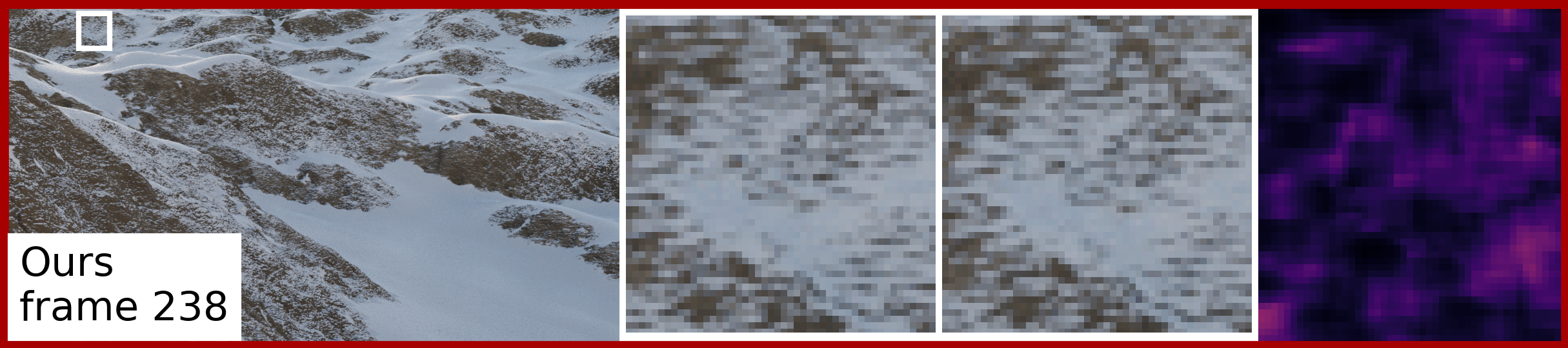}
}
\includegraphics[width=\textwidth]{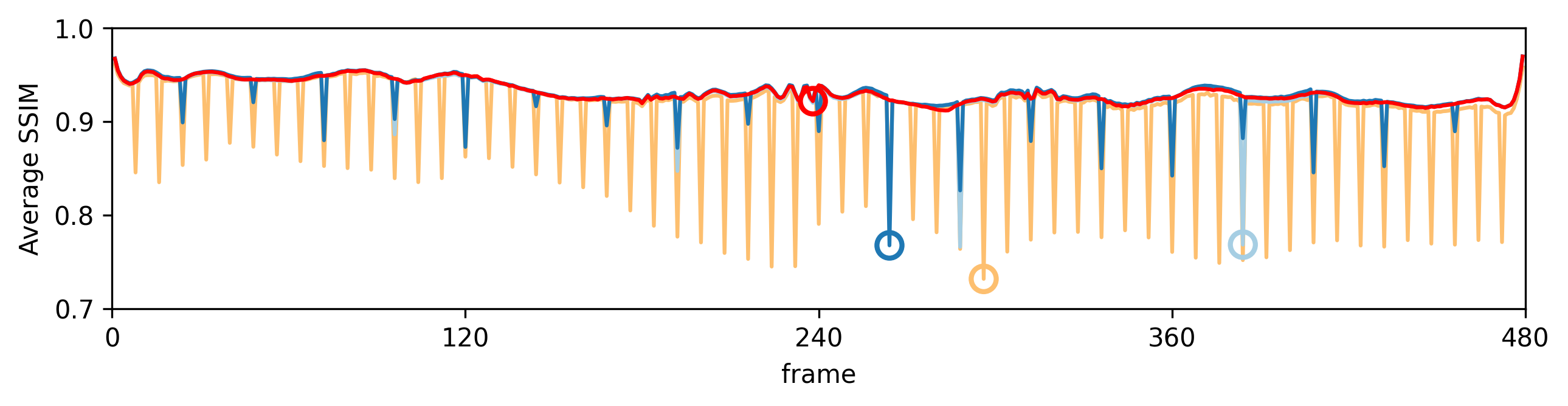}
{
\centering
\includegraphics[width=\linewidth]{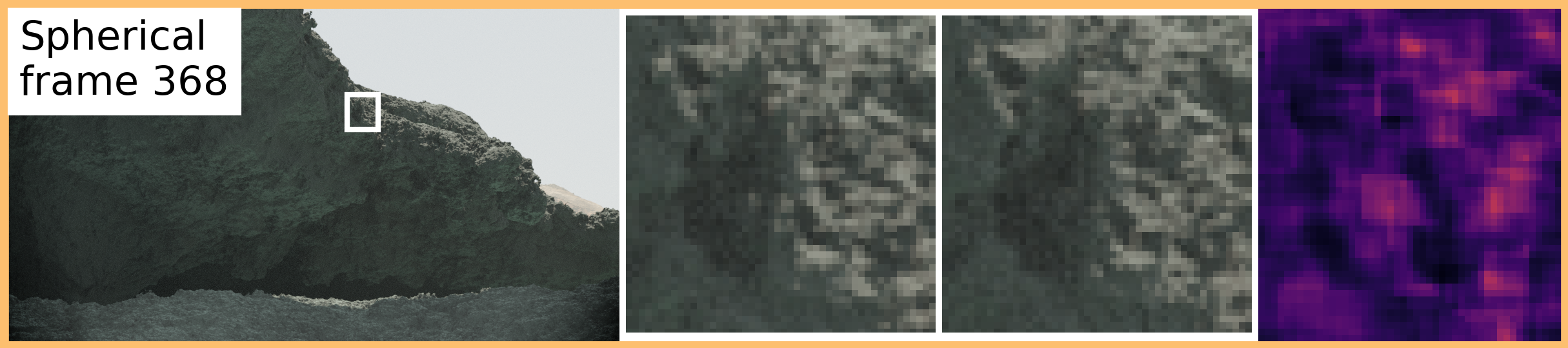}
\includegraphics[width=\linewidth]{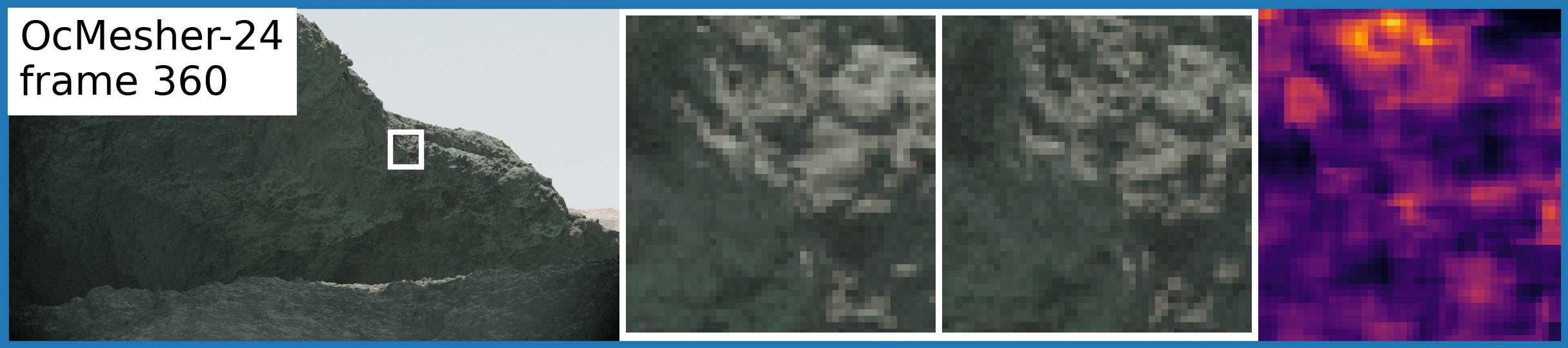}
\includegraphics[width=\linewidth]{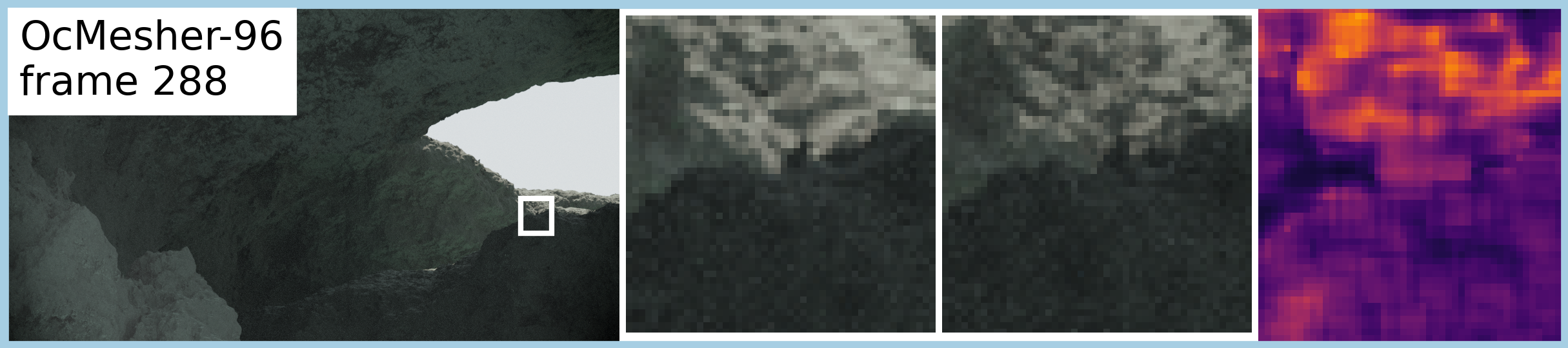}
\includegraphics[width=\linewidth]{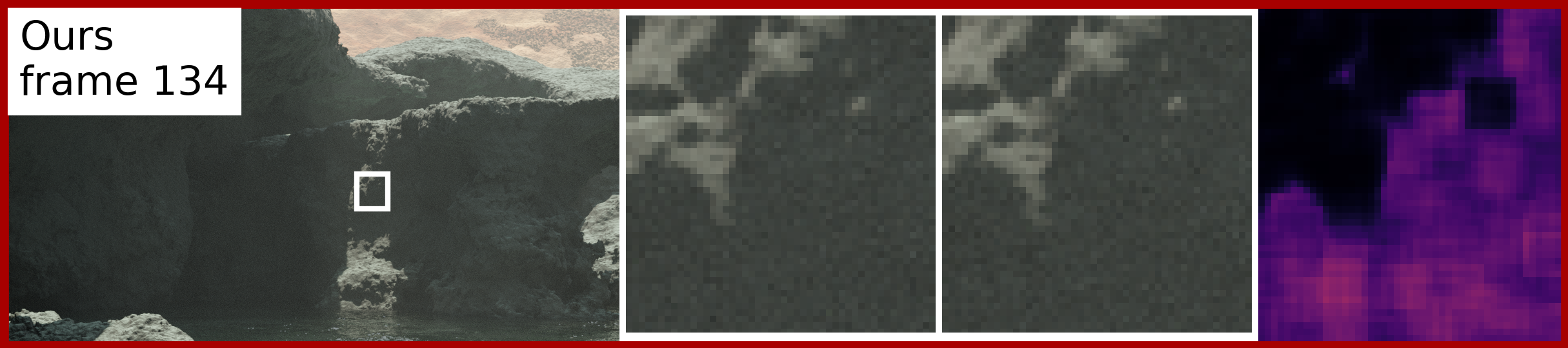}
}
\includegraphics[width=\textwidth]{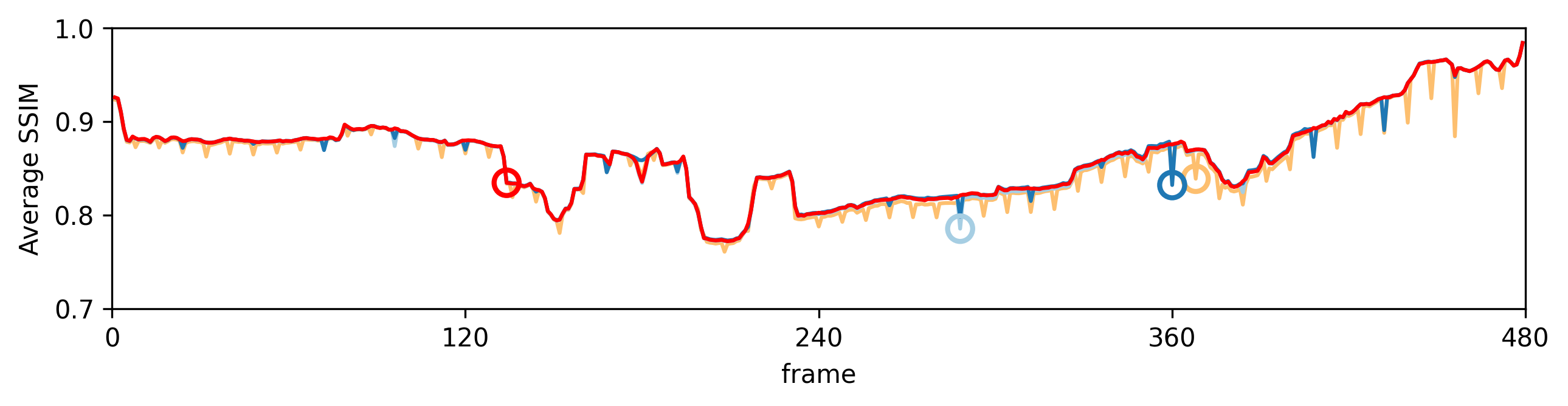}
\end{minipage}
\caption{The \emph{worst} frames by SSIM for each method in the \textit{Forest}, \textit{Mountain}, \textit{Arctic}, and \textit{Cave} scenes. 
See \sect{ssec:ssim} and \fig{fig:ssim} for explanation of figure components.
Note that insufficient sampling in dark regions produces considerable visual noise in the \textit{Cave} scene, which contributes to degraded SSIM scores for all methods in parts of this scene. While the advantage of our method is not as pronounced for \textit{Cave} as for other scenes, it remains apparent in the plot.
}
\label{fig:extra1}
\end{figure*}

\begin{figure*}[!htbp]
\begin{minipage}{0.47\textwidth}
{
\centering
\includegraphics[width=\linewidth]{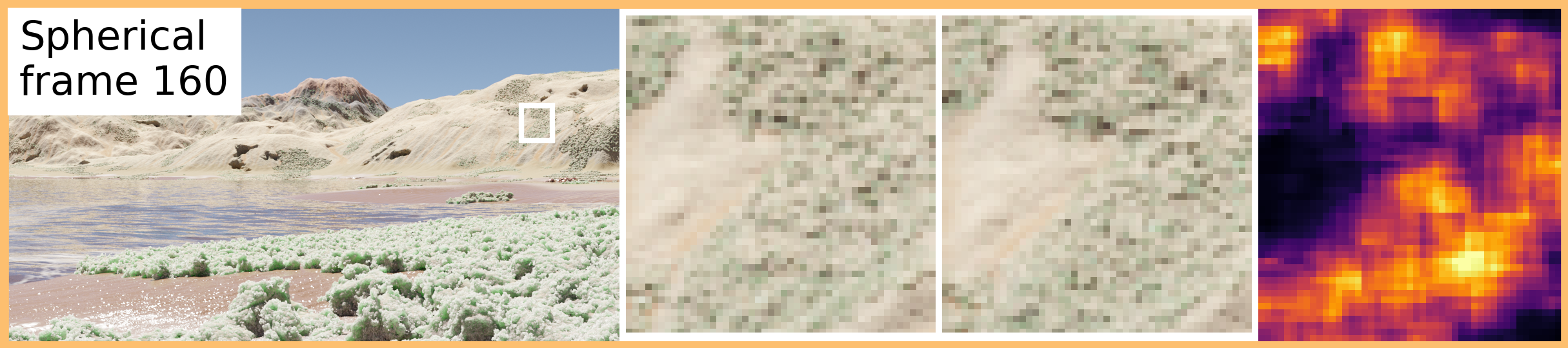}
\includegraphics[width=\linewidth]{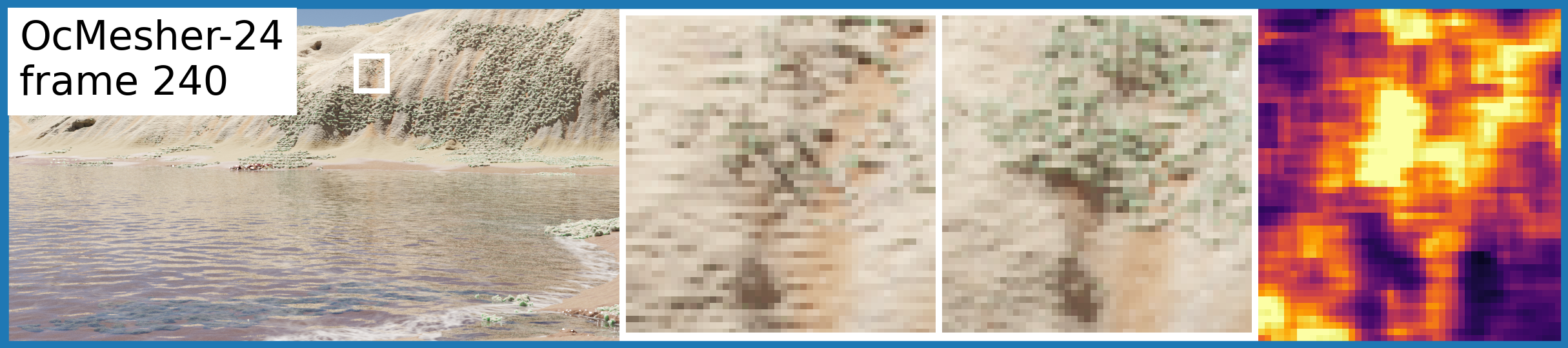}
\includegraphics[width=\linewidth]{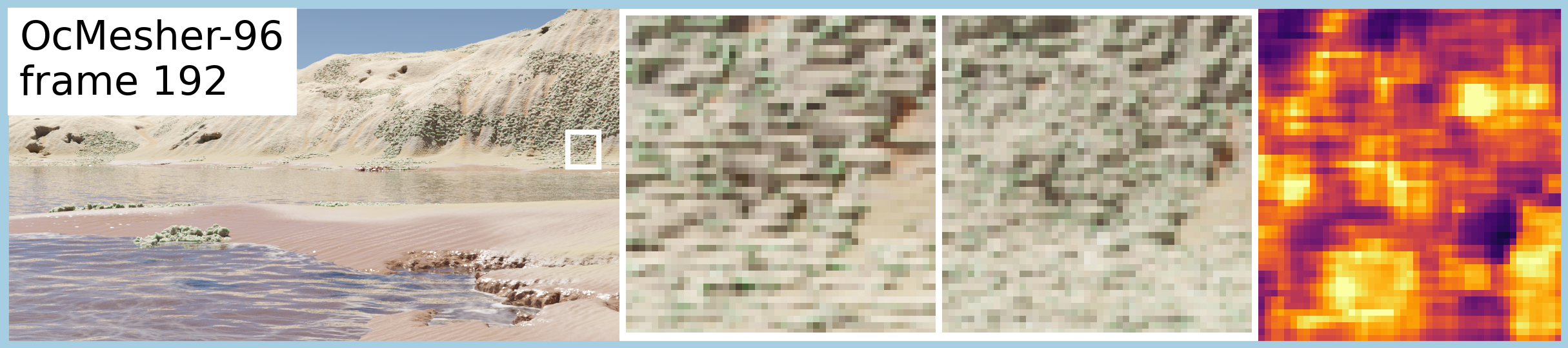}
\includegraphics[width=\linewidth]{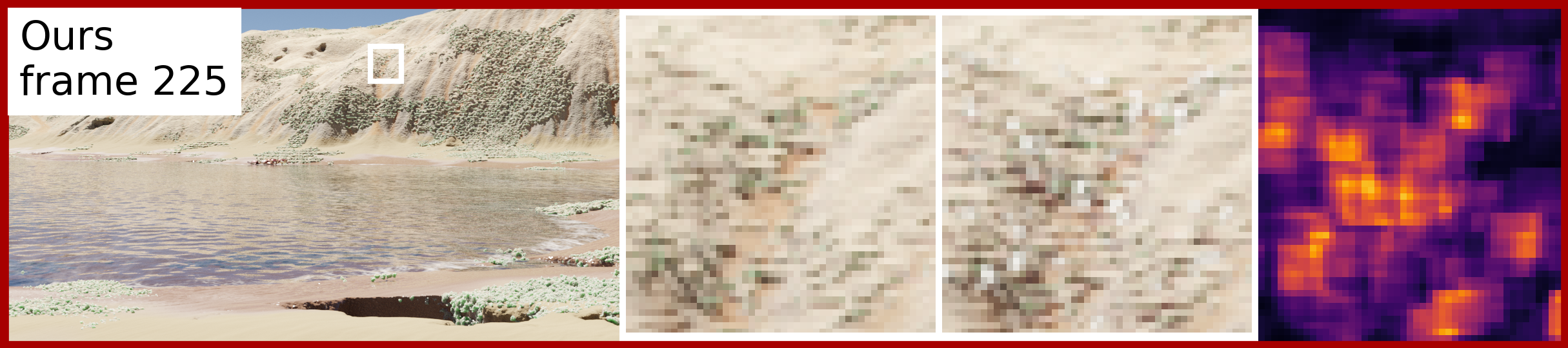}
}
\includegraphics[width=\textwidth]{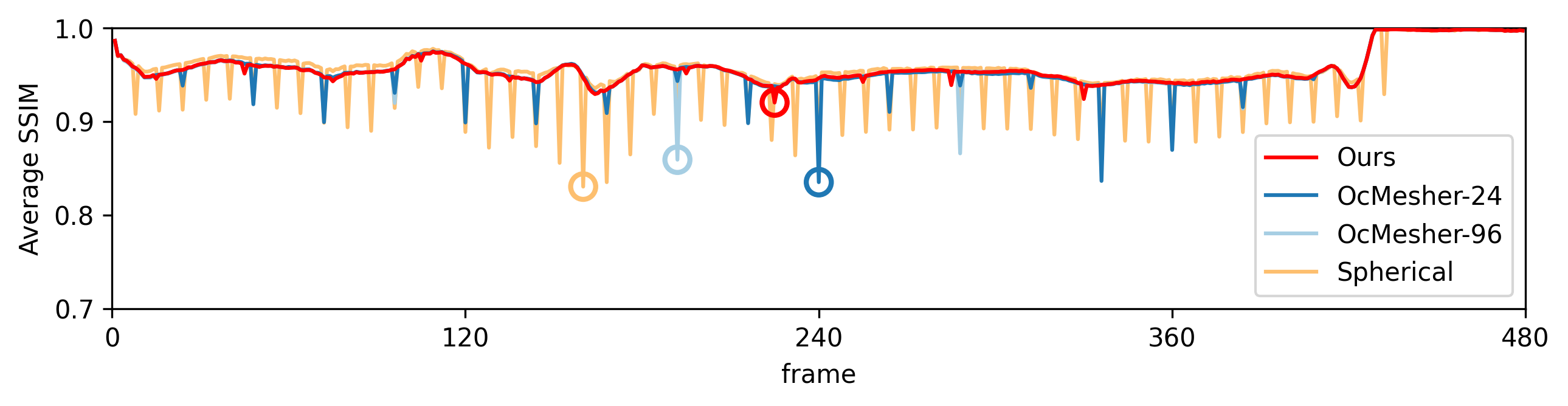}
\end{minipage} \hfill \begin{minipage}{0.47\textwidth}
{
\centering
\includegraphics[width=\linewidth]{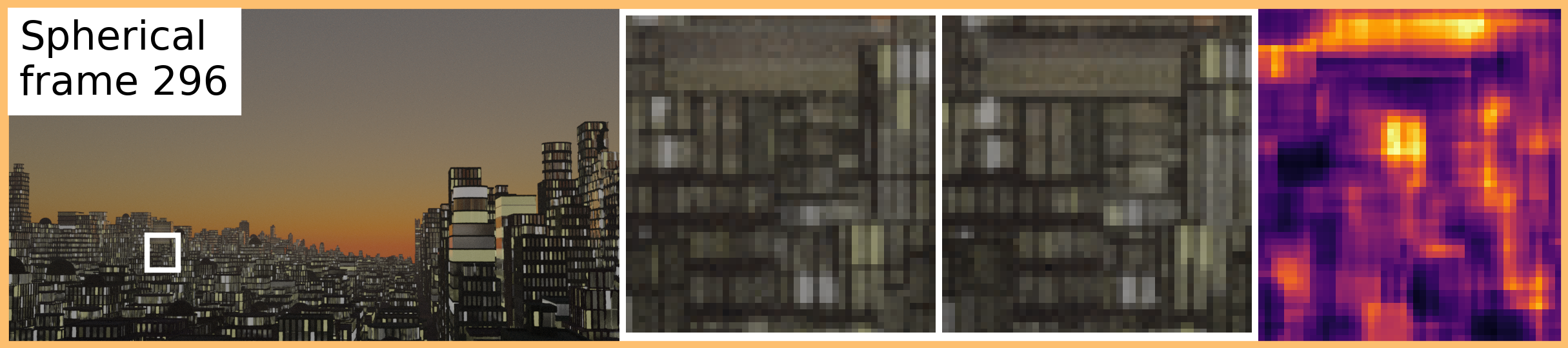}
\includegraphics[width=\linewidth]{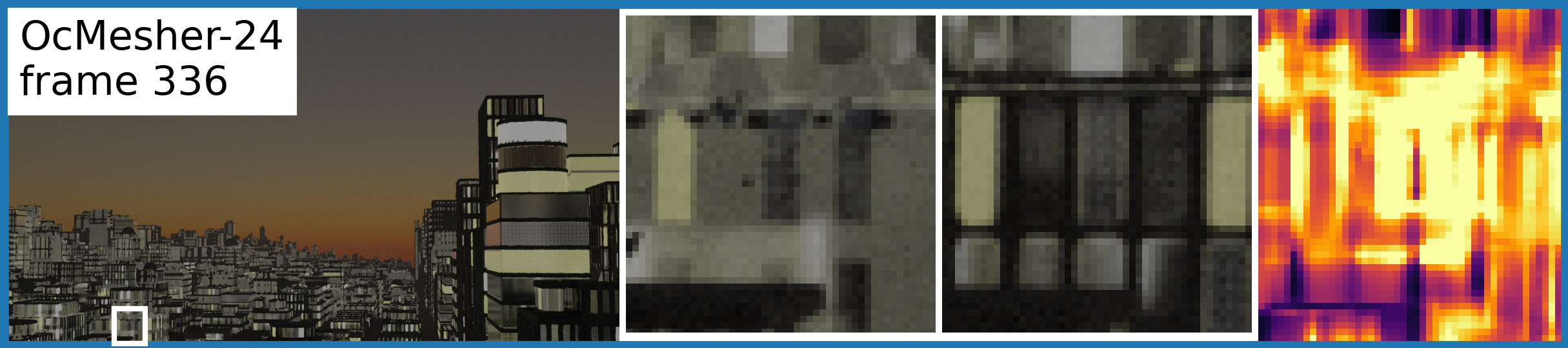}
\includegraphics[width=\linewidth]{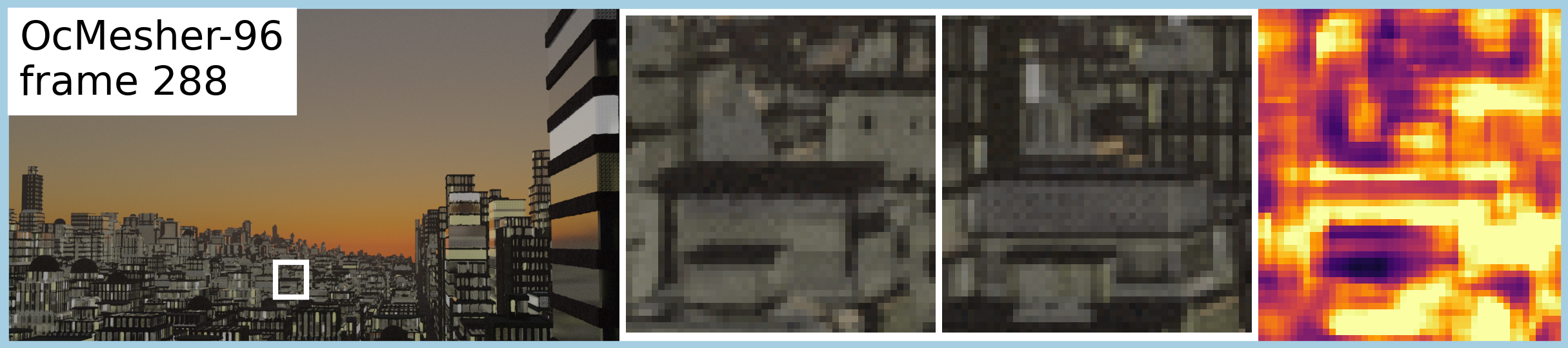}
\includegraphics[width=\linewidth]{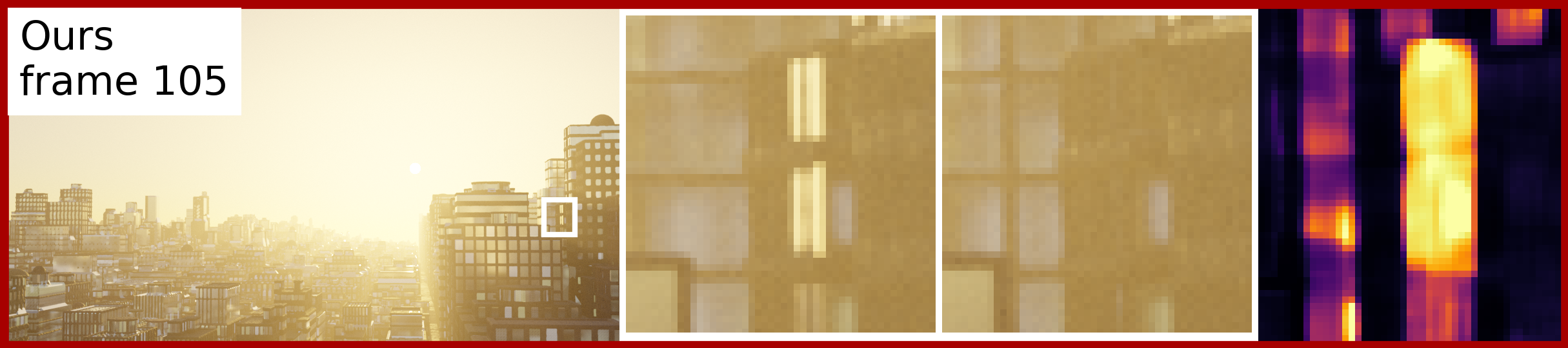}
}
\includegraphics[width=\textwidth]{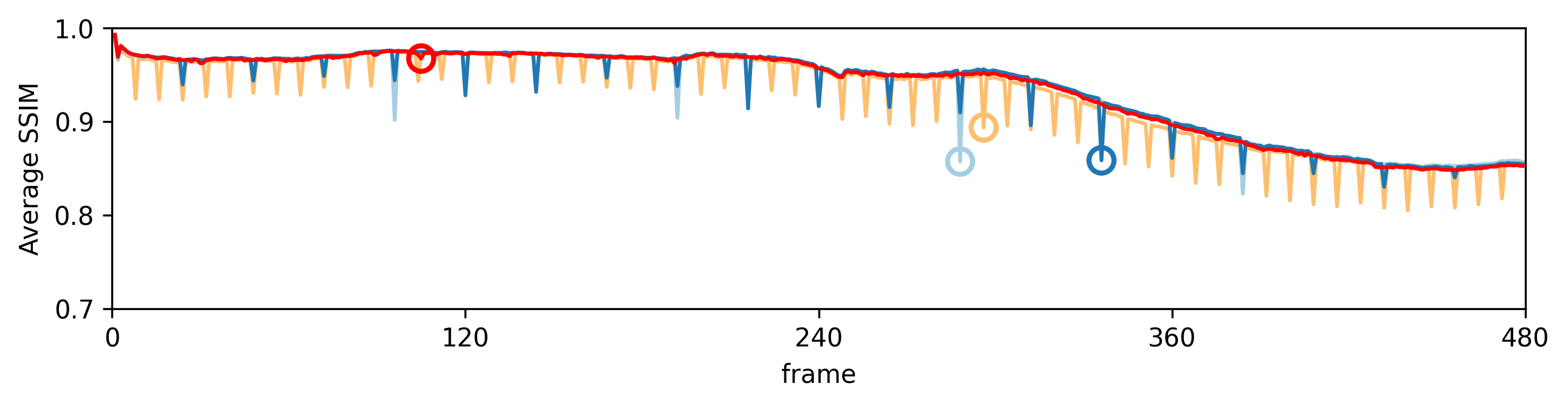}
\end{minipage}
\caption{
The \emph{worst} frames by SSIM for each method in the \textit{Beach} and \textit{City} scenes. 
See \sect{ssec:ssim} and \fig{fig:ssim} for explanation of figure components.
Note that the ocean waves are added as animated displacement and texture over the static water surface (see the supplementary video for the animated scene). Urban scenes are also modeled using occupancy functions. In these two challenging scenes, while the advantage of our method is not pronounced in the zoomed-in window, it remains apparent in the plot.
}
\label{fig:extra2}
\end{figure*}

\begin{figure*}[!htbp]
\begin{minipage}{0.47\textwidth}
{
\centering
\includegraphics[width=\linewidth]{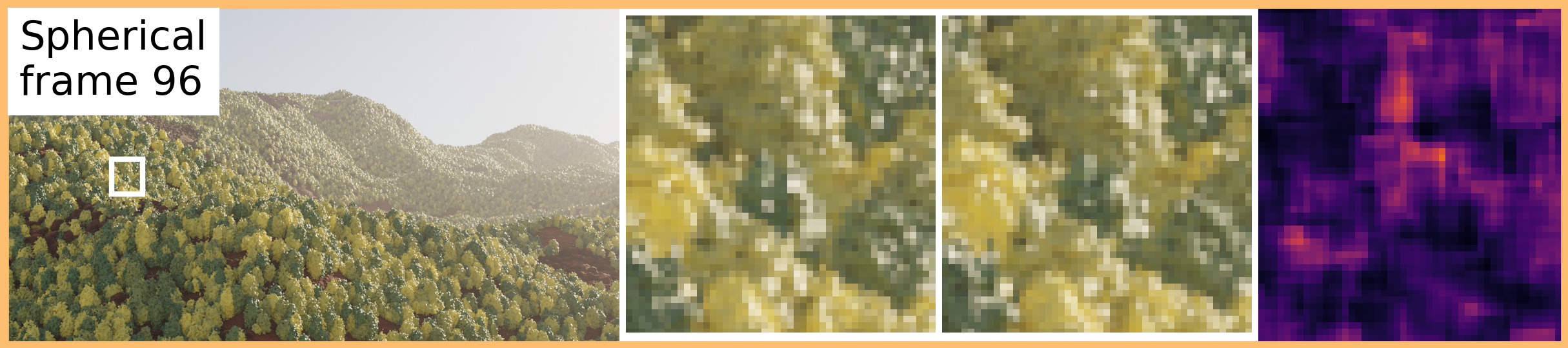}
\includegraphics[width=\linewidth]{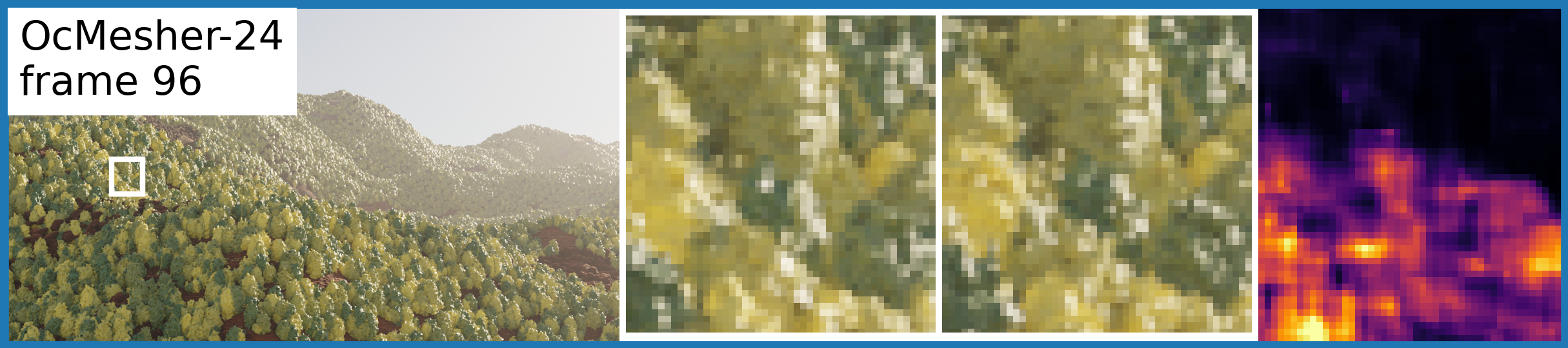}
\includegraphics[width=\linewidth]{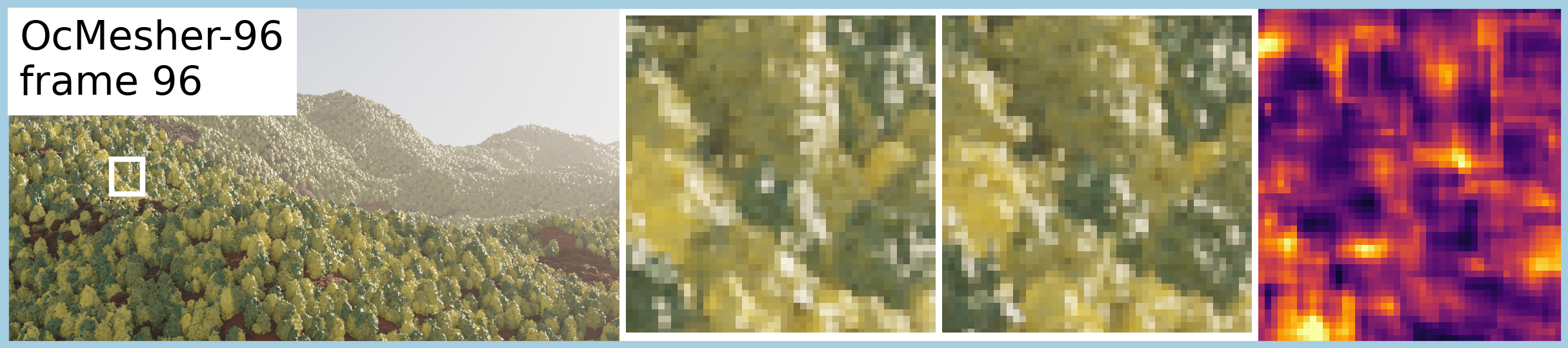}
\includegraphics[width=\linewidth]{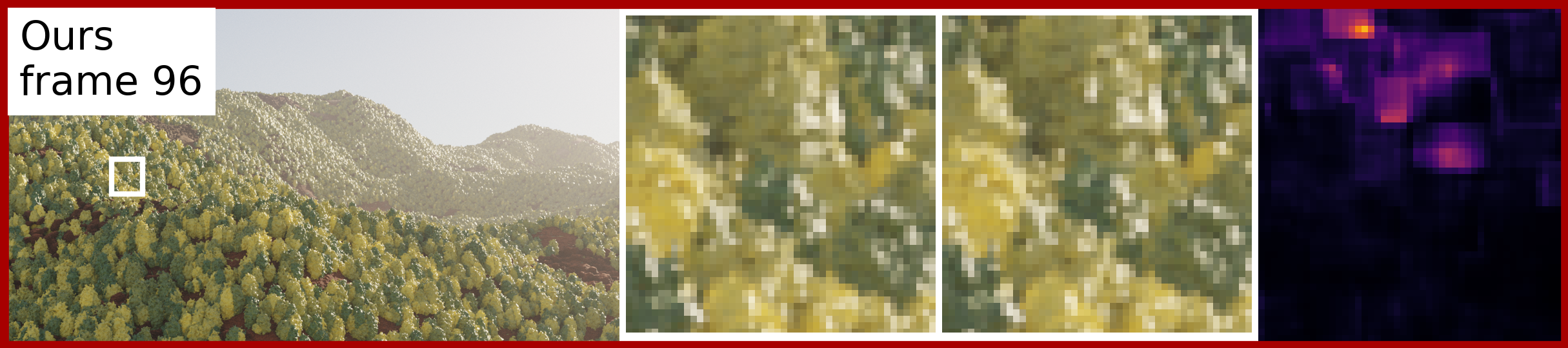}
}
\end{minipage} \hfill \begin{minipage}{0.47\textwidth}
{
\centering
\includegraphics[width=\linewidth]{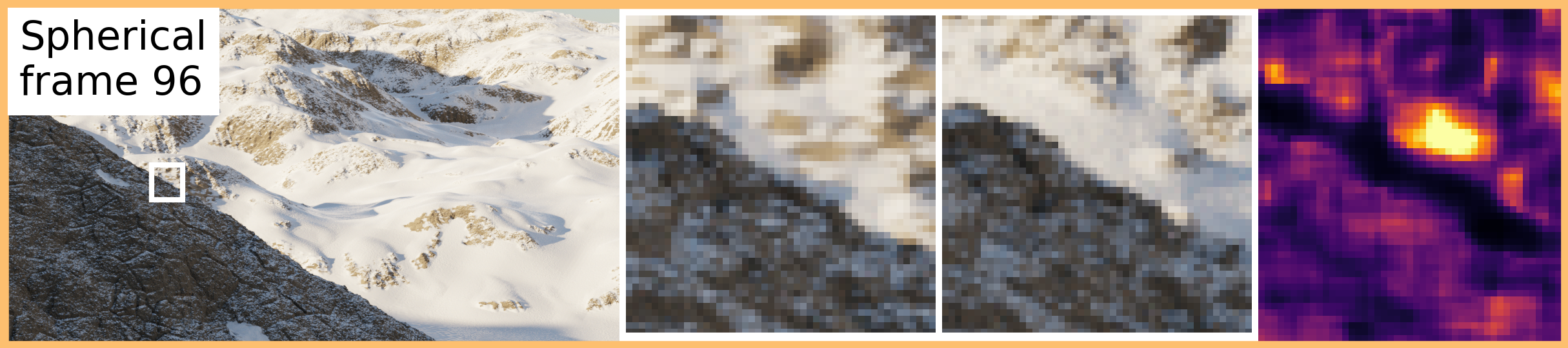}
\includegraphics[width=\linewidth]{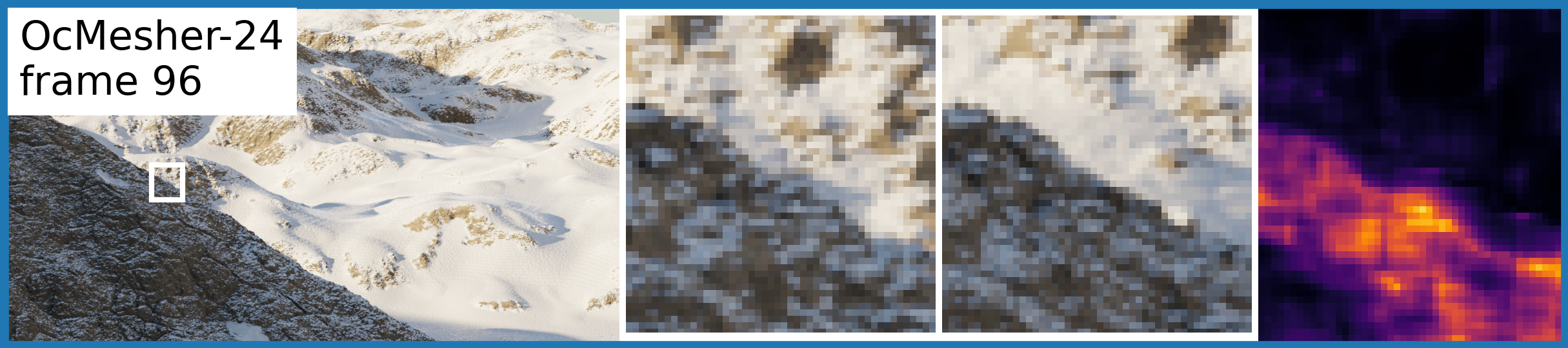}
\includegraphics[width=\linewidth]{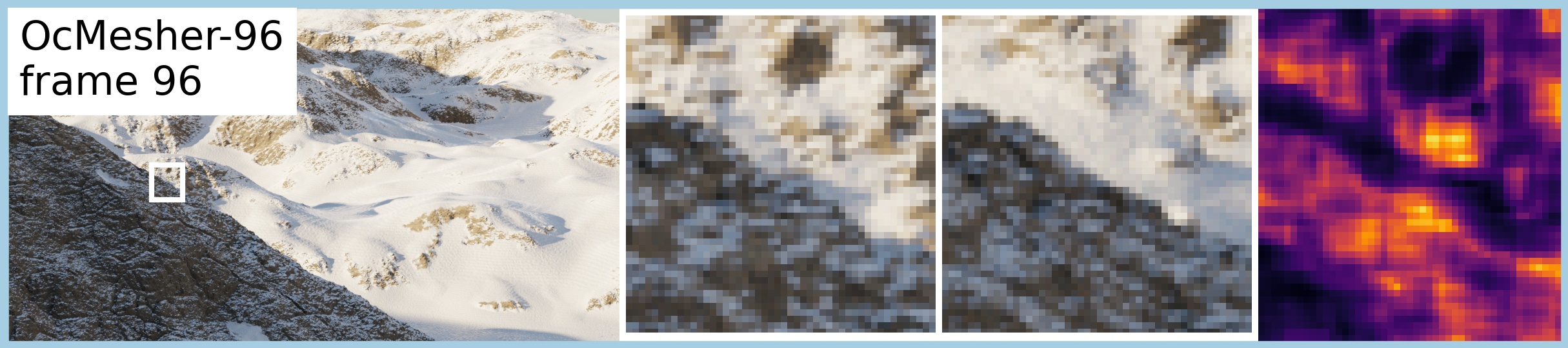}
\includegraphics[width=\linewidth]{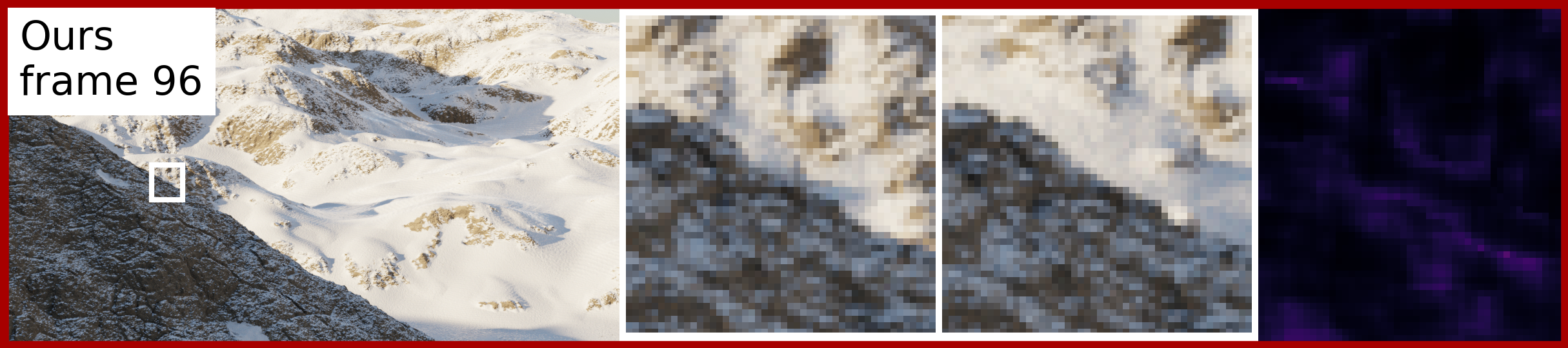}
}
\end{minipage}
\caption{
For a more direct comparison, we align the time splitting (temporal discontinuities) in our method with the others, and compare the same zoomed-in window at a specific frame in the \textit{Forest} and \textit{Mountain} scenes. This may not be ``fair'' because the choice of frame or zoom can always favor one method. This is the 96th frame, when all four methods suffer from temporal discontinuities, and the zoom is chosen to show differences between the OcMesher versions.}
\label{fig:sameframe}
\end{figure*}

\newpage
\bibliographystyle{ACM-Reference-Format}
\bibliography{bibliography}

\end{document}